\documentclass{aa}  

\usepackage{graphicx}
\usepackage{empheq}
\usepackage{amsmath}    
\usepackage{amssymb}    
\usepackage{bm}
\usepackage{nccmath}
\usepackage{varioref}
\usepackage{hyperref}
\hypersetup{
    colorlinks=true,                         
    linkcolor=blue,
    citecolor=red,
    urlcolor=blue  }
    
\usepackage[capitalize]{cleveref}
\crefname{section}{Sect.}{Sects.}
\Crefname{section}{Section}{Sections}

\usepackage{import}
\usepackage{transparent}

\usepackage{svg}
\newcommand{\orcid}[1]{\href{https://orcid.org/#1}{\includegraphics[width=10pt]{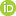}}}

\usepackage{txfonts}
\newcommand{\dd}{\mathrm{d}}

\begin{document} 

  \title{The RayGalGroupSims cosmological simulation suite for the study of relativistic effects: An application to lensing-matter clustering statistics}

  \titlerunning{The RayGalGroupSims simulation suite}

 \author{Y. Rasera\orcid{0000-0003-3424-6941} \inst{1} \thanks{\email{yann.rasera@obspm.fr}} \and  M-A. Breton\orcid{0000-0003-4391-8869} \inst{2} \thanks{\email{michel-andres.breton@lam.fr}} \and  P-S. Corasaniti\orcid{0000-0002-6386-7846}  \inst{3,4} \and J. Allingham\orcid{0000-0003-2718-8640} \inst{5} \and  F. Roy \inst{3} \and  V. Reverdy\orcid{0000-0002-5781-4107} \inst{6,3,7} \and  T.Pellegrin \inst{8,9,3} \and   S. Saga\orcid{0000-0002-7387-7570} \inst{3} \and A. Taruya \orcid{0000-0002-4016-1955} \inst{10} \and S. Agarwal\orcid{0000-0002-5356-1253} \inst{11} \and S. Anselmi \orcid{0000-0002-3579-9583} \inst{12,3}  }

\authorrunning{RayGal team}

   \institute{Laboratoire Univers et Théories, Université de Paris, Observatoire de Paris, Université PSL, CNRS, F-92190 Meudon, France
              \and
              Aix Marseille Univ, CNRS, CNES, LAM, Marseille, France 
              \and
              Laboratoire Univers et Théories, Observatoire de Paris, Université PSL, Université de Paris, CNRS, F-92190 Meudon, France
              \and
              Sorbonne Universit\'e, CNRS, UMR 7095, Institut d'Astrophysique de Paris, 98 bis bd Arago, 75014 Paris, France
              \and
              Sydney Institute for Astronomy, School of Physics, A28, The University of Sydney, NSW 2006, Australia
              \and
            Department of Mathematics (DMA), École Normale Supérieure, 45, rue d'Ulm, 75005 Paris, France
              \and
              National Center for Supercomputing Applications, University of Illinois at Urbana-Champaign, Illinois, USA
              \and
             LESIA, Observatoire de Paris, Université PSL, CNRS, Sorbonne Université, Université de Paris, France 
             \and
             Laboratoire de Physique des Plasmas (LPP), École Polytechnique, IP Paris, Sorbonne Université, CNRS, Observatoire de Paris, Université PSL, Université Paris Saclay, Paris, France
             \and
             Center for Gravitational Physics, Yukawa Institute for Theoretical Physics, Kyoto University, Kyoto 606-8502, Japan
             \and 
              African Institute for Mathematical Sciences, 6 Melrose Road, Muizenberg 7945, Cape Town, South Africa
              \and
              INFN, Sezione di Padova, via Marzolo 8, I-35131, Padova, Italy
             }

   \date{Received ---; accepted ---}

  \abstract
   {General relativistic effects on the clustering of matter in the Universe provide a sensitive probe of cosmology and gravity theories that can be tested with the upcoming generation of galaxy surveys. These will require the availability of accurate model predictions, from large linear scales to small non-linear ones.}
   {Here, we present a suite of large-volume high-resolution N-body simulations specifically designed to generate light-cone data for the study of relativistic effects on lensing-matter observables without the use of simplifying approximations. As a case study application of these data, we perform an analysis of the relativistic contributions to the lensing-matter power spectra and cross-power spectra.}
  {The \emph{RayGalGroupSims} suite (\textsc{RayGal} for short) consists of two N-body simulations of $(2625\,h^{-1}\,{\rm Mpc})^3$ volume with $4096^3$ particles of a standard flat $\Lambda$CDM model and a non-standard $w$CDM phantom dark energy model with a constant equation of state. Light-cone data from the simulations have been generated using a parallel ray-tracing algorithm that has integrated more than 1 billion geodesic equations without the use of the flat-sky or Born approximation.}
   {Catalogues and maps with {relativistic weak lensing} that include post-Born effects, magnification bias (MB), and {redshift-space distortions} (RSDs) due to gravitational redshift, Doppler, transverse Doppler, and integrated Sachs-Wolfe--Rees-Sciama effects are {publicly released}. Using this dataset, we are able to reproduce the linear and quasi-linear predictions from the \textsc{Class} relativistic code for the ten power spectra and cross-spectra (3$\times$2 points) of the matter-density fluctuation field and the gravitational convergence at $z=0.7$ and $z=1.8$. We find a 1-30\% level contribution from both MB and RSDs to the  matter power spectrum, while the fingers-of-God effect is visible at lower redshift in the non-linear regime. Magnification bias also contributes at the $10-30\%$ level to the convergence power spectrum, leading to a deviation between the shear power spectrum and the convergence power spectrum. Magnification bias also plays a significant role in the galaxy-galaxy lensing by decreasing the density-convergence spectra by $20\%$ and coupling non-trivial configurations (such as the configuration with the convergence at the same redshift as the density, or at even lower redshifts). }
   {The cosmological analysis shows that the relativistic 3$\times$2 points approach is a powerful cosmological probe. Our unified approach to relativistic effects is an ideal framework for the investigation of gravitational effects in galaxy studies (e.g. clustering and weak lensing) as well as their effects in galaxy cluster, group, and void studies (e.g. gravitational redshifts and weak lensing) and cosmic microwave background studies (e.g. integrated Sachs-Wolfe--Rees-Sciama and weak lensing).} 
   

   \maketitle

%

\section{Introduction}

Observations of the distribution of galaxies in the Universe carry a wealth of cosmological information pertaining to the cosmic matter content, the state of expansion, the amplitude of matter-density fluctuations, and the nature of dark energy and dark matter.\ They also enable general relativity (GR) to be tested on large cosmic scales. This is because, in addition to the cosmological imprint arising from the dynamical processes that lead to the formation of cosmic structures, observables of the clustering of galaxies also carry the signature of relativistic effects that alter the path of photons propagating between emitting galaxies and the observer.

Relativistic effects modify the apparent distribution of galaxies. They leave distinct imprints on galaxy and lensing two-points statistics through two processes: (i) the weak-lensing effect, which alters the apparent angular position, size, and shape of galaxies in the sky; and (ii) the redshift-space distortions (RSDs) due to the proper motion of the emitting galaxies, the local gravitational potentials, and the propagation of light rays through time-varying potentials.

The current as well as upcoming generation of galaxy surveys, such as eBOSS, HSC, DES, KIDS, DESI, LSST, Euclid, and SKA, will provide a precise mapping of the distribution of galaxies across a wide range of scales with measurements that are sensitive to such effects. These observational programmes will increase the number of galaxies with measured redshifts and/or ellipticities by several orders of magnitude, thus enabling accurate analyses of weak-lensing and clustering data to be performed (as well as their cross-spectra, the so-called 3$\times$2 points analysis). This will allow strong constraints on the cosmological model parameters to be inferred and alternative models of dark energy and modified gravity theories to be tested, provided accurate theoretical model predictions are available.

The effects of GR on the galaxy two-points statistics beyond the RSD signal caused by the peculiar velocity of galaxies have been investigated in the linear regime in numerous works \citep[see e.g.][]{yoo2009new,yoo2010general,bonvin2011what, challinor2011linear}, and extensions to the quasi-linear and non-linear regime have been primarily carried out using approximate analytical approaches \citep{didio19,taruya2020,saga2020,didio20,beutler2020}.  Similarly, studies of GR effects on the weak-lensing shear have been investigated under various approximations. As an example, GR corrections up to second order in the gravitational potential of the lensing shear were computed in \citet{bernardeau2010fullsky}, and the imprint on the shear power spectrum in the linear regime was evaluated under the Born approximation in \citet{didio13}. 

Analytical and numerical studies of the GR contributions to the gravitational potentials of matter-density fluctuations have found that they leave a small imprint on the clustering of matter \citep{chisari2011connection,adamek2016general}. That is why Newtonian N-body simulations are still a valid tool for investigating relativistic kinematics effects \citep{fidler2015general,fidler2016relativistic}, which, on the other hand, leave observational imprints that cannot be neglected \citep[see e.g][]{Ghosh2018,Breton2019imprints}. 

Several general relativistic N-body and/or ray-tracing codes have been developed to study these effects 
in the non-linear regime of the clustering of matter \citep{adamek2016general,borzyszkowski2017liger,giblin2017general,sgier20}. Nevertheless, the use of approximations at different stages of the numerical computation prevents the possibility of inferring cosmological model predictions that are of general validity. For instance, a standard approach to generating light-cone data from high-resolution N-body simulations consists in assuming the Born approximation \citep{borzyszkowski2017liger} while possibly building the light cone from snapshots with the replica method \citep{sgier20}. A ray-tracing method based on the calculation of the lensing distortions of a ray bundle via the solving of the geodesic equations was presented in \citet{killedar2012gravitational}. However, the ray-tracing algorithm is limited to a fixed grid and thus does not enable the high spatial resolution that can be achieved by N-body codes with adaptive mesh refinement (AMR) to be exploited.

\citet{giblin2017general} performed fully non-linear general relativistic simulations for an ideal fluid; however, the resolution remains limited. \citet{adamek2016gevolution} developed a numerical GR N-body code in the weak-field limit without AMR. \citet{adamek2019hubble} and \citet{lepori20} implemented a ray-tracing algorithm similar to that of \citet{reverdy2014propagation} and \citet{Breton2019imprints}, which solves the geodesic equations without relying on the Born approximation; it was used to investigate several relativistic effects on the weak-lensing convergence power spectrum \citep{lepori20}, the matter power spectrum multipoles \citep{guandalin20}, and the cross-correlation power spectrum between the matter-density and lensing convergence \citep{coates20}. 

Here, we describe and publicly release the numerical data generated from the \emph{RayGalGroupSims}  (which stands for Ray-tracing Galaxy Group Simulations, hereafter \textsc{RayGal})   simulation suite\footnote{\href{https://cosmo.obspm.fr/public-datasets/}{https://cosmo.obspm.fr/public-datasets/}}, a set of simulations specifically designed to study the relativistic effects on the clustering of matter observables. In particular, the light-cone data have been generated using a ray-tracing algorithm that solves the geodesic equations with the gravitational potential evaluated along the different AMR levels of the N-body solver \citep{reverdy2014propagation, Breton2019imprints,breton2020}. Thus, these numerical datasets benefit from a large dynamical range of simulations with a large volume and a high spatial resolution without the need to rely on the Born approximation or other limiting assumptions.
In \citet{Breton2019imprints}, the light-cone data from the \textsc{RayGal} simulation of a $\Lambda$-cold dark matter model (hereafter $\Lambda$CDM) were used to perform a study of the asymmetry of the halo cross-correlation function due to the relativistic effects. 

In this work, we introduce the complete simulation suite, including data from the simulation of a non-standard dark energy model, and present a case study application of the dataset. In particular, we present the evaluation of relativistic effects on the angular matter-density power spectrum, lensing convergence power spectrum, and their cross-spectra. We find a good agreement in the linear and quasi-linear regimes with the analytical predictions from the relativistic code \textsc{Class}, while deviations occur in the non-linear regime. Furthermore, we find a non-negligible imprint of RSDs and magnification bias (MB) on the harmonic matter power spectrum (1-30\%). At low redshift, the damping due to the fingers-of-God effect is clearly visible. The MB contributes to the lensing convergence power spectrum to the $10-30\%$ level. As the shear power spectrum is largely insensitive to the MB, this results in important differences with respect to the convergence power. Magnification bias  also decreases the matter overdensity--gravitational convergence spectra (i.e. galaxy-galaxy lensing) by $20\%$. By coupling the signal between different redshift shells, relativistic effects boost the correlations between lensing-matter clustering observables. This is the case for the correlation between the matter-density field and the convergence evaluated at the same redshift as well as the correlation between the matter-density field  and the convergence measured at a redshift smaller than that of the density field. The cosmological evolution shows that the relativistic contributions to the 3$\times$2 points statistics make it promising probe of dark energy and modified gravity. 

The paper is organised as follows: We describe the \textsc{RayGal} simulations and the associated numerical datasets in \cref{methodology}; we present a case study of the imprint of relativistic effects on weak-lensing observables using the \textsc{RayGal} data in \cref{sec:results_application_3x2pt}; and finally we discuss the conclusions in \cref{conclusions}.

\section{Methods: Simulations, ray tracing, and data}
\label{methodology}
\subsection{Cosmological models}
\label{cosmomod}

\begin{figure}
\centering
      \includegraphics[width=0.9\hsize]{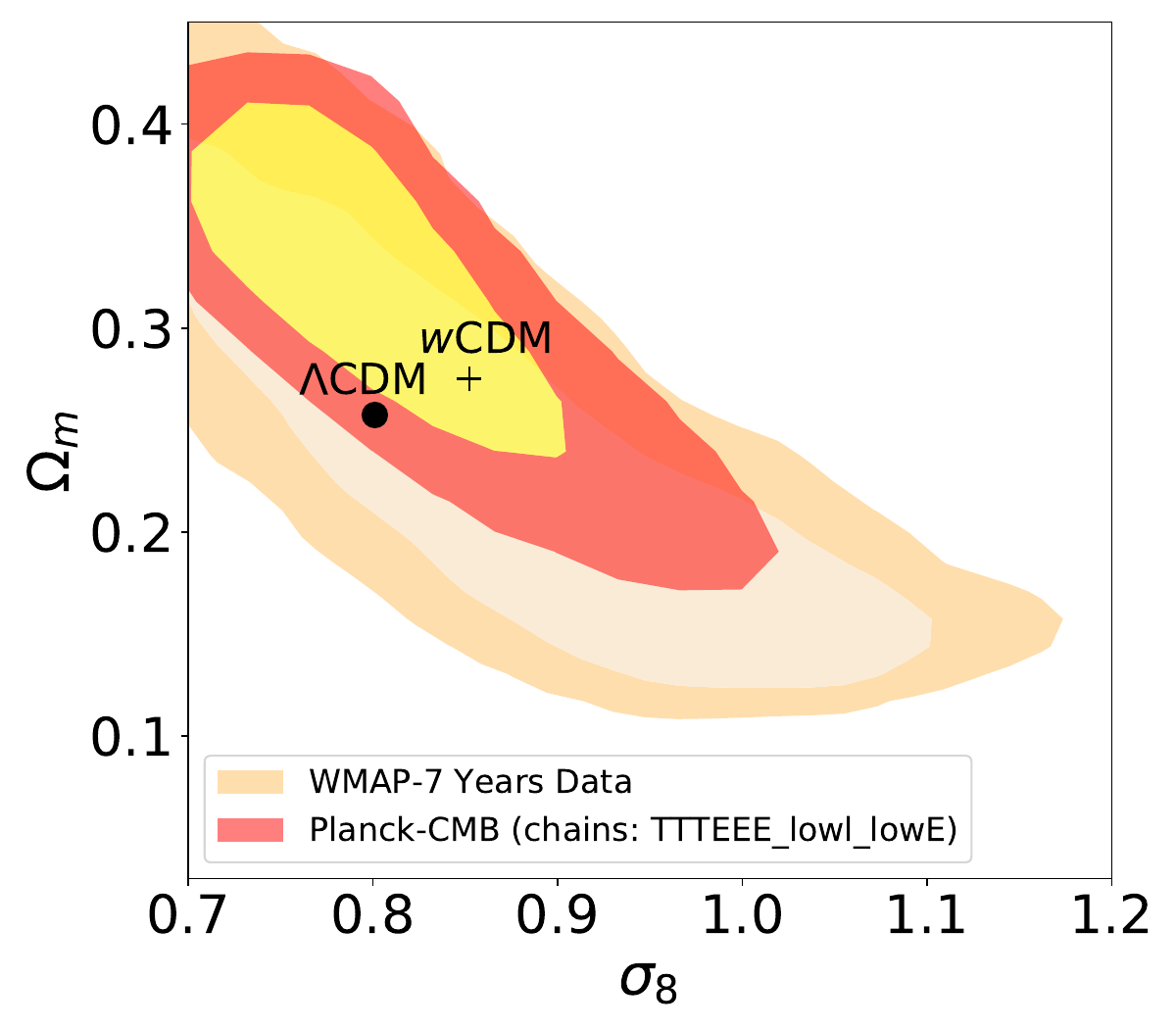}  
      
       \includegraphics[width=0.96\hsize]{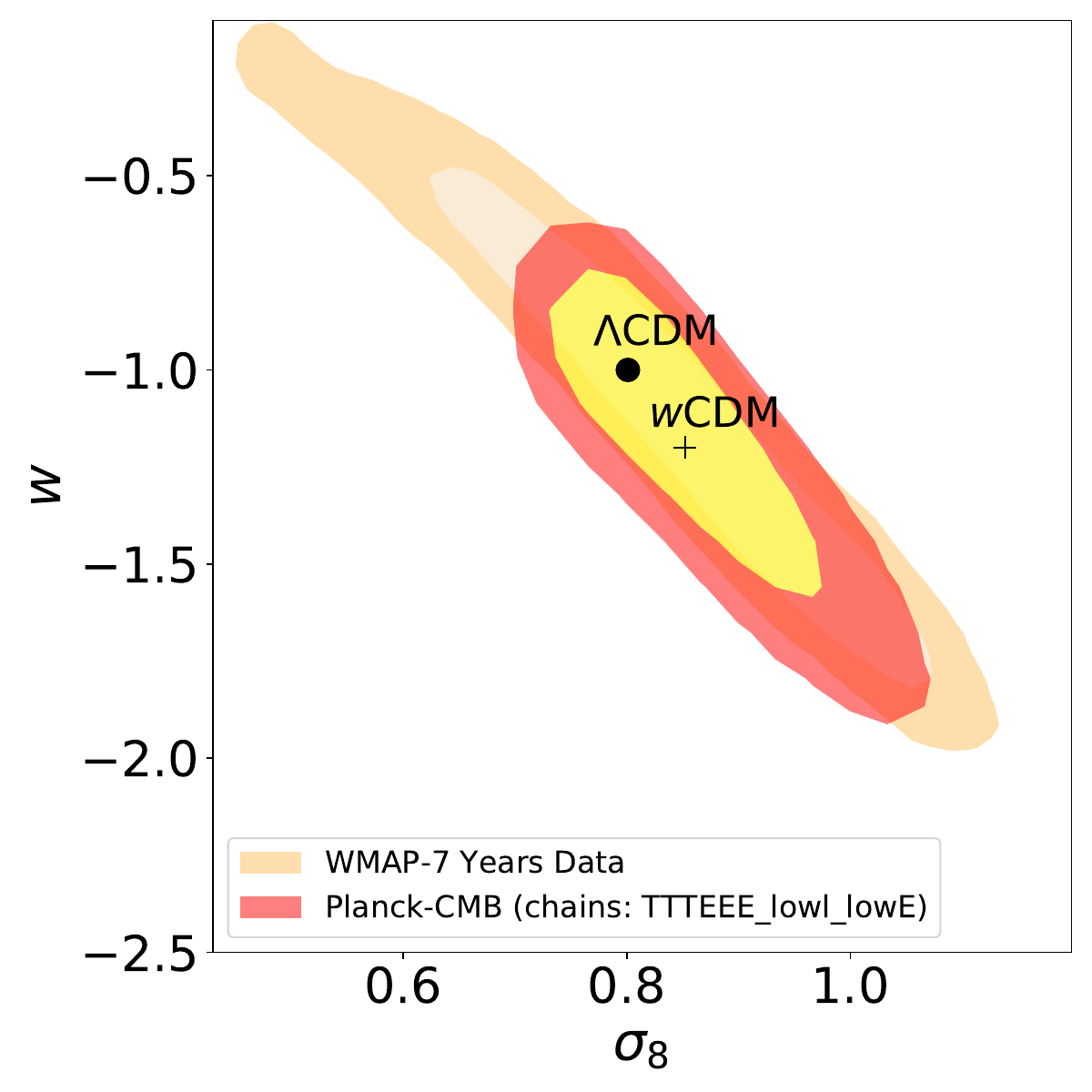} 
      \caption{$1$ and $2\sigma$ credible regions from the WMAP 7 yr data analysis (beige filled contours) and the {\it Planck} primary CMB analysis of the TT, TE, and EE anisotropy spectra \citep{2020A&A...641A...6P} (red and yellow filled contours). The parameter values of the $\Lambda$CDM and $w$CDM models correspond to the dot and the cross markers, respectively.}
      \label{fig:CMBcontours}
\end{figure}

The nature of dark energy still remains unknown. In order to test the cosmological dependence of relativistic effects on the clustering of matter observables, we simulated two different spatially flat cosmological models: a standard $\Lambda$CDM model and a `phantom' dark energy $w$CDM model with constant equation of state $w=-1.2$. The cosmological parameters of these models have been calibrated against the cosmic microwave background (CMB) anisotropy power spectra from the WMAP 7 year data \citep{komatsu2011seven}, these are summarised in~\cref{tab:cosmologicalparameters}. The values of $\Omega_m$ and $\sigma_8$ of the $w$CDM model have been chosen so as to be within the $1\sigma$ confidence region of the WMAP inferred constraints in the $\Omega_m-\sigma_8$ plane and along the degeneracy line of $\sigma_8-w$ contours (see beige filled contours in \cref{fig:CMBcontours}). Henceforth, this model is statistically indistinguishable at $1\sigma$ from $\Lambda$CDM best-fit model to the WMAP 7 year data. The simulated models are the same as those selected for the DEUS-Full Universe Runs  \citep{alimi2012,rasera2014cosmic,bouillot2015}. These are also consistent with the constraints from the latest {\it Planck} primary CMB analysis within $2\sigma$ \citep[TT, TE, and EE, the temperature and polarisation spectra][]{2020A&A...641A...6P} (see red-yellow filled contours in \cref{fig:CMBcontours}). We notice that values of the equation of state varying in the range $-1.3\lesssim w\lesssim -1$, being compatible with Planck data, are often considered in the literature (e.g. \citealt{alestas2020} and reference therein), for instance in the context of the `H$_0$ tension' \citep{riess2019}. Since these models are hard to discriminate at the homogeneous and linear perturbation level, it is important to find new cosmological probes, for instance in the non-linear regime of cosmic structure formation.

\begin{table}
\caption{Cosmological parameter values of the $\Lambda$CDM and $w$CDM simulated models, respectively.} 
        \centering
        \begin{tabular}{cccc} 
                \hline\hline
                Model&$\Omega_m$&$\sigma_8$&$w$\\
                \hline\hline
                $\Lambda$CDM&0.25733&0.80101&-1.0\\
                \hline
                $w$CDM&0.27508&0.85205&-1.2\\
                \hline
        \end{tabular}
        \tablefoot{ The columns from left to right give the values of the cosmic matter density, $\Omega_m$, the linear root-mean-square fluctuations smoothed on the $8$~$h^{-1}$Mpc scale, $\sigma_8$, and the redshift-independent equation of state, $w$. In all models the baryon density is set to $\Omega_b=0.04356$, the density of the relativistic species to $\Omega_r \simeq 0.00008$, the curvature to $\Omega_k=0$, the reduced Hubble constant to $h=0.72$, and the slope of the primordial spectrum to $n_S=0.963$.}
        \label{tab:cosmologicalparameters}
\end{table}

\subsection{N-body simulations}
\label{subsec:simulations}
The \textsc{RayGal} suite was run at the French TGCC and CINES super-computing centres. It consists of $N$-body simulations of the $\Lambda$CDM and $w$CDM models, respectively, spanning a volume of $(2625~h^{-1}\textrm{Mpc})^3$ sampled by $4096^3$ dark-matter particles. The optimal compromise between mass resolution (of order $\sim 2\times10^{10}~h^{-1}$M$_{\odot}$ depending on the simulated cosmology) and large statistics makes them ideal to investigate the cosmology-dependent distribution and properties of halos from Milky Way size to galaxy-cluster size (including galaxy-group size halos, which are the main deflectors in weak-lensing studies). The main novelty of these simulations is the relativistic ray tracing within the gravity light cones, which allows us to build realistic catalogues of dark-matter particles and halos that contain relativistic effects.

Initial conditions were generated with a modified version of MPGRAFIC \citep{prunet2008initial}. In this code, the density field is generated as a Gaussian random field, while the associated displacement field is computed (in the version we developed) using second-order Lagrangian perturbation theory (2LPT). The starting redshift was chosen so as to ensure that the maximum displacement (among all particles) is of order one coarse cell (i.e. $\sim 0.64~h^{-1}\textrm{Mpc}$). The use of 2LPT minimises the effects of transients \citep{Crocce2006}, while assuming a late starting redshift minimises discreteness errors \citep{Michaux2020}.

Simulations were performed with (an optimised version of) the parallel AMR $N$-body code \textsc{RAMSES}\footnote{\href{https://bitbucket.org/rteyssie/ramses/src/master/}{https://bitbucket.org/rteyssie/ramses/src/master/}} \citep{teyssier2002cosmological}, which makes use of a multi-grid Poisson solver \citep{Guillet2011}. A triangular shaped cloud (TSC) assignment scheme was used (instead of the default cloud-in-cell assignment) so as to increase the isotropy of the gravitational field. 
The characteristics of the simulations are summarised in \cref{tab:raygalparam}.   

\begin{table*} 
\caption{\textsc{RayGal}  $\Lambda$CDM simulation parameters ($w$CDM simulation parameters are given in parentheses when they differ).}
    \begin{tabular}{ccccccccccc}
    \hline\hline
    L$_{\textrm{box}}$ (h$^{-1}$Mpc) &n$_\textrm{part}$ & n$_x$ & n$_{\textrm{ref}}$&m$_\textrm{ref}$&n$_\textrm{cell}$ & m$_p$ (h$^{-1}$M$_\odot$) & $\Delta_x$ (h$^{-1}$kpc)& $z_i$ &C$_{\textrm{dt}}$& P$_\textrm{lin}(k)$  \\
    \hline
    2625 &$6.87\cdot 10^{10}$&4096& 7 &8&$4.41(4.46)\cdot 10^{11}$&$1.88(2.01)\cdot 10^{10}$&5 &46(51)&0.5&CAMB\\
    \hline
    \end{tabular}
    \tablefoot{L$_{\textrm{box}}$ is the box length, n$_{\textrm{part}}$ the number of particles, n$_x$ the grid size, n$_{\textrm{ref}}$ the number of refinements, m$_\textrm{ref}$ the refinement threshold, n$_{\textrm{cell}}$ the final number of AMR cells, m$_p$ the particle mass, $\Delta_x$ the spatial resolution, $z_i$ the starting redshift, C$_{dt}$ the Courant-like factor that determines the size of the integration time step (with respect to the largest stable time step), and P$_\textrm{lin}(k)$ specifies the source of the linear power spectrum used to generate initial conditions. All calculations have been performed in double precision.}  
    \label{tab:raygalparam}
\end{table*}

While storing all particles and gravity cells at all time steps would be ideal, in practice it is unfeasible since it would overflow all the available storage devices (each snapshot occupies approximately 25~TB of memory). Therefore, we opted for a hybrid strategy and stored from 40 to 50 redshift snapshots (depending on the simulated cosmological model). On the other hand, particles and gravity cells are also stored at every coarse time step (thus ensuring good time resolution) in concentric shells located at the light-travel distance from an observer located at the centre of the box at $z=0$: these are the background light-cone data. We saved three light cones with different depths (with $z_{\textrm max}=0.5$, 2, and 10, respectively) and apertures (full sky,  2500, and $400$~deg$^2$, respectively) using the onion-shells techniques of \citet{fosalba2008onion,teyssier2009fullsky}. Unlike many other studies, the maximum redshift $z_{max}$ of the cones was chosen in a conservative way to make sure that the comoving volume of the cone is smaller than the comoving volume of the simulation: this is required to minimise the effects of replica. The full-sky light cone does not rely on any replica techniques, while the narrow light cones are tilted with respect to the $x$-axis of the simulations (by 17 or 25 degrees along the two spherical coordinates angles depending on the light cones) to minimise replica effects. Rather than storing the AMR cells that are the closest to the past null Friedmann-Lemaître-Robertson-Walker (hereafter FLRW) light cone of the observer, we stored two cells at each spatial location: one slightly ahead of time and one slightly behind in time. Thanks to this `double-layer' strategy, it is then possible to explore different approaches for building the light cones: by taking the closest cell (in time), by averaging between the two cells, or by interpolating between the cells. It is also possible to compute time derivatives, such as the derivative of the gravitational potential $\dot{\phi}$, the key quantity of integrated Sachs-Wolfe--Rees-Sciama (ISW-RS) effect. It is also worth noticing that, in principle, light propagates on the null real light cone and not on the null FLRW light cone (i.e. at a given spatial location the gravitational field can be different for different light rays because of the variation in the potential during the time delay between light-ray arrivals) . While in this article we neglect this effect in our ray tracing, very advanced studies could in principle account for this very subtle effect by interpolating between the two cells for each individual light ray.

In RAMSES, each task from the MPI library writes a binary file that corresponds to a segment of the Peano-Hilbert space filling curve. Since these numerous raw binary files are not user friendly, they are rewritten as a small number of HDF5 files (with the tools provided by \textsc{pFoF}\footnote{\href{https://gitlab.obspm.fr/roy/pFoF}{https://gitlab.obspm.fr/roy/pFoF}}). For snapshots, each HDF5 file corresponds to a cubic sub-volume of the whole simulation. Particle files contain information about the dark matter particles. For light cones, each HDF5 file corresponds to a given shell. The particle files contain information about particles while the gravity files contain information about the AMR cells. Within each file, particles and cells are reorganised in small cubes (corresponding to an HDF5 group) to make further selection faster. Finally, the HDF5 files are self-documented to facilitate their usage. 

Halos were detected in all the snapshots using a parallel friend-of-friends algorithm (as implemented in pFoF as described in \citealt{roy2014pfof}) with various linking lengths ranging from $b=0.05$ to $b=0.4$. A variety of classical halo properties were computed and the constituent halo particles were also stored for each halo in the catalogues. Since the volume spanned by halos defined with $b=0.4$ is much larger than the volume of most commonly defined halos, users are free to re-compute their favourite halo definition from halo particles and to evaluate any physical quantity of interest (with the parallelisation on a per halo basis being straightforward). We also provide halo properties for halos detected with a spherical overdensity halo finder using a range of overdensity thresholds (with respect to the mean matter density) from $\Delta_m=50$ to $\Delta_m=12800$ (as well as other popular definitions based on critical density or virial overdensity). Halos are also detected on the light cone with pFoF for some selected linking lengths, while the rest of the detections are currently in progress.

\subsection{Relativistic ray tracing and catalogue creation}
\label{subsec:ray-tracing}

\begin{figure*}
\centering
      \includegraphics[width=0.9\hsize]{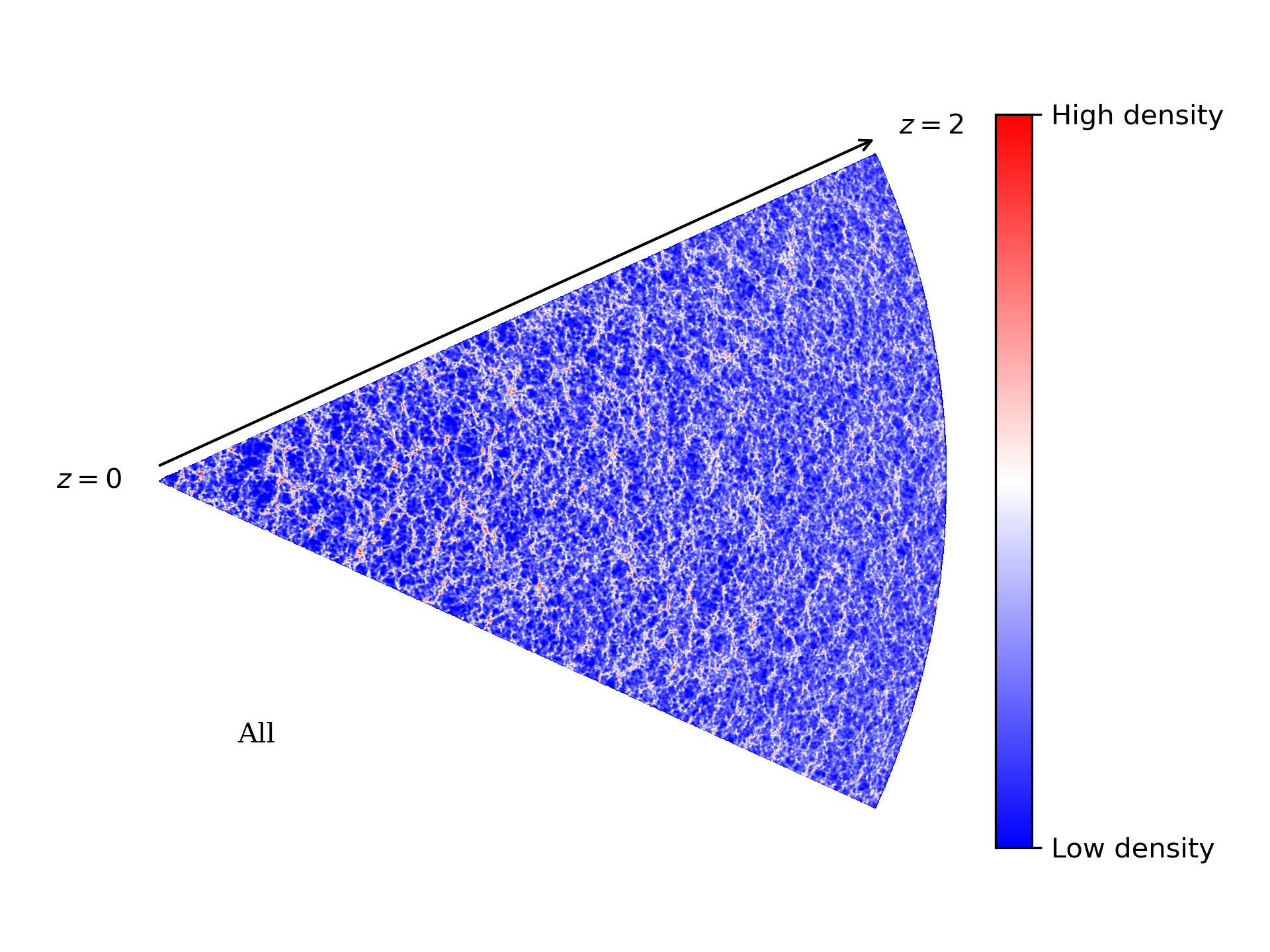}
      \caption{Apparent matter overdensity within  the \textsc{RayGal} intermediate light cone called `Narrow 2'. This includes `all' weak-field relativistic contributions.  }
\label{coneraygal}

\end{figure*}

\begin{figure*}
\centering
\begin{tabular}{c|c}
 \includegraphics[width=0.45\hsize]{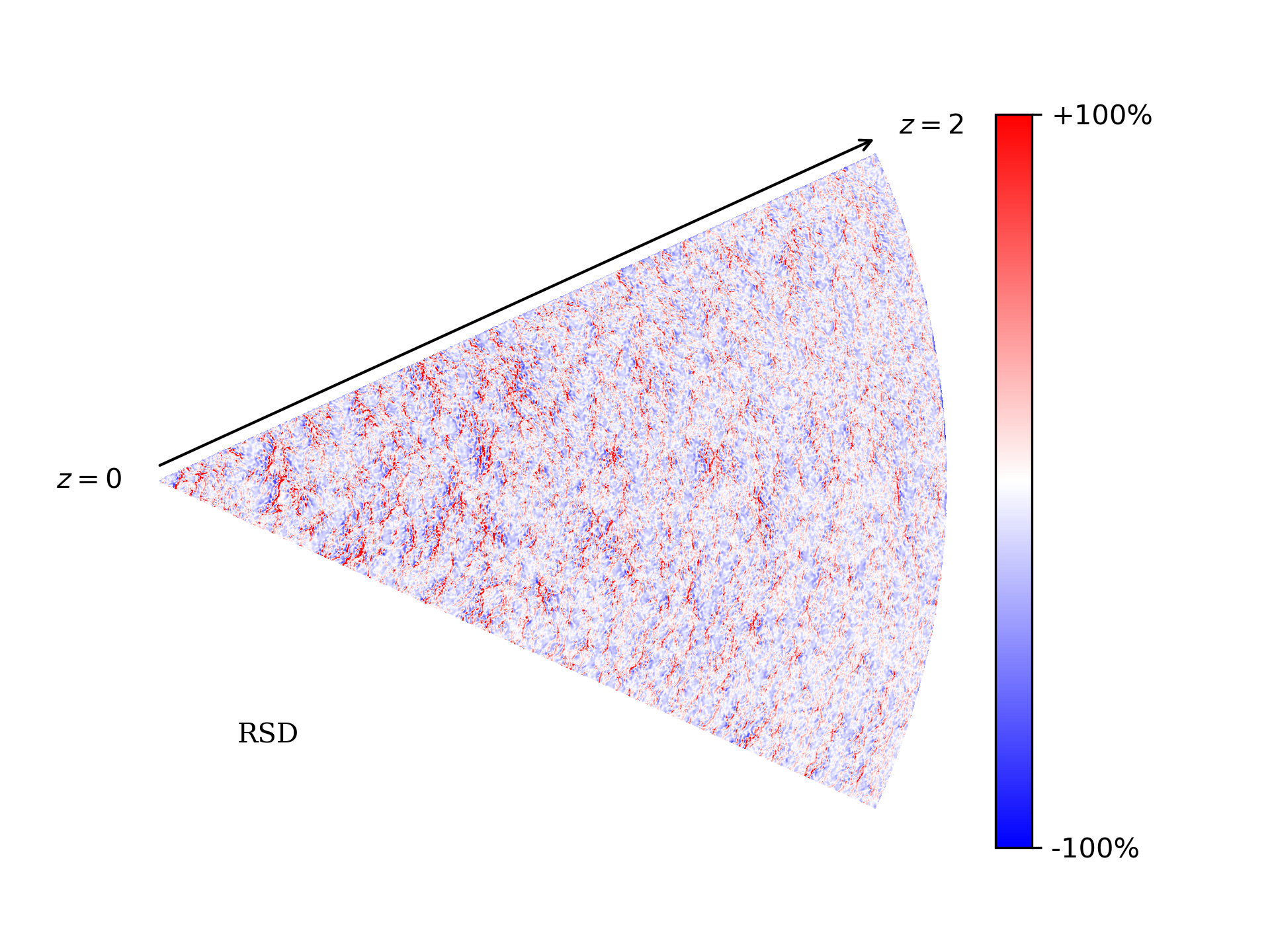} &  \includegraphics[width=0.45\hsize]{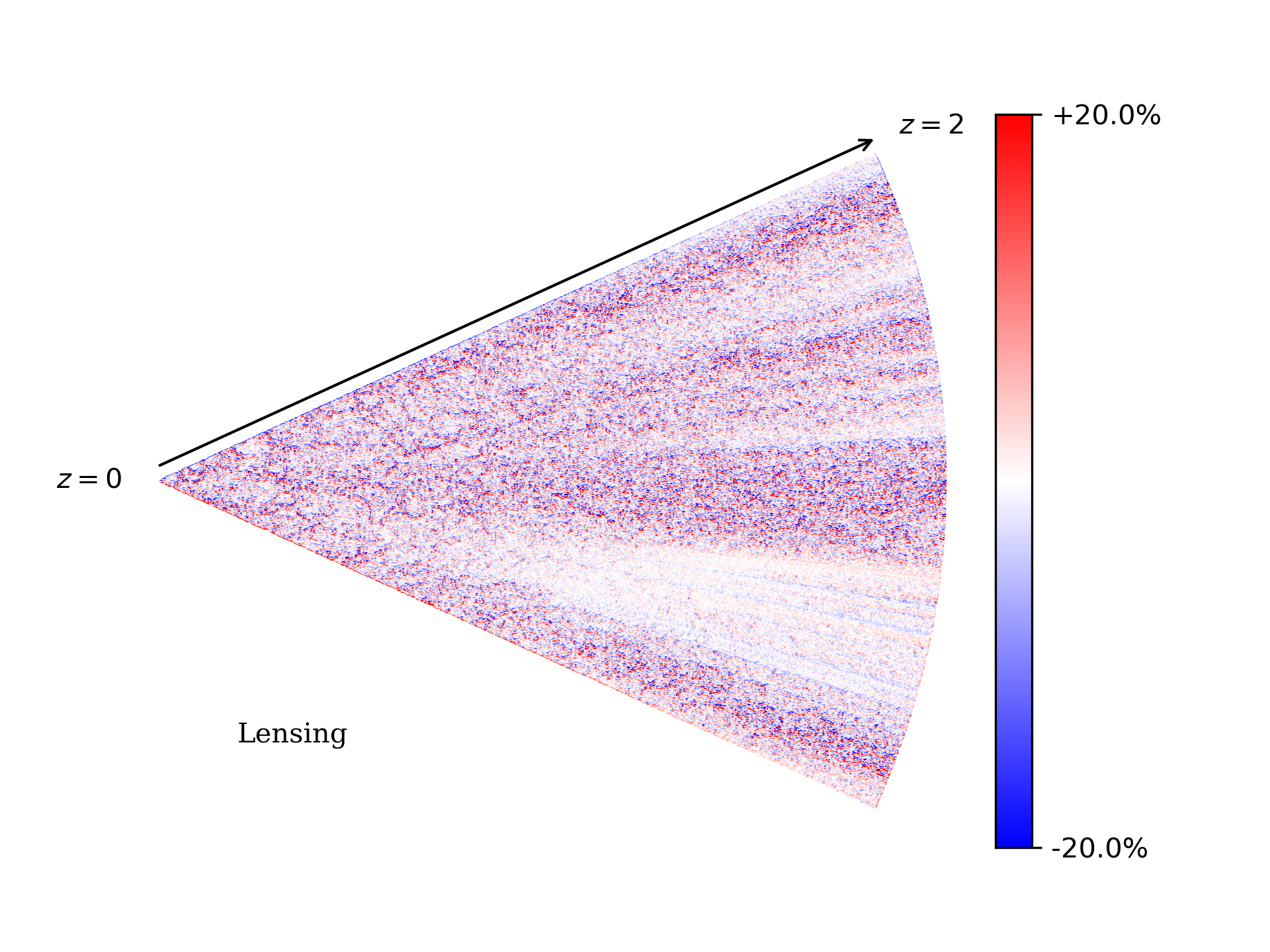} \\
  \includegraphics[width=0.45\hsize]{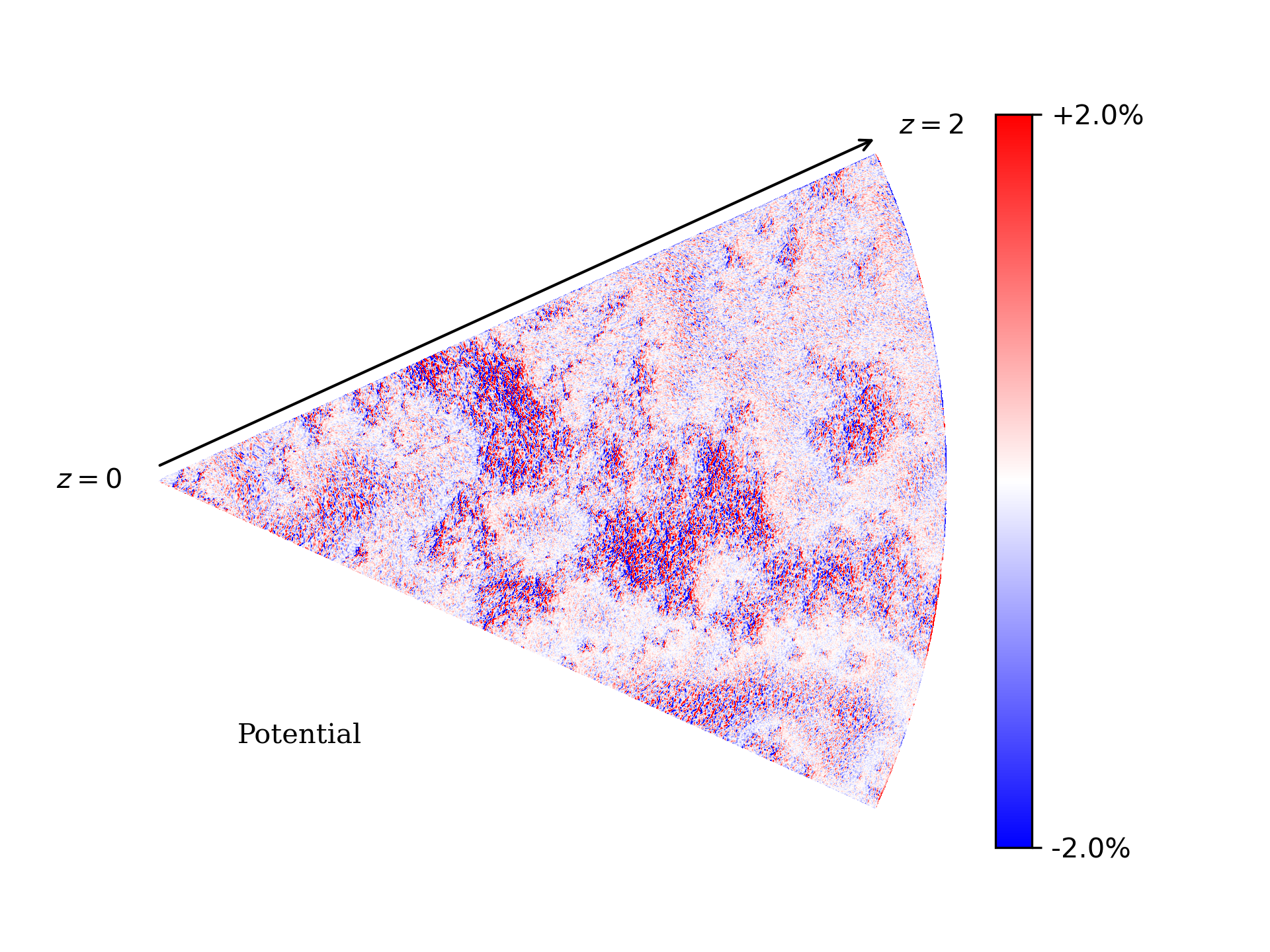}    &  \includegraphics[width=0.45\hsize]{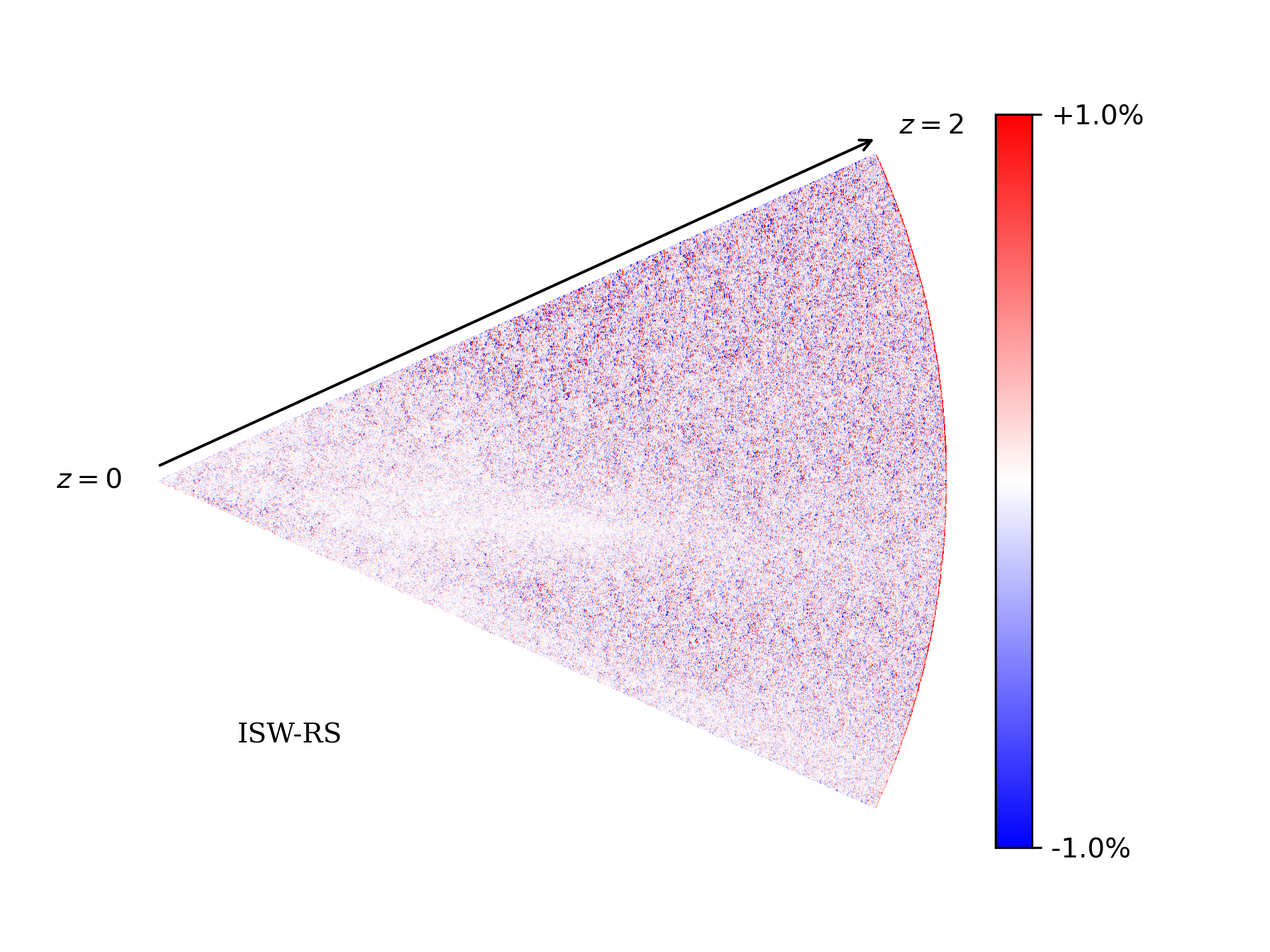}
\end{tabular}
\caption{Various contributions to the apparent matter overdensity within the \textsc{RayGal} intermediate light cone: Doppler effect (top left), weak lensing (top right), gravitational redshift (bottom left), and ISW-RS effect (bottom right).}
\label{coneraygal_contrib}
\end{figure*}

To produce realistic observables, we use the ray-tracing framework \textsc{Magrathea-pathfinder} \citep{Breton2021magrathea} based on the AMR library \textsc{Magrathea} \citep{reverdy2014propagation}, which computes the propagation of light rays within the 3D AMR structure of light cones (such as the ones described in \cref{subsec:simulations}).

We consider a weakly perturbed FLRW metric given by
\begin{equation}
 \label{eq:metric}
   g_{\mu\nu}\textrm{d}x^{\mu}\textrm{d}x^{\nu} = a(\eta)^2 \left[-(1 + 2\phi/c^2)c^2\textrm{d}\eta^2 + (1 - 2\phi/c^2)\delta_{ij}\textrm{d}x^i\textrm{d}x^j \right],
\end{equation}
with $\phi$ the gravitational potential, $\eta$ the conformal time, $x^i$ the comoving position and $c$ the speed of light. The integration is performed backwards in time, starting from the observer at $z=0$. Photons are initialised with $k^0 = 1$ and $k_\nu k^\nu = 0$, where $k^\alpha = (\dd\eta/\dd\lambda, \dd x^i/\dd\lambda)$ and $\lambda$ is the affine parameter. We interpolate the gravity information from the light cone to the photon position using an inverse TSC scheme (so that to be consistent with the interpolation procedure used by the $N$-body solver) at the finest refinement level that contains the photon. If there are not enough neighbours at the same level, we perform an interpolation on a coarser grid and eventually repeat this procedure until we find enough neighbours to compute the interpolation. If we cannot interpolate at the coarse level, the propagation stops (it means that the photon is out of the numerical light cone). Once all the gravity information at the photon location has been evaluated, we solve the geodesic equations: 
\begin{eqnarray}
   \label{eq:geodesic_equation}
              \frac{\textrm{d}^2 \eta}{\textrm{d}\lambda^{2}} &=& -\frac{2a'}{a}\frac{\textrm{d}\eta}{\textrm{d}\lambda}\frac{\textrm{d}\eta}{\textrm{d}\lambda} - \frac{2}{c^2}\frac{\textrm{d}\phi}{\textrm{d}\lambda}\frac{\textrm{d}\eta}{\textrm{d}\lambda} + 2\frac{\partial\phi}{\partial\eta}\left(\frac{\textrm{d}\eta}{\textrm{d}\lambda}\right)^2. \\
              \frac{\textrm{d}^2 x^i}{\textrm{d}\lambda^2} &=& -\frac{2a'}{a}\frac{\textrm{d}\eta}{\textrm{d}\lambda}\frac{\textrm{d}x^i}{\textrm{d}\lambda} + \frac{2}{c^2}\frac{\textrm{d}\phi}{\textrm{d}\lambda}\frac{\textrm{d}x^i}{d\lambda} - 2\frac{\partial\phi}{\partial x^i}\left(\frac{\textrm{d}\eta}{\textrm{d}\lambda}\right)^2,
\end{eqnarray}
with $\phi$ the gravitational potential and $a' = \dd a/\dd\eta$ the time derivative of the scale factor. In particular, we use the precise value of $\partial\phi/\partial\eta$ estimated from the double-layer strategy adopted to build the light cones, thus ensuring that the ISW-RS effect is correctly implemented even in the strongly non-linear regime.

Our method fully takes advantage of the adaptive mesh. In particular, we use an adaptive step equal to one-fourth of the finest cell at the photon location. This allows for a precise light propagation in high-density regions. \\

To emulate what we really observe, we found the null geodesics connecting the observer to each individual source (the dark matter particle, halo, etc.). To do so, we used the methodology described in \citet{Breton2019imprints}, which consists of the following steps. First, we launch a light ray towards the comoving direction of the source. However, due to deflections induced by the matter-density field, the photon will probably not hit the source. If that is the case, we then iterate on the initial conditions using a Newton-like method until the angle subtended by the source's comoving position and the photon at the same distance is smaller than a given threshold (in our case, set to 10~milliarcsecond). Once the null geodesic connecting the observer and a source is found, we estimate all the contributions to the observed redshift at first order in metric perturbations, accounting for local and integrated terms using either a linear decomposition of the redshift or its exact definition. We also provide several (partial) redshift definitions that progressively incorporate part of the relativistic effects in order to disentangle the various contributions\footnote{Although we implemented the possibility to give any peculiar velocity to the observer $\bm{v}_0$, for simplicity we decided to set $\bm{v}_0=\bm{0}$ in this release.}:
\begin{eqnarray}
\label{eq:firstredshift}
        && z_0 = \frac{a_0}{a}-1, \\
        &&z_1 = \frac{a_0}{a} \left( 1 + \frac{\phi_o - \phi_s}{c^2}\right)-1, \\
        && z_2 = \frac{a_0}{a} \left( 1 + \frac{\bm{v}_s\cdot\bm{n}}{c} + \frac{\phi_o - \phi_s}{c^2}\right)-1, \\
        &&z_3 = \frac{a_0}{a} \left( 1 + \frac{\bm{v}_s\cdot\bm{n}}{c} + \frac{\phi_o - \phi_s}{c^2} + \frac{1}{2}\left(\frac{v_s}{c}\right)^2\right)-1, \\
        && z_4 = \frac{a_0}{a} \left( 1 + \frac{\bm{v}_s\cdot\bm{n}}{c} + \frac{\phi_o - \phi_s}{c^2} + \frac{1}{2}\left(\frac{v_s}{c}\right)^2- \frac{2}{c^2}\int^{\eta_o}_{\eta_s} \dot{\phi}\textrm{d}\eta\right)-1, \nonumber \\ \\
\label{eq:lastredshift}
    &&z_5=\frac{(g_{\mu\nu}k^{\mu}u^{\nu})_s}{(g_{\mu\nu}k^{\mu}u^{\nu})_o}-1, 
\end{eqnarray}
where $g_{\mu\nu}k^{\mu}u^{\nu} = -ack^0 \left( 1 + \phi/c^2 + \bm{v}\cdot\bm{n}/c + \frac{1}{2}\left(\frac{v}{c}\right)^2 \right)$, the subscripts `$o$' and `$s$' denote an evaluation at the observer and the source, respectively, and $z_n$ gives the different redshifts provided for each single source. $z_0$ is the comoving redshift (from background expansion), $z_1$ also includes gravitational redshift, $z_2$ includes the Doppler effect and $z_3$ adds the transverse Doppler effect. The $z_4$ includes the contribution to the final redshift from all the effects, including the ISW-RS effect. The $z_5$ is the exact redshift definition and should match $z_4$ (except for some small higher-order contributions).  

Furthermore, given the fact that the observed angular position of the source, $\bm{\theta}$, is known, as is the comoving angular position, $\bm{\beta}$, we can compute the deflection angle and evaluate the lensing distortion matrix, $\bm{\mathcal{A}}$, along the true photon trajectory. This reads as
\begin{equation}
\label{eq:distortion_matrix}
\bm{\mathcal{A}}
\equiv
\frac{\partial\bm{\beta}}{\partial\bm{\theta}}
=
\begin{pmatrix}
    1-\kappa-\gamma_1 & -\gamma_2 + \omega \\
    -\gamma_2 - \omega    & 1-\kappa + \gamma_1  
\end{pmatrix} ,
\end{equation}
where $\kappa$ and $\gamma = (\gamma_1, \gamma_2)$ are the weak-lensing quantities related, respectively, to the convergence and shear, while $\omega$ is the (small) image rotation.
This matrix can be estimated using either an infinitesimal-beam or finite-beam methods (both are described in \citealt{breton2020}). For the present release, we use the finite-beam method with a bundle semi-aperture of $10^{-5}$ radians (which corresponds for instance to the apparent angle of a ruler of comoving size $\sim 10~h^{-1}\textrm{kpc}$ located at a comoving distance of $\sim 1~h^{-1}\textrm{Gpc}$ in a flat universe). It is important to stress that there are two possible ways to define the beam cross-sectional area: perpendicular to the radial direction or perpendicular to the light-ray wave vector. Here we consider the second option (i.e. the so-called Sachs basis), which is more closely related to observable quantities: we account for the tilt of the cross-sectional area.

One has to be careful since the link between these lensing quantities and actual observables is not necessarily straightforward. The magnification $\mu$ is directly given by the inverse of the determinant of the distortion matrix. For a comoving source, this gives the amplification of the flux by weak lensing, where the flux is indeed an observable. In virtue of the distance-duality relation, $\mu$ also provides the fractional change in the apparent area of the source where the area is also an observable. However, for a non-comoving source, one should also account for redshift perturbations to evaluate the change in flux and area: this is called Doppler convergence \citep{bonvin2008effect}. The relative variation in the flux is given by the relative variation (induced by the redshift perturbations) of the square of the luminosity distance. These effects play a subdominant role at high redshift (z>0.5) but may play a role at lower redshift.  These subtleties are well described in \citet{breton2020}. Another observable in weak-lensing studies is the apparent shape of the sources (i.e. their ellipticity $e$). The average ellipticity is related to the reduced shear $g=\gamma/(1-\kappa)$ as $<e>=<g>$ (if $\kappa<<1$ one finds $<e>\approx <\gamma$> ). The role of redshift perturbations on the shape is negligible. Textbooks often assume that the shear power spectrum and the gravitational convergence power spectrum are the same (thus highlighting the central role of the convergence in weak-lensing studies): while this is true at first order, we discuss this point in this article. Finally, the orientation of the source is an observable that is a probe of the rotation. However, it is a very subtle effect and it is also very challenging to know the original orientation of the sources. There are however some proposals using the polarisation of radio-galaxies \citep{Francfort21}. Here we provide both the fundamental lensing quantities and the detailed redshift perturbations so that the user can reconstruct any physical or observable quantities of interest. In any case, the gravitational convergence $\kappa$ play an important role since $\kappa > |\gamma| > \omega$ and the Doppler corrections are subdominant at $z>0.5$, this is why we focus on this key quantity in this article.

It should be noted that in the current implementation, we only detect one image per source. This is a valid approximation in the weak-lensing regime (although convergence and shear are not necessarily small). A possible extension of our work would be the use a multiple-root finder to account for multiple images. However, there are multiple reasons as to why the weak-lensing regime should still be a reasonable approximation for \textsc{RayGal}. First, for most practical cases the number of strong lensing events for galaxy-size lenses is small. Even if the probability to have strong-lensing events increase with redshift, for a realistic radial selection function the number of observable sources heavily decreases beyond $z \approx 2-3$. Second, strong lensing is usually relevant at very small angular scales, which is not necessary the focus of the present simulation. Third, when we perform {source-averaging}, the tail of the magnification's probability distribution function is damped with respect to the {directional-averaging} case (see e.g. \citealt{takahashi2011probability,breton2020}, and references therein). Moreover, this effect is further enhanced when using the finite-beam method \citep{fleury2017weak, fleury2019cosmic, breton2020}. Finally, even if strong lensing events occur, our geodesic finder would most likely find the principal image of the source.

\subsection{Maps}

Maps were built by projecting the matter distribution and its properties onto a sphere. We discretised the maps by considering the \textsc{Healpix} (Hierarchical Equal Area isoLatitude PIXelation) pixelation scheme \citep{Gorski2005}.

The matter distribution can be weighed by radial selection functions of different type. Here, we generated two type of maps: (i) Dirac radial selection maps, \textsc{Healpix} maps generated assuming a Dirac selection function, and (ii) top-hat radial selection maps, \textsc{Healpix} maps generated assuming a top-hat selection function.

Dirac radial selection maps are computed by stopping all the light rays at a fixed comoving distance from the observer. The quantity of interest (for instance the distortion matrix) is then interpolated from the closest values on the light cone.  Top-hat radial selection maps are built from the relativistic catalogues. Particles (or halos) are selected within a shell of mean redshift $z_{\textrm mean}$ and (half-)width $ \Delta z$. Particles (or halos) are then sorted according to Healpix while keeping track of the empty pixels (no particles) and the masked pixels (outside of the cones). The surface overdensity is evaluated by computing the excess of surface density in pixels with respect to the mean surface density of the shell in the unmasked region. The distortion matrix is evaluated as an average over all particles. Empty cells are filled with the default undefined value provided by Healpix.

\subsection{\textsc{RayGal} data: Snapshots, light cones, relativistic catalogues, and maps}

\begin{table*}
\caption{Data available for \textsc{RayGal} snapshots in $\Lambda$CDM and $w$CDM cosmologies.}
        \centering
        \begin{tabular}{cccccccccccccc} 
                \hline\hline
                Data | z&$z_i$&3.00&2.33&2.00&1.50&1.00&0.67&0.50&0.43&0.25&0.11&0.00&other z\\
                \hline\hline
                Part. &x&&&x&&x&&x&&&&x&\\ 
                \hline
                Halos part. &&x&x&x&x&x&x&x&x&x&x&x&\\ 
                \hline
            Halos prop. &x&x&x&x&x&x&x&x&x&x&x&x&x\\ 
        \hline
        \end{tabular}
        \tablefoot{ We note that halo properties are available for about 30 to 40 `other $z$' per cosmology, but the redshifts may differ from one cosmology to another. $z$ is the redshift, $z_i$ the starting redshift, `Part.' stands for particles, and `prop.' stands for properties.}
        \label{tab:snapshots}
\end{table*}

\begin{table*}
\caption{Data available for \textsc{RayGal} light cones in $\Lambda$CDM and $w$CDM cosmologies ($w$CDM simulation parameters are given in parentheses when they differ).}
        \centering
        \begin{tabular}{cccccccc} 
                \hline\hline
                Type | charac.&$z_{\textrm max}$&aperture (deg$^2$)&gravity&part.&rel. cat. part(\#)&rel. cat. halo (\#)&double layer\\
                \hline\hline
                Narrow 1 &10.00&400&x&x&$10^8$&$2.5\times 10^6$ (on going)&x\\ 
                \hline
                Narrow 2 &2.00&2500&x&x&$3.5\times 10^8$&$1.2\times 10^7$ ($1.3 \times 10^7$)&x\\ 
                \hline
            Full sky &0.48&41253&x&x&$2\times 10^8$&$1.3\times 10^7$ ($1.5\times 10^7$)&x\\ 
        \hline
        \end{tabular}
        \tablefoot{`Charac.' stands for characteristics, $z_{\rm max}$ is the maximum redshift, aperture is the angular aperture, gravity indicates the presence of gravity information, `part.' indicates the presence of particle information, `rel. cat. part(\#)' indicates the number of particles in the relativistic catalogues, and `rel. cat. halo (\#)' indicates the number of halos. `Double-layer' indicates the presence of two light cones so as to compute time derivatives or perform time interpolation. The particle density in cone Narrow 2 is similar to the galaxy density of present and future photometric surveys, while the halo density is similar to the galaxy density of present and future spectroscopic surveys.}
        \label{tab:lightcones}
\end{table*}

\textsc{RayGal} data are divided into four categories: snapshots, FLRW light cones, relativistic catalogues, and maps. Most of the \textsc{RayGal} data are available on the website of the COS team at LUTH \footnote{ \href{https://cosmo.obspm.fr/public-datasets/}{https://cosmo.obspm.fr/public-datasets/}}. 

Snapshot data, corresponding to particles at a given time, consist of: (i) snapshots of all particles at a given time (position, velocity, and identity),
(ii) all particles within halos (position, velocity, and identity),
and (iii) the properties of all the halos (identity, number of particles, most-bounded particle position, centre of mass position, mean velocity, maximum radius, moment of inertia, circular velocity, velocity dispersion, angular momentum, kinetic energy, self-binding energy, density profile, and circular velocity profile).

The FLRW light-cone data (corresponding to particles and cells near the past FLRW light cone) consist of:
(i) the particle light cone, all particles just before or just after the past null light cone (position, velocity, redshift) and
(ii) the gravity light cone, all gravity cells  just before and just after the past null light cone (position, potential, gravitational field, redshift).
These snapshot and light-cone data are built from the simulations and post-processing as described in \cref{subsec:simulations}.

Relativistic catalogues, corresponding to sources as seen by the observer and accounting for relativistic effects such as weak lensing and RSDs, consist of:
(i) the halo catalogue (un-lensed and lensed angular positions, background, and perturbed redshifts (from \cref{eq:firstredshift} to \cref{eq:lastredshift}), weak-lensing distortion matrix, and halo mass),
and (ii) the particle catalogue, a random subset of dark matter particles (same properties as for the halo catalogue, except mass). They represent a population of sources with a bias $b=1$.
These catalogues are built from the ray-tracing techniques described in \cref{subsec:ray-tracing}.

Maps, corresponding to the projection of the matter distribution and its properties, consist of Dirac radial selection maps ($\kappa$, $\gamma_1$, $\gamma_2$, and $1/\mu$, respectively the convergence, shear components, and inverse magnification).

We provide Dirac radial selection maps for several redshifts within the range of our light cones. Top-hat radial selection maps (for $\kappa$, $\gamma_1$, $\gamma_2$ and $\delta$) can be computed from the released catalogues using \textsc{Healpix} tools. For the narrow cones, pixels outside the footprint are masked. In that case, we only provide the relevant pixels within the angular selection function, therefore  producing lighter files.

The characteristics of \textsc{RayGal} data are summarised in \cref{tab:snapshots} and \cref{tab:lightcones}. 
An illustration of the apparent matter distribution and the various relativistic effects within the light cone `Narrow 2' are shown in \cref{coneraygal} and \cref{coneraygal_contrib}  \footnote{We made use of Py-SPHViewer \citep{alejandro_benitez_llambay_2015_21703} for the realisation of these images.}. {In this cone, the source angular density ($\sim 40~\rm{arcmin}^{-2}$) is similar to the one expected from present and future photometric surveys such as DES, KIDS, HSC, Euclid,   and LSST while the halo density ($\sim 10^{-3}~h^{3}~\mathrm{Mpc}^{-3}$) is also similar to the one expected in spectroscopic surveys such as BOSS, eBOSS, DESI, Euclid, and SKA}.

\begin{figure*}
\centering
\begin{tabular}{c|c}
      \includegraphics[width=0.45\hsize]{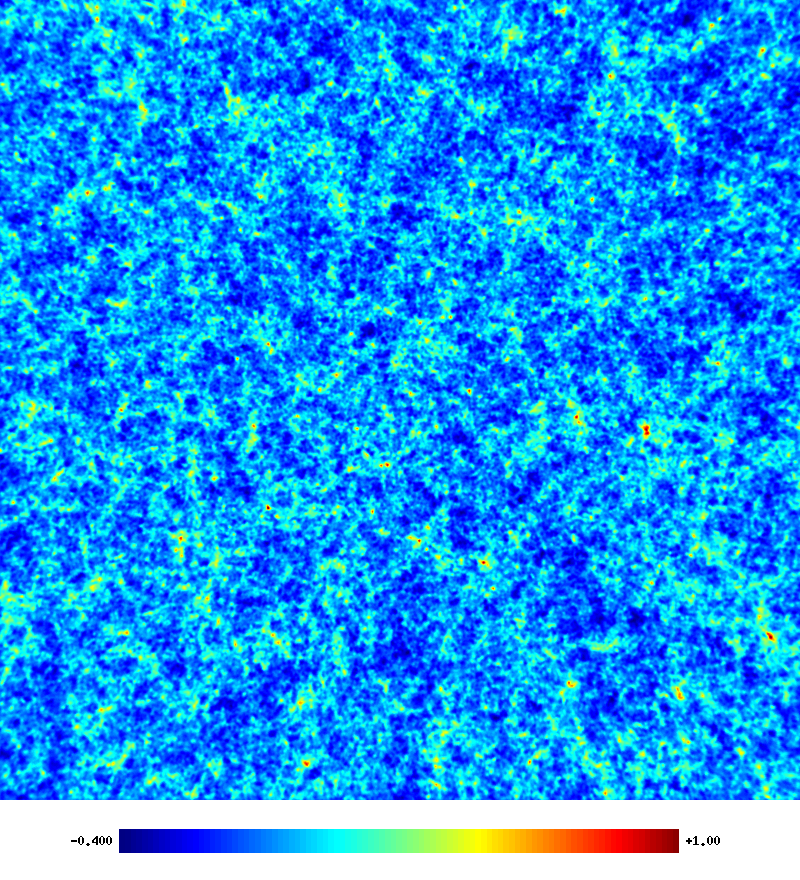}&\includegraphics[width=0.45\hsize]{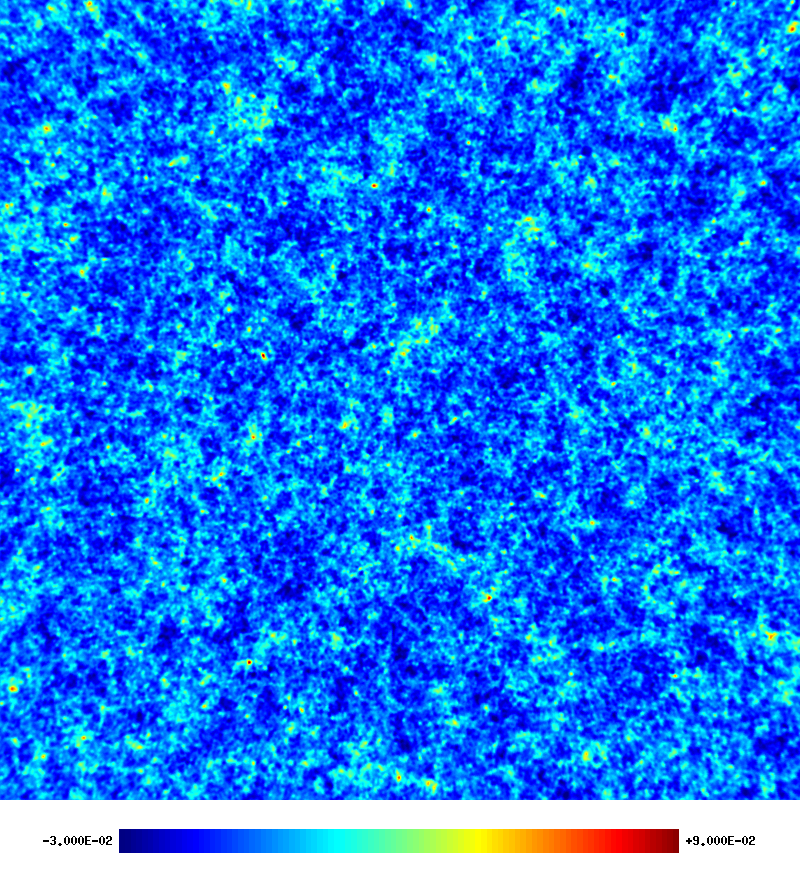}
\end{tabular}
      \caption{\textsc{Healpix} maps from \textsc{RayGal}. Left: Overdensity map at $z=0.7$ including relativistic effects. Right: Convergence map at $z=1.8$ including relativistic effects. The size of the field extracted from cone Narrow 2 is about $2500$~deg$^{2}$. A gnomonic projection has been used, and the maps have been smoothed with a Gaussian beam of 10~arcmin full width at half maximum for representation purposes. We note that some large overdensities at $z=0.7$ can be seen in the convergence map at $z=1.8$.} 
\label{mapraygal}

\end{figure*}

\begin{figure*}
\centering
\begin{tabular}{c|c}
\includegraphics[width=0.45\hsize]{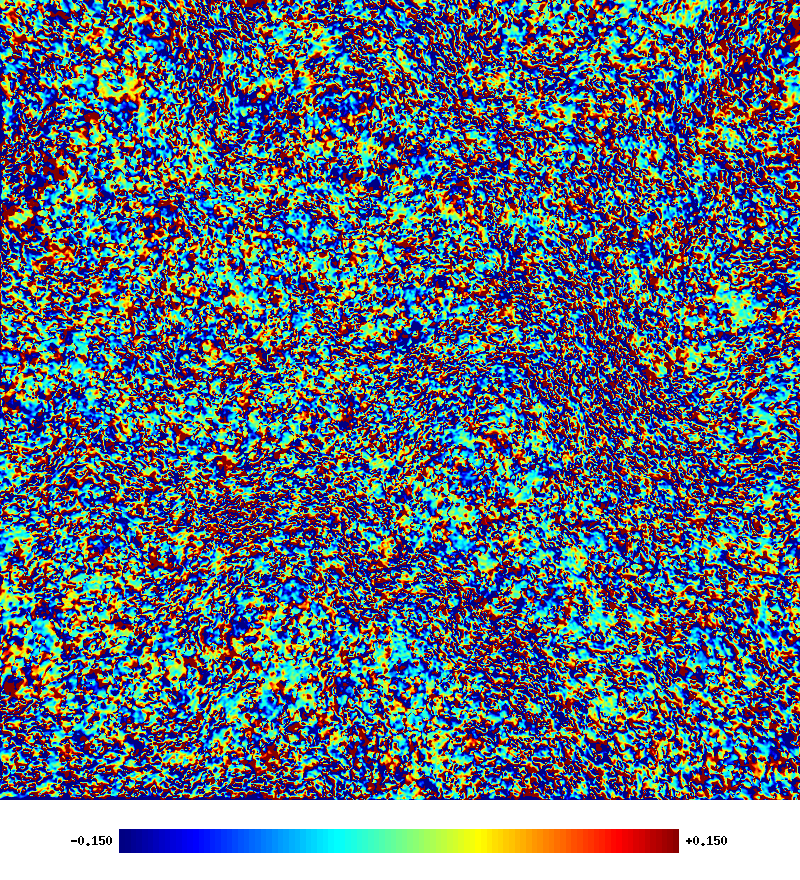}&\includegraphics[width=0.45\hsize]{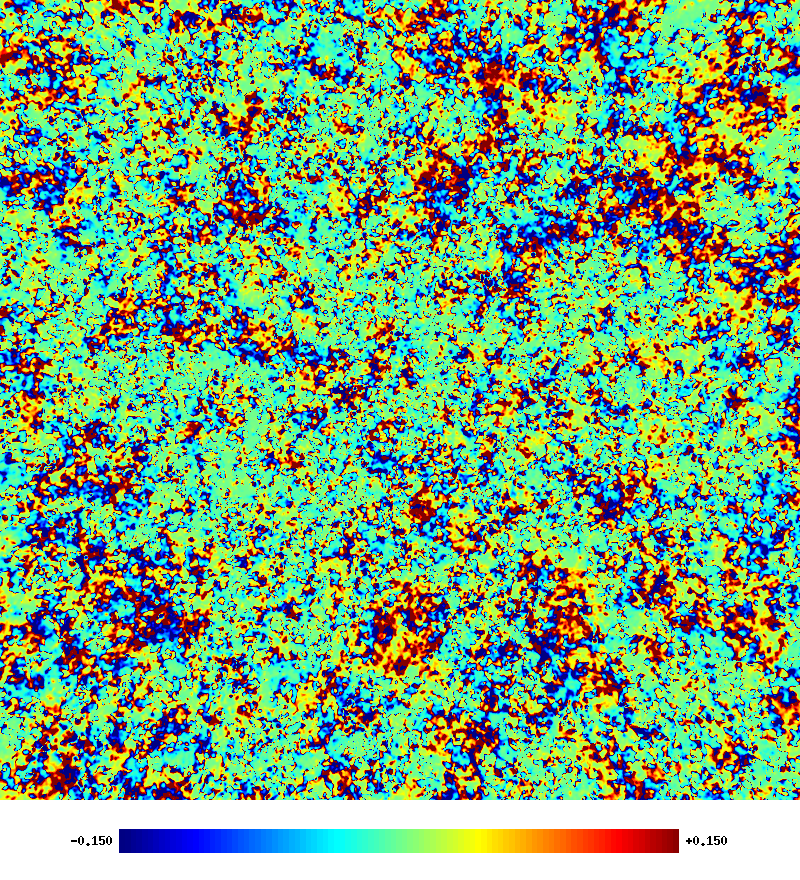}   \\
\includegraphics[width=0.45\hsize]{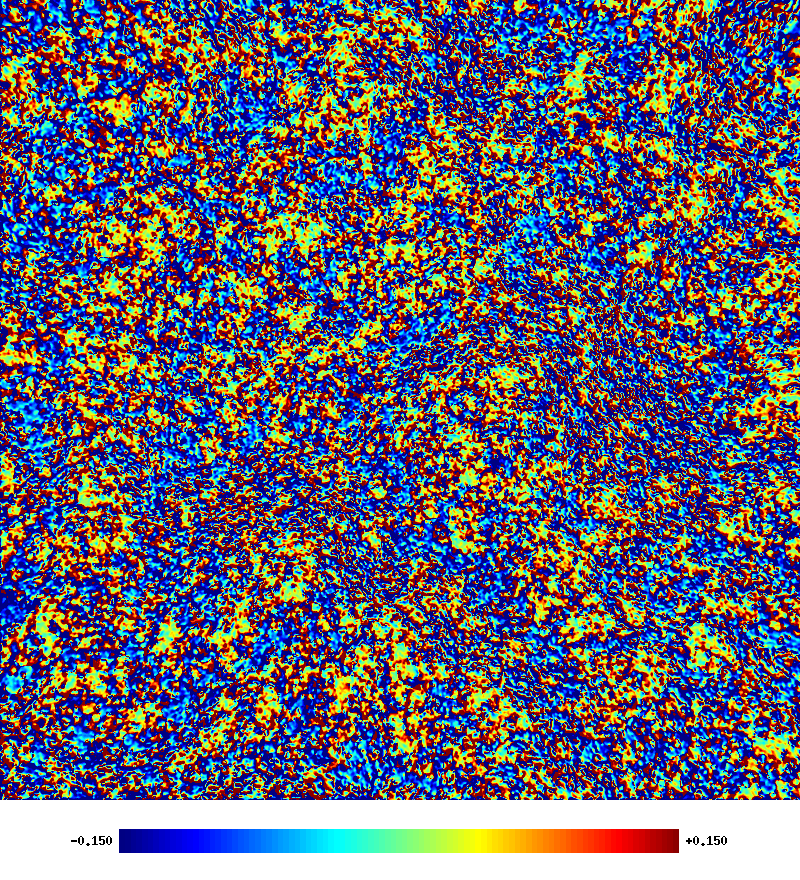}&\includegraphics[width=0.45\hsize]{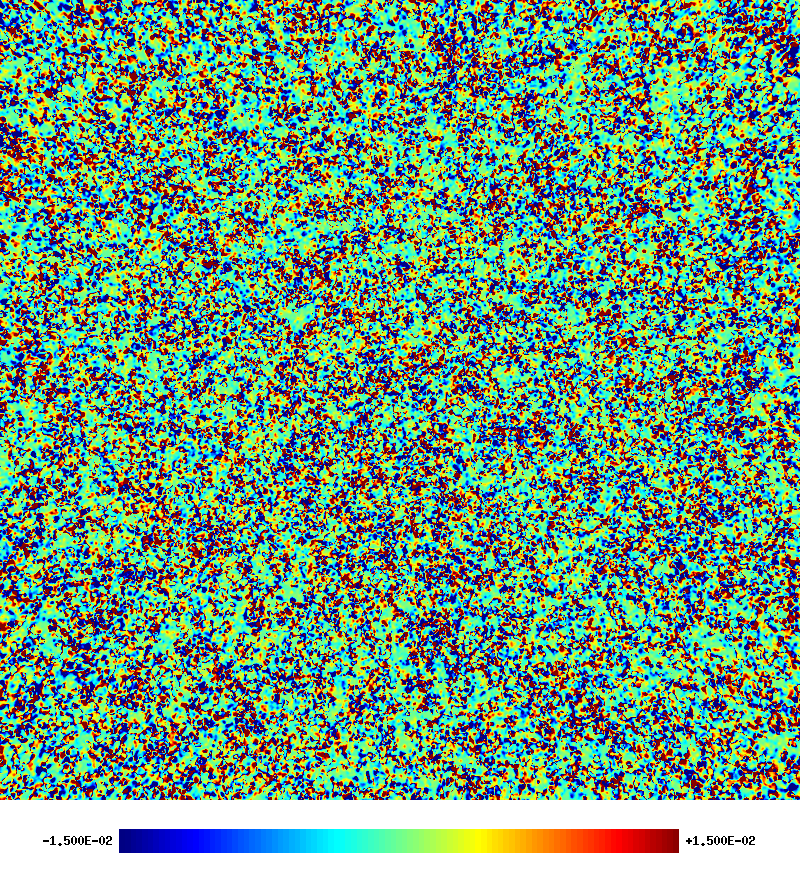}
\end{tabular}
\caption{Relativistic contributions in the maps shown in \cref{mapraygal}. Top left: Relative difference between the overdensity maps at $z=0.7$ with and without MB. Top right: Relative difference between the overdensity maps at $z=0.7$ with and without RSDs. Bottom left: Relative difference between the convergence maps at $z=1.8$ with and without the MB effect. As expected, the MB effect is anti-correlated with the value of the convergence. Bottom right: (Small) relative difference between the convergence maps at $z=1.8$ with and without RSDs. Relativistic effects play a non-trivial role at the ten percent level in the density and convergence maps.} 

\label{mapraygal_contrib}
\end{figure*}

\subsection{Case study: Configuration, definition, angular cross-spectrum evaluation, and analytical predictions}

As already mentioned, in this work we present a case study application of the \textsc{RayGal} light-cone datasets to the analysis of relativistic effects on the angular cross-spectra. In particular, we correlate the key quantity of clustering (i.e. the matter overdensity, $\delta$) and that of lensing (i.e. the gravitational convergence, $\kappa$) using \textsc{Healpix} maps. It is worth noting that these are not directly observable quantities (such as the galaxy overdensities or the galaxy ellipticities), nonetheless they are the central quantities in clustering and lensing studies from which one can deduce the correlation between observables. The goal here is not to focus on specific observables for a given survey, rather to provide general results about the imprint of relativistic effects on the density and lensing angular power spectra and cross-spectra.   

\subsubsection{Two-shell configuration}
  
The matter overdensity and the gravitational convergence are averaged within 3D pixels defined by the radial extension of the shell and the \textsc{Healpix} pixelation scheme for the angular extension. The number of pixels is equal to $12 \times N_{\rm side}^2$ and we chose $N_{\rm side}=4096$ for top-hat radial selection map. Given the number of sources in each shell, using a larger $N_{\rm side}$ increases the number of empty pixels and the gain in precision is limited. For test maps under Born approximation, we first computed a map with $N_{\rm side}=8192$ (resulting in a total number of light rays of the same order of magnitude as the number of sources in catalogues shells), which was then degraded (i.e. each pixel is computed as an average over the 4 smaller pixels within it) down to $N_{\rm side}=4096$ so as to use the same resolution in all maps.
 
We considered two shells (1 and 2) extracted from the $\Lambda$CDM 2500~deg$^2$ light cone. We isolated two possible redshifts for each shell: $z=0.7 \pm 0.2$ and $z=1.8 \pm 0.1$. We chose $z=1.8$ since it is one of the largest possible redshift in the cone while the comoving distance to $z=0.7$ is about half the comoving distance to $z=1.8$, thus corresponding to a maximum of the lensing kernel for a source located at $z=1.8$. Given that relativistic effects correlate distant shells, we investigated all the ten possible auto- and cross-power spectra for these two shells. We first investigated the clustering (three density-density spectra), then the weak lensing (three convergence-convergence spectra), and finally the galaxy-galaxy lensing (four density-convergence spectra). 

\subsubsection{Relativistic effect definition}

Weak lensing as a relativistic effect has already been widely studied in the literature, especially under the Born approximation (i.e. by computing lensing quantities and in particular the convergence as an integral of the transverse Laplacian along the un-deflected light-path). Henceforth, in the context of weak lensing, we use as a reference case the convergence computed using the Born approximation ($\kappa_{\rm Born}$). The mean Born convergence for a population of source galaxies (see \citealt{kilbinger2015}) is given by
\begin{equation}
\label{eq:kappaborneq}
    \kappa_{\rm Born}(\bm{\theta})= \int_{0}^{\chi_{\rm lim}} n(\chi_S) d \chi_S \int_{0}^{\chi_S}  d\chi  \frac{\chi_S-\chi}{\chi_S} \frac{\Delta_{\perp}\phi(\chi \bm{\theta} ,a(\chi))}{c^2} \chi  ,
\end{equation}
with $\chi$ the comoving distance (in a flat universe), $\chi_S$ the comoving distance of the source, $\chi_{\rm lim}$ the limiting comoving distance of the galaxy sample,  $n(\chi_S)$ the source probability distribution, and $\Delta_{\perp}\phi$ the Laplacian perpendicular to the line of sight.
We considered a top-hat radial selection function (i.e. a shell). For a shell of volume $V_{\rm shell}$  and constant density, $n(\chi_S)\propto4 \pi \chi_S^2/V_{\rm shell}$. It is worth noting that a common additional approximation is to assume
\begin{equation}
    \Delta_{\perp}\phi = \frac{3H_0^2 \Omega_m \delta}{2 a}
,\end{equation}
with $H_0$ the Hubble constant. We do not make this assumption in our Born calculation. Moreover in this expression $\delta$ is often estimated within shells of typical size between $10-100$~$h^{-1}$Mpc. Here we estimate $\Delta_{\perp}\phi$ at the resolution of the simulation (5-600~$h^{-1}$kpc). We checked that using such approximations leads to similar results in our article. A detailed study of the impact of these two approximations on various statistics would deserve a dedicated paper. Our reference is therefore an accurate Born convergence calculation along the un-deflected path.

In the context of clustering or number count, our reference is the comoving matter overdensity $\delta_{\rm com}$. The overdensity is estimated as the local excess of particles (in real space) with respect to the mean matter density,
\begin{equation}
    \delta_{\rm com} (\bm{\theta})= \frac{N_{\rm com}(\bm{\theta})}{\langle N_{\rm com}(\bm{\theta})\rangle}-1,
\end{equation}
where $N_{\rm com}(\bm{\theta})$ is the number density of source. Here again, light rays are assumed to propagate in a straight line. The Doppler effect can be considered as one of the standard effect in RSD studies (involving 3D correlation). However, the Doppler effect is often neglected in the context of angular correlation (see \citealt{grasshorn20}) as well as in weak-lensing and galaxy-galaxy lensing studies (see \citealt{Ghosh2018}).  

{We therefore call `relativistic effects' all the non-trivial GR effects beyond these two references ($\delta_{\rm com}$ and $\kappa_{\rm Born}$) cases.}
To gain further understanding on the nature of relativistic effects, we further decomposed them into two pieces.

   The first, MB effects, are related to the decrease or increase in the apparent source overdensity due to weak-lensing magnification or de-magnification as introduced by \citet{turner1984magbias}. The apparent density is related to the comoving density by $1+\delta=(1+\delta_{\rm com}) \mu^{-1}\mu^{2.5 s}$ with $\mu$ the magnification and $s$ the logarithmic slope of the luminosity function. The first term $\mu^{-1}$ is called the dilution term and is always present. The second term is only present for flux limited sample. Because we consider a volume and mass-limited sample of sources, we are only sensitive to the dilution effect (corresponding to a flux limited sample with a logarithmic slope of the luminosity function of $s=0$). This also biases the estimate of the convergence of a sample of galaxies as highly magnified region (with large convergence) are poorly sampled while de-magnified region are very well sampled. Moreover, widely used source-weighted estimator are even more biased by this effect. We also include in this category other (subdominant) weak-lensing effects, such as post-Born effects related to the propagation of light, which does not happen in a straight line. 
   
The second, RSDs, are related to redshift perturbations, which cause a change in the apparent density. This includes the Doppler effect, gravitational redshift effect, transverse Doppler effect, and ISW-RS effect. We note that lensing deflections effect are already included in MB.

All these effects do not exactly add up linearly. We therefore did not include them one by one (e.g. none, MB alone, RSDs alone) but progressively (e.g. none, MB, and MB+RSDs). We note that even though the way to decompose the full signal in various components depends on which phenomenon we are interested in, the observed signal including all contributions is unique and well defined. 

An illustration of the overdensity and convergence within the two shells and the importance of relativistic effects is shown \cref{mapraygal} and \cref{mapraygal_contrib}.

\subsubsection{Angular cross-spectrum estimator}

 The angular cross-spectra between two Healpix maps is estimated with \textsc{Polspice}\footnote{\href{http://www2.iap.fr/users/hivon/software/PolSpice/}{http://www2.iap.fr/users/hivon/software/PolSpice/}} \citep{Szapudi2001,Chon2004}. The spectra are computed with fast spherical harmonic transforms. They are corrected from angular selection function effects (i.e. mask effects) as well as pixel effect (i.e. deconvolution from the pixel window function)\footnote{In practice, for the Narrow 2 cone we used the options: \texttt{-apodizetype 0 -apodizesigma 25. -thetamax 50. -pixelfile YES}.}. The density auto-spectra are polluted by shot noise at large $\ell$. To correct this effect, we generated a noise map with the same number of randomly placed particles as in the data from which we compute the shot noise spectra. We then subtract these shot noise spectra from the measured spectra. This is more accurate than using the analytical formula since it removes the possible non trivial bias from the estimator in presence of a mask. In the context of weak-lensing and galaxy-galaxy lensing studies, we follow \citet{schmidt2009}. Since we use a pixel-based estimator we weight each $\kappa$ pixel with the density $1+\delta$ (this is equivalent to an inverse variance weight). As shown in \cref{appendix:2PCF_convergence_z1p8}, such a weight allows a good match with the direct pair-based estimator. 
 
\subsubsection{Angular cross-spectrum predictions (with \textsc{Class})} 

Analytical predictions usually rely on the linear mapping assumption. Especially in the case of the Doppler contribution at small scale this is an approximation. However, non-linear corrections are less visible than in RSD studies because here we consider the angular power spectrum (which is indirectly sensitive to radial motions in a way that depends on the thickness of the shell). Therefore, under this assumption one can decompose the overdensity as 
\begin{equation}
\label{eq:delta_decomposition}
    \delta_i \approx \delta_i^{\rm com}+\delta_i^{\rm MB}+\delta_i^{\rm RSD},
\end{equation}
where $\delta_i^{\rm com}$ is the comoving overdensity, $\delta_i^{\rm MB}\approx -2 \kappa_i$ is the MB\ contribution dominated by the dilution term ($\kappa_i$ is the gravitational convergence and we assumed a logarithmic slope $s=0$ for the source count in the context of our mass selected sample), and $\delta_i^{\rm RSD}$ is the contribution from redshift perturbations. In the context of source averaging (as in galaxy surveys) one can define an average convergence \footnote{We omit the average in the following for clarity.} as  
\begin{equation}
\label{eq:kappa_decomposition}
    \langle\kappa_i\rangle_s \approx \kappa_i^{\rm Born}+\kappa_i^{\rm MB}+\kappa_i^{\rm RSD},
\end{equation}
where $\kappa_i^{\rm Born}$ is the convergence according to Born approximation, $\kappa_i^{\rm MB}\approx -2 (\kappa_i^{\rm Born})^2$ is the MB contribution (assuming $s=0$) and $\kappa_i^{\rm RSD}$ is the contribution from redshift perturbation.  More details as well as the expression of the angular spectra under the linear mapping assumption are provided in \cref{appendix:linear_mapping}.

By including relativistic effects in our simulations, we follow a strategy similar to that of studies of the CMB. The two most widely used CMB codes in the literature are \textsc{Camb} \citep{lewis2000efficient} and \textsc{Class} \citep{lesgourgues11}, which provide predictions for these effects \citep{challinor2011linear,didio13}. Here, we confront our results to analytical predictions from the latter code, since by adjusting the name list\footnote{Beyond trivial modification of redshifts and cosmological parameters, we set the parameter \texttt{selection\_tophat\_edge=0.01} to ensure the top-hat selection profile has a sharp edge.} it automatically computes the ten cross-spectra we are interested in.

\textsc{Class}\footnote{\href{https://lesgourg.github.io/class\_public/class.html}{https://lesgourg.github.io/class\_public/class.html}} is a linearised Einstein-Boltzmann solver, very accurate in the linear regime. It predicts lensing, density, and density-lensing angular power spectra without relying on Limber approximation. Lensing predictions are performed at first order (Born without MB) and without RSDs. However, density and density-lensing predictions include MB and RSDs (Doppler effect, gravitational redshift, ISW-RS).
\textsc{Class} can also make predictions in the non-linear regime and we use them as our reference predictions. In the non-linear regime it relies on several assumptions, such as:
(i) weak lensing under the Born approximation (MB cannot be accounted for, and post-Born effects are neglected), (ii) RSDs with the linear mapping assumption from real space to redshift space and approximate treatment of the Doppler effect in the non-linear regime (i.e. no finger-of-God effect), and (iii) the \textsc{Halofit} prescription \citep{takahashi12} for the non-linear matter power spectrum $P(k)$.
  
Using the \textsc{Halofit} prescription induces up to 5\% errors in the predicted matter power spectrum around $k=1~h$~Mpc$^{-1}$ \citep{heitmann14}. The question of the error of $P(k)$ due to the \textsc{Halofit} prescription is beyond the scope of this work, which is focused on relativistic effects. We therefore prefer to feed \textsc{Class} with a matter power spectrum interpolated from the \textsc{RayGal} suite. The precision of the fit becomes $\sim 1$\% between $k=0.002~h$~Mpc$^{-1}$ and $k=20~h$~Mpc$^{-1}$ (beyond which we still use \textsc{Halofit} prescription) for redshift between $z=0$ and $z=2.33$. Below $k=0.2~h$~Mpc$^{-1}$, the effect of cosmic variance was reduced by subtracting the difference between the linear spectrum in the simulation (i.e. a specific realisation of initial conditions) and the linear spectrum from the \textsc{Camb} code (i.e. ensemble average). We notice that at large wavenumbers we did not remove the shot noise because its effect was found negligible in the relevant range of scales for the current study ($\ell<5000$).

As we will see in the following, one of the main effects that cannot be captured by \textsc{Class} (beyond non-linear RSDs at low redshift) is the effect of MB on convergence-convergence spectra and density-convergence spectra since it involves the bi-spectrum $\langle\kappa \kappa \kappa\rangle$. To cross-check the validity of our results (which is obtained from an accurate but complicated ray-tracing procedure) we computed from our Born maps, the absolute value of the inverse magnification $|\mu_{\rm Born}^{-1}|$. From this we computed a (rough) overdensity map including MB and a (rough) convergence map including MB by weighting the comoving density and the Born convergence by $|\mu_{\rm Born}^{-1}|$. More precisely the approximate density and convergence maps are given by $\delta^{\rm weight}_{\rm MB}=|\mu_{\rm Born}^{-1}|(1+\delta_{\rm com})-1$ and $\kappa^{\rm weight}_{\rm MB}=|\mu_{\rm Born}^{-1}| \kappa_{\rm Born}$. We then computed the `$|\mu_{\rm Born}^{-1}|$ weight MB spectra' by cross-correlating these maps. They are not analytical predictions but rather an estimate of the MB effect directly from the Born maps allowing for further cross-check of our ray-tracing results.

{To conclude, we computed the ten auto- and cross-spectra for two redshift shells both analytically and using the \textsc{RayGal} data. For each case we computed a reference spectrum (comoving and/or Born), a lensed spectrum accounting for MB correction and a relativistic spectrum accounting for both MB corrections and relativistic redshift perturbations (MB+RSDs). Additionally, we computed a rough estimate of MB effect from Born inverse magnification maps: this $|\mu_{\rm Born}^{-1}|$ weight MB spectrum serves as a cross-validation of our MB results when analytical predictions are not available.}

\section{Results: Analysis of density-lensing angular power spectra and cross-spectra (3$\times$2 points)}
\label{sec:results_application_3x2pt}

Here, we present the results of the computation of the angular matter-density and gravitational convergence auto- and cross-power spectra, $C_{\delta, \delta} ; C_{\delta, \kappa} ; C_{\kappa, \kappa}$, from the \textsc{RayGal} data. 

\begin{figure*}
   \centering
   \begin{tabular}{cc}
      \includegraphics[width=0.5\hsize]{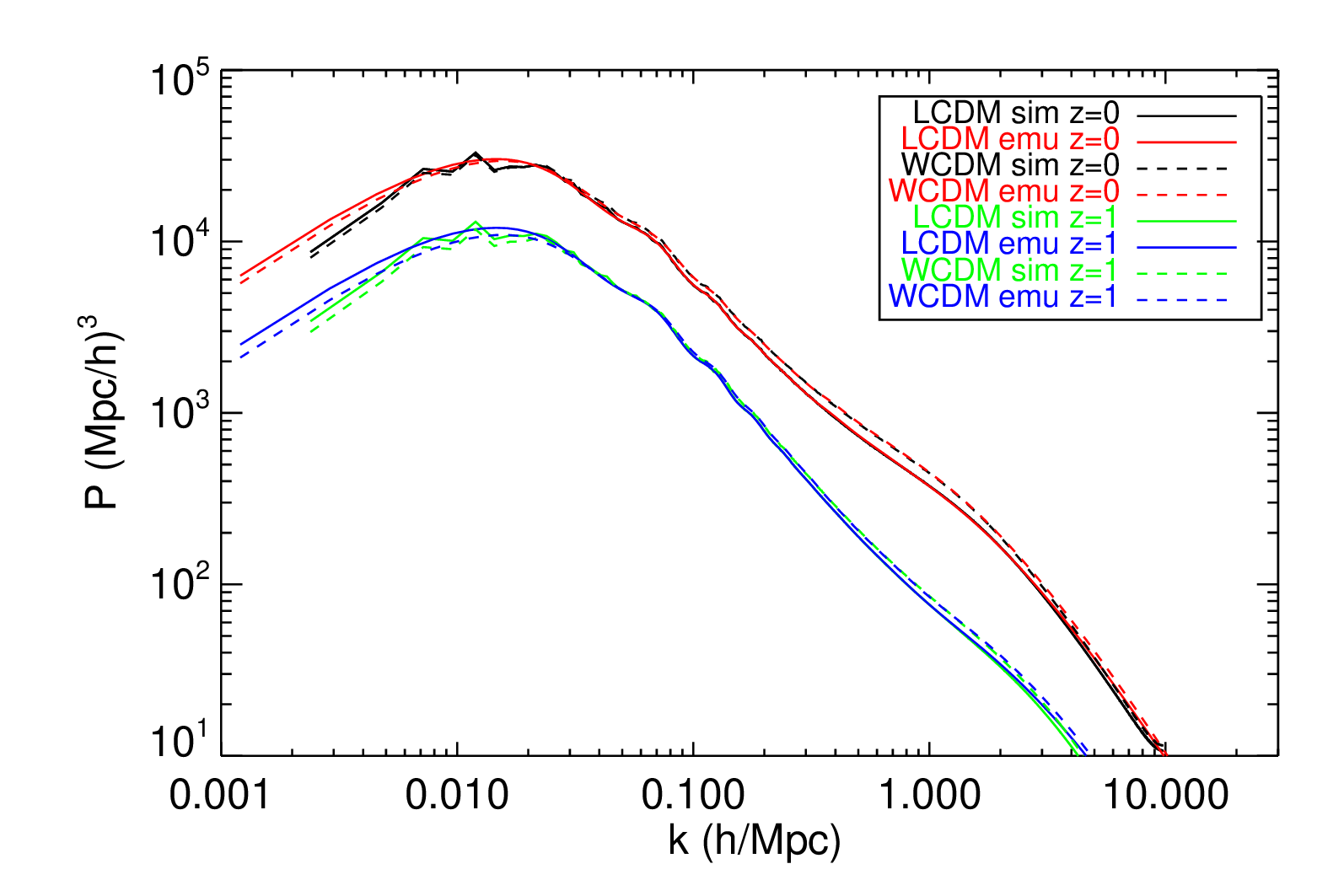}  & \includegraphics[width=0.5\hsize]{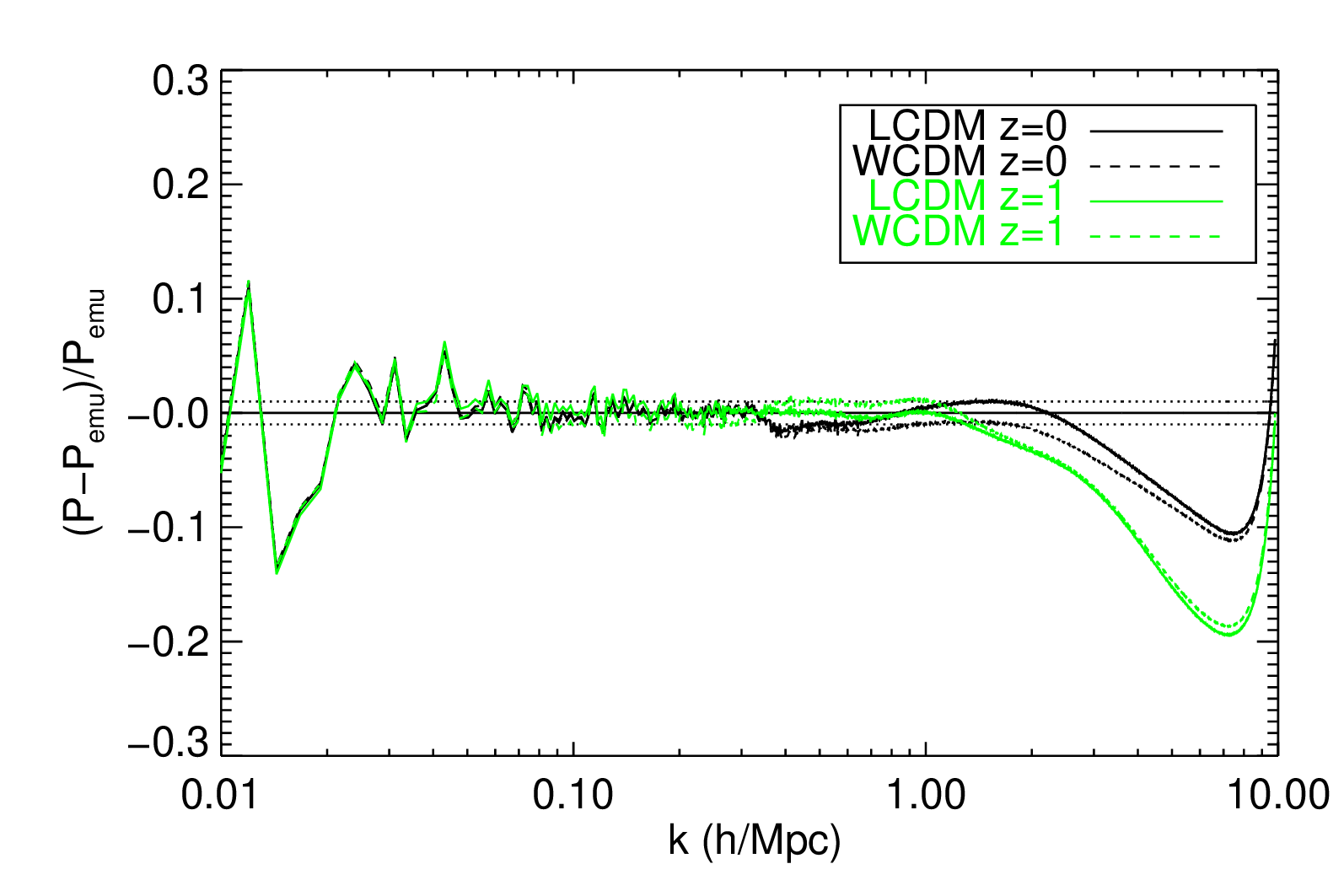} \\
   \end{tabular}
      \caption{\textsc{RayGal} real-space matter power spectra. Left: 3D matter power spectra at $z=0$ (top lines in black and red) and $z=1$ (bottom lines in green and blue) for $\Lambda$CDM (continuous lines) and $w$CDM (dashed lines) cosmologies. \textsc{RayGal} power spectra are shown in black ($z=0$) and green ($z=1$), and emulator power spectra are shown in red ($z=0$) and blue ($z=1$). Right: Relative deviations between simulations and emulator at $z=0$ (black) and $z=1$ (green) in $\Lambda$CDM (continuous lines) and $w$CDM (dashed lines) cosmologies. Percent level agreement is shown up to $k=1-2~h$~Mpc$^{-1}$, where resolution effects damp the power spectrum. Baryonic effects are also expected to damp the power spectrum near this wavenumber. 
      }
      \label{Fig:pk_raygalgroupsims_vs_coyote}
\end{figure*}

\subsection{Validation of the 3D Cartesian real-space matter power spectrum}

Although the real-space distribution is not the target of this study, we perform an analysis of the 3D matter power spectrum (computed from a snapshot of periodic-boundary-conditions simulations at a fixed redshift) as a test of robustness of the numerical analysis. First, we determine the range of validity of the matter power spectrum. Second, the evaluated matter power spectrum is used as input for our analytical predictions of relativistic effects. 

We compute the 3D matter power spectrum in real space at $z=0$ and $z=1$ (for both the $\Lambda$CDM and $w$CDM runs) with the fast Fourier transform based parallel estimator \textsc{Powergrid} \citep{prunet2008initial}. The density is computed using a Cloud-In-Cell assignment scheme within a cartesian grid with $8192^3$ elements (the power spectrum is then deconvolved from the assignment window). The comparison with spectra from the emulator \textsc{Cosmicemu} \citep{heitmann2016mira} shows a 1\% agreement up to $k=1$-$2~h$~Mpc$^{-1}$ for both cosmologies (see \cref{Fig:pk_raygalgroupsims_vs_coyote}). We notice that the location of the drop in power, due to finite mass resolution effect, is similar to that induced by baryonic physics and in particular active galactic nuclei feedback \citep{Chisari2018} in cosmological simulations with hydrodynamics. It indicates that a higher-resolution simulation would not necessarily improve matter power spectra predictions unless (costly) baryonic physics is properly and accurately taken into account (which is a non-trivial task). The increase in power near $k=7 - 10~h$~Mpc$^{-1}$ is due to both shot noise and aliasing, which have not been subtracted. The oscillations at small wavenumbers are mostly related to sample variance: they are limited to the $2-3\%$ level down to $k=0.02~h$~Mpc$^{-1}$ where linear theory starts to reach percent level precision. The percent level agreement between \textsc{Cosmicemu} and \textsc{RayGal} spectra over 2 decades of wavenumbers ($k=0.02$ to $2~h$~Mpc$^{-1}$) is a good cross-check of the precision of our real-space data.

\subsection{Clustering: Density-density angular spectra}
\label{subsec:result_deltadelta}

In this section we focus on the angular matter clustering, $C_{\delta_1 \delta_2} (\ell,z_1,z_2)$, for shells 1 and 2 at redshift  $z_1 \in \{0.7,1.8\}$ and $z_2 \in \{0.7,1.8\}$, respectively.
\subsubsection{Comoving density}
 \begin{figure*}
   \centering
   \begin{tabular}{cc}
      \includegraphics[width=0.5\hsize]{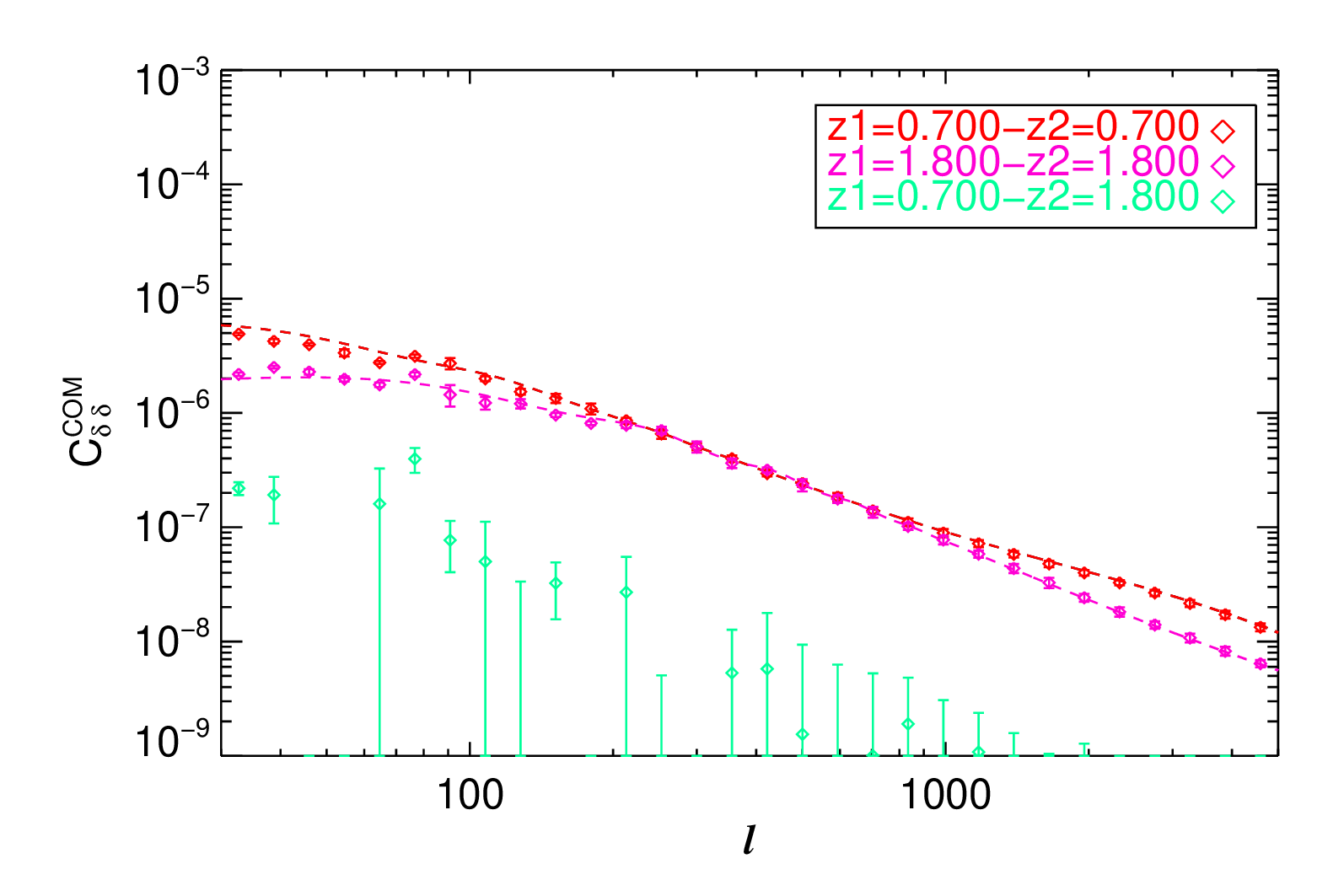}  & \includegraphics[width=0.5\hsize]{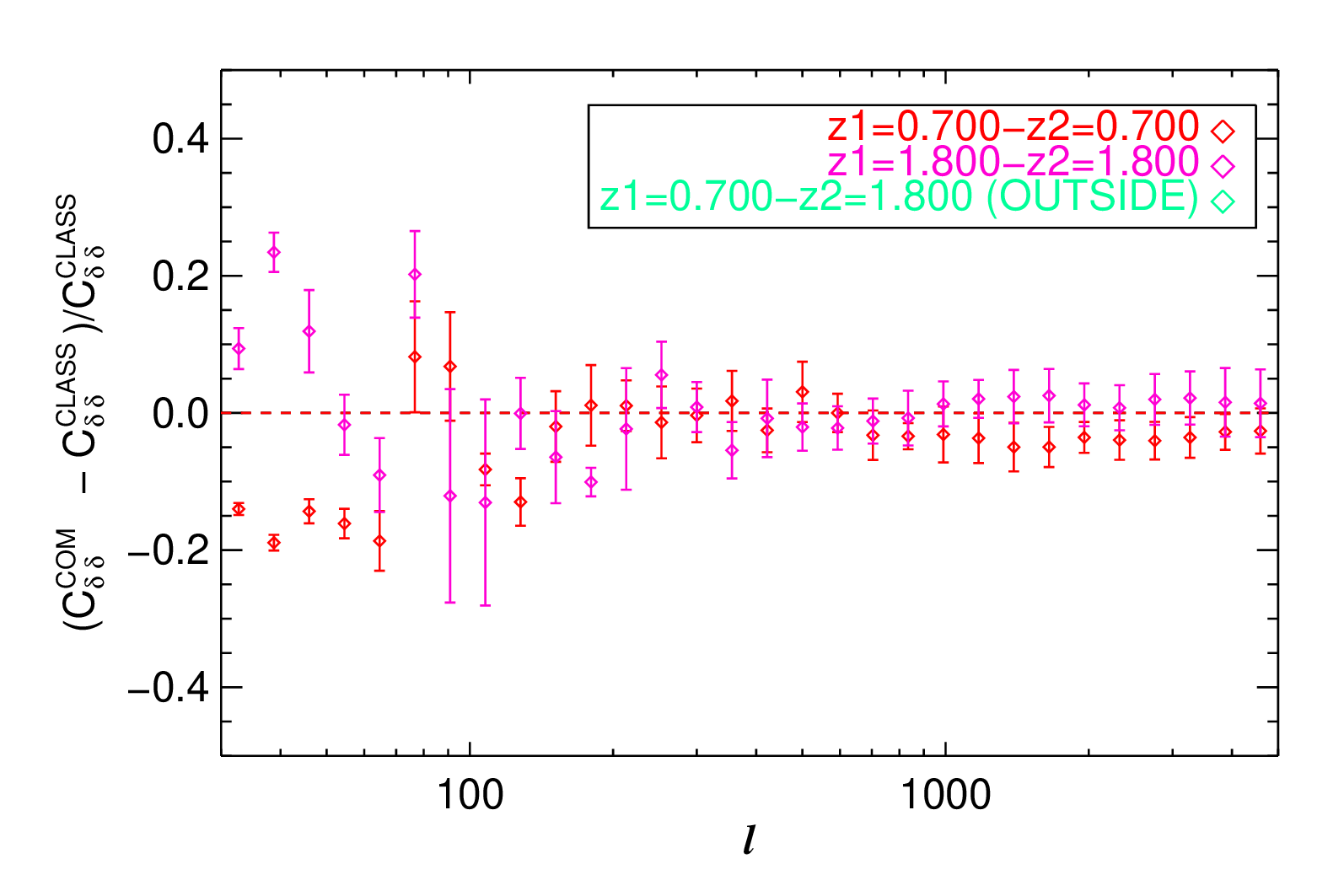} \\
   \end{tabular}
      \caption{{Comoving} matter-density angular auto-spectrum in a single shell at $z=0.7$ (red) and $z=1.8$ (pink) in $\Lambda$CDM cosmology. The cross-spectrum between two shells at $z=0.7$ and $z=1.8$ is shown in green. Left: Spectrum from the \textsc{RayGal} 2500~deg$^2$ light cone (diamonds with error bars) and the spectrum from \textsc{CLASS} (dashed line). We note that the \textsc{CLASS} prediction for the cross-spectrum is not visible since it is oscillating around zero with a small amplitude, reaching at maximum $10^{-10}$ at $\ell=30$. The \textsc{RayGal} cross-spectrum is also oscillating around zero but with a larger amplitude  (very likely due to sample variance), thus explaining the missing points, which are negative (or smaller than $10^{-9}$). Right: Relative deviation from the \textsc{CLASS} prediction. The cross-spectrum is nearly zero in real space, and the relative deviation is therefore outside the graph, with very large error bars (that are omitted for readability). Overall, we find a good agreement for the auto-spectrum, except at large scales (small $\ell$) due to the finite area of the light cone.} 
         \label{Fig:cl_lcdmw7_narrow_00002_delta_com_delta_com}
   \end{figure*}
   \begin{figure*}
   \centering
   \begin{tabular}{cc}
      \includegraphics[width=0.5\hsize]{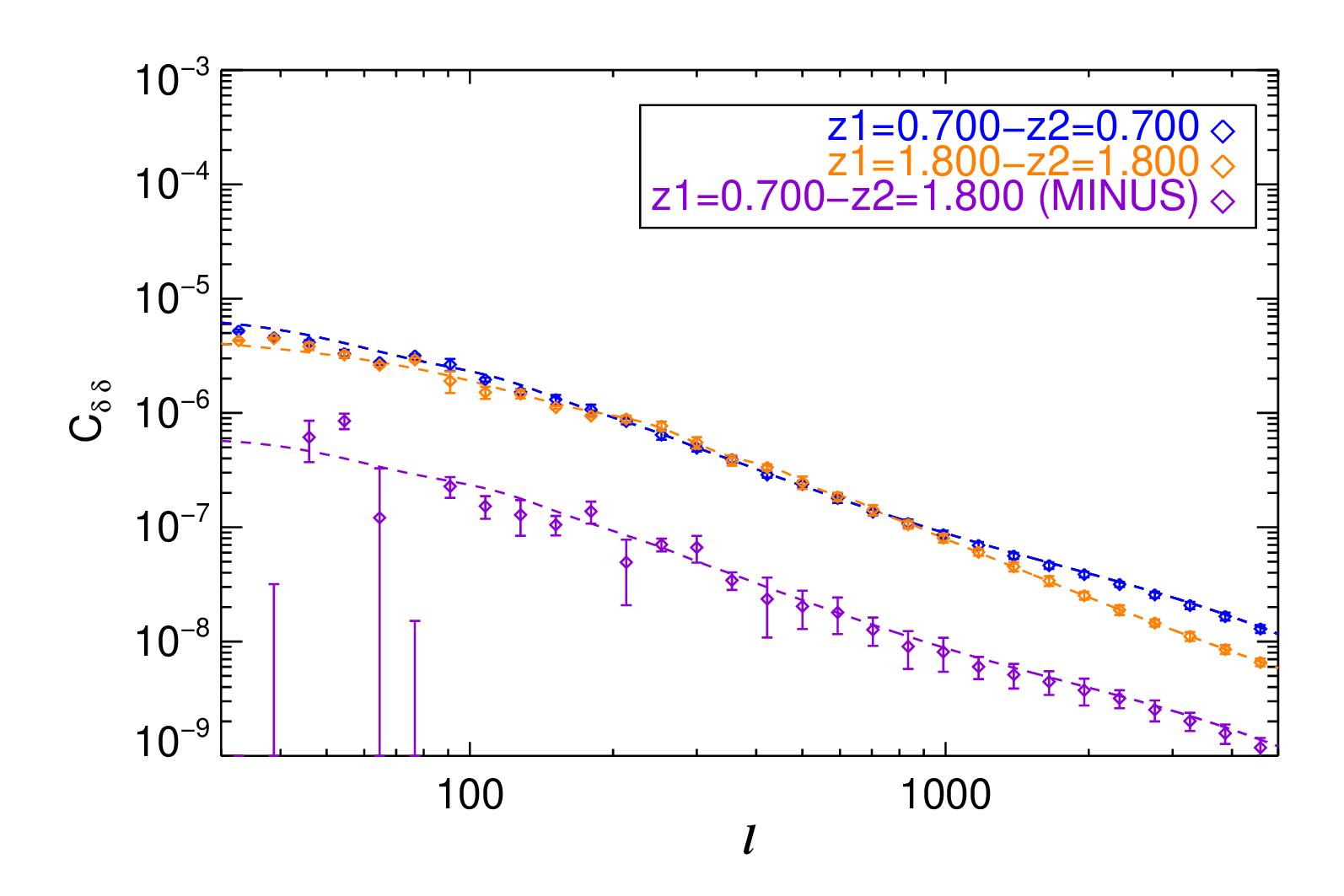}& \includegraphics[width=0.5\hsize]{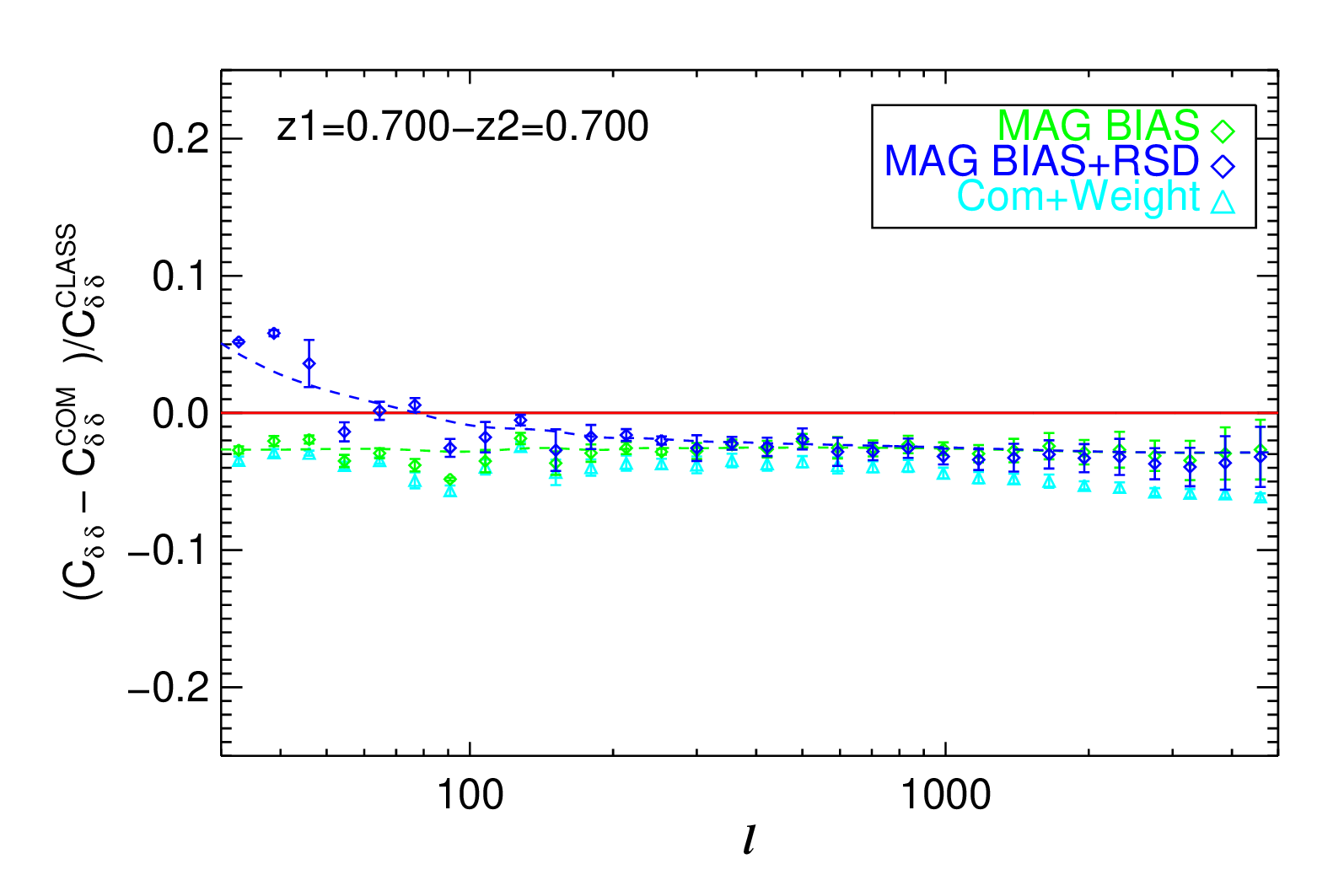}\\
      \includegraphics[width=0.5\hsize]{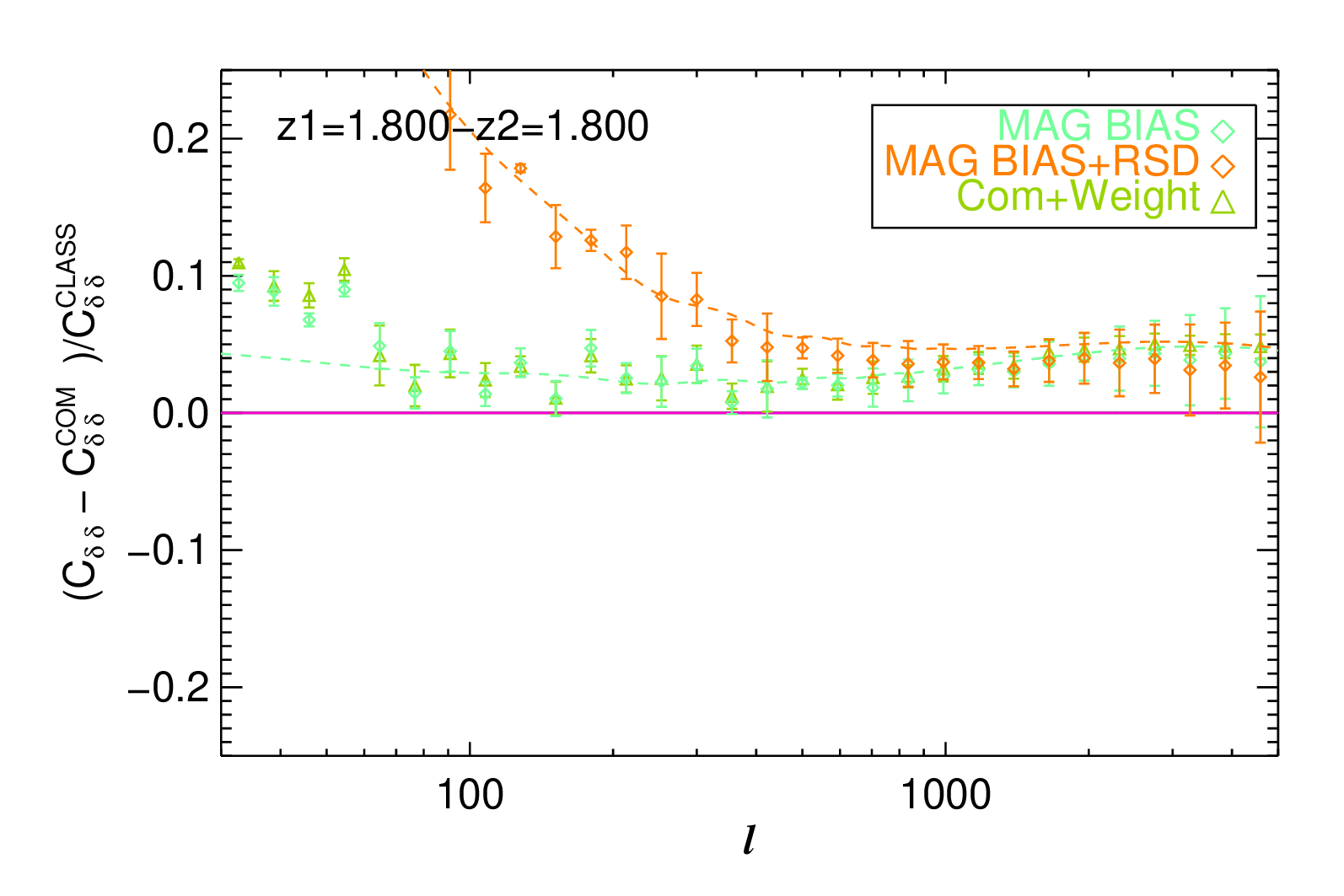}&\includegraphics[width=0.5\hsize]{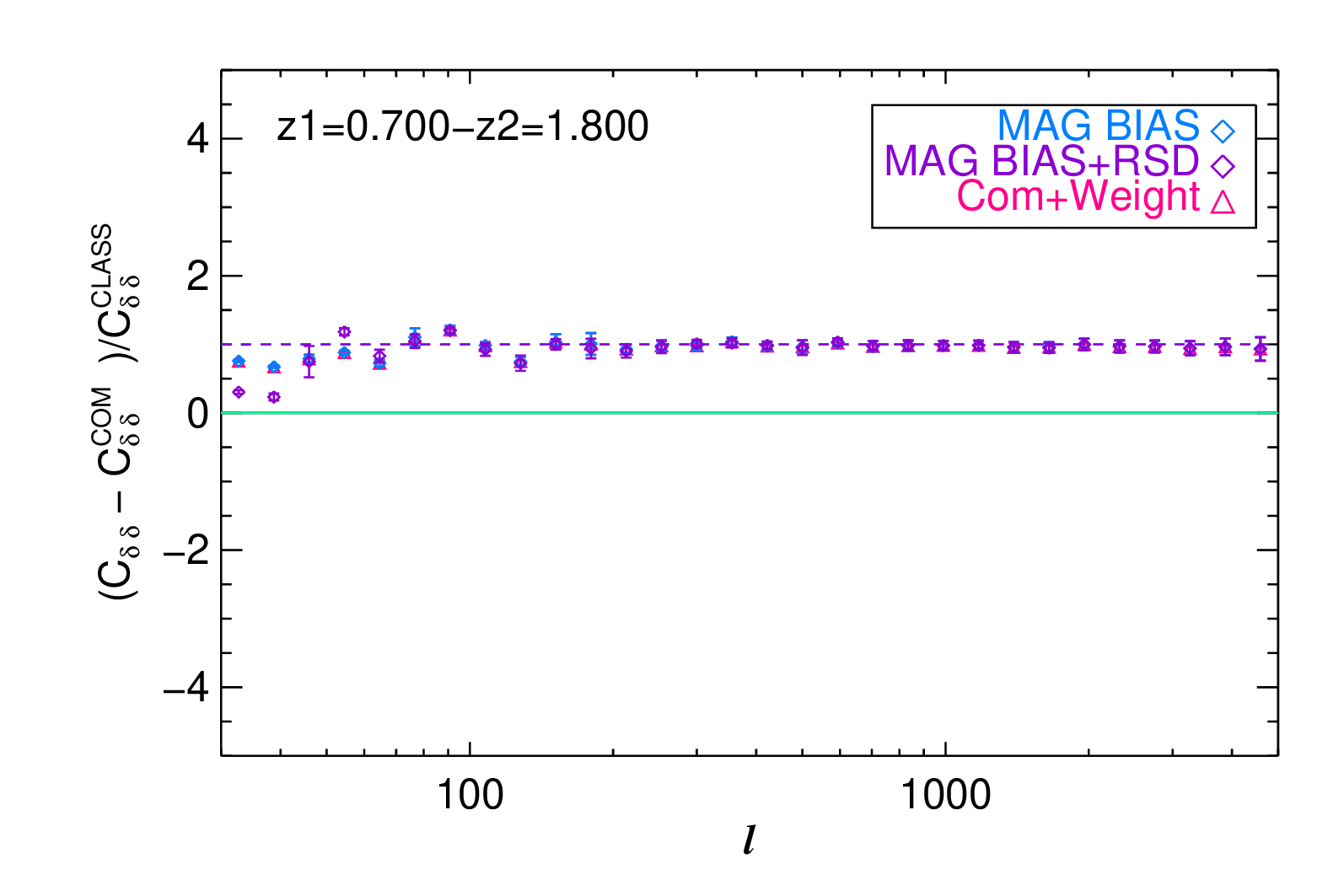}\\
     \end{tabular}
      \caption{Measurements from the \textsc{RayGal} 2500~deg$^2$ light cone (diamonds) and the \textsc{CLASS} predictions (dashed lines). Top left: Matter-density angular auto-spectrum {with relativistic corrections (MB+RSDs)} in a single shell at $z=0.7$ (blue) and $z=1.8$ (orange) in $\Lambda$CDM cosmology. The opposite of the cross-spectrum between two shells at $z=0.7$ and $z=1.8$ is shown in purple. 
      Top right:  Relative deviation from comoving matter-density angular auto-spectrum at $z=0.7$ in $\Lambda$CDM cosmology due to {relativistic effects}.  The effects of the dilution term of MB (i.e. $s=0$) on the observed matter-density spectrum is shown in green, and the effect of the dilution term of MB plus RSDs is shown in dark blue. An estimate of the MB effect using an inverse magnification ($|\mu_{\rm Born}|^{-1}$) weight is shown as light blue triangles (see text for details).
      Bottom left: Same but for $z=1.8$, with the MB effect in light green, MB+RSDs in orange, and the $|\mu_{\rm Born}|^{-1}$weight MB estimate in green.
      Bottom right: Same but for the $z=0.7-z=1.8$ cross-spectrum with MB in light blue, MB+RSDs in purple, and the $|\mu_{\rm Born}|^{-1}$weight MB estimate in pink. Overall, there is an impressive agreement with \textsc{Class} even though there is a combination of MB and RSD effects. For the cross-spectrum, relativistic effects largely dominate. These matter cross-spectra become sensitive probes of the density-convergence spectra.}
         \label{Fig:cl_lcdmw7_narrow_00002_delta_delta}
   \end{figure*}

   The comoving matter-density angular auto- and cross-spectra are shown \cref{Fig:cl_lcdmw7_narrow_00002_delta_com_delta_com}, left panel. The spectra are binned with 30 bins in the multipole interval $30\le \ell \le 5000$. Symbols represent the mean value of the spectra in the bin, while the error bars represent the standard deviation of the spectra within the same bin. The continuous lines show the analytical predictions. The auto-spectra $C_{\delta_1 \delta_2} (\ell,z_1=0.7,z_2=0.7)$ and $C_{\delta_1 \delta_2} (\ell,z_1=1.8,z_2=1.8)$  show an excellent agreement with the \textsc{Class} analytical prediction across the entire multipole interval. At large scale ($\ell<50$), the fluctuations we see are very likely to be related to the finite size of the light-cone angular aperture (sample variance). The good agreement in the linear regime between $\ell=50$ and $\ell=500$ validates our simulation measurement. The non-linear regime begins near $\ell\sim 500$, corresponding approximately to $k \sim$ $0.2~h$~Mpc$^{-1}$ at $z\sim1,$ where $P(k)$ is known to deviate from linear theory at the $\sim$10\% level (see e.g. \citealt{rasera2014cosmic}). As we calibrated the $P(k)$ used in the analytical predictions to that from the \textsc{RayGal} simulation, the agreement still holds in the non-linear regime between $\ell=500$ and $\ell=5000$. We halt at $\ell=5000$, which corresponds approximately to $k \sim$ $2~h$~Mpc$^{-1}$ at $z=1$, where finite resolution effects in our simulation (and possibly baryonic effects in the Universe) can no longer be neglected. Since the maximum $\ell$ we consider is of order of $N_{\rm side}$, we also expect the aliasing effect to be small. The green curve shows the cross-spectrum $C_{\delta_1 \delta_2} (\ell,z_1=0.7,z_2=1.8),$ which is completely negligible in real space both in numerical data and analytical prediction. This is because the shells are separated by about 1750~$h^{-1}$Mpc. The right panel shows the relative deviation from \textsc{Class}. It illustrates the agreement with the analytical prediction to be better than the $5\%$ level and within error bars for the auto-spectra. The cross-spectrum lies outside the plot because of a division by a nearly zero \textsc{Class} prediction. Overall, this is a good cross-validation of \textsc{Class} and \textsc{RayGal}'s random subset matter angular spectra.

\subsubsection{Relativistic effects}

In \cref{Fig:cl_lcdmw7_narrow_00002_delta_delta}, we plot the matter overdensity auto- and cross-spectra in the presence of the relativistic effects (upper left panel). We can see a noticeable change compared to the case without relativistic effects. shown in \cref{Fig:cl_lcdmw7_narrow_00002_delta_com_delta_com}. The agreement between \textsc{RayGal} measurements and \textsc{Class} prediction (including RSDs and MB) is as good as before from the linear to the non-linear regime. The cross-spectrum between the $z=0.7$ and $z=1.8$ shells is now very different from zero: it reaches about $10$\% of the auto-spectra, which is not negligible at all. This is due to weak lensing that couples the clustering signal between two density shells: the density in the low-redshift shell is a lens for the density of the high-redshift shell. Here again there is a good agreement between predictions and simulations except below $\ell=50$ where the simulation spectrum strongly underestimates the analytical predictions. This is likely related to the limited aperture of the narrow light cone, which affects this subtle correlation.
   
The other panels allow us to investigate relativistic effects in more details: they show the relativistic contributions normalised by the \textsc{Class} spectra including all relativistic effects (we chose this normalisation since this is a smooth and non-zero spectrum in all our plots). The upper right and bottom-left panels show the relativistic contributions to the auto-spectra. The MB contribution is mostly constant of order $+3$\% at $z=0.7$ and $-3$\% at $z=1.8$. This signature of lensing on the matter distribution is a subtle effect, well captured by \textsc{RayGal}, in good agreement with \textsc{Class} expectation from $\ell=50$ to $\ell=5000$. This effect has been investigated in several works (e.g. \citealt{fosalba2015miceA} and reference therein). It is related to the so-called MB, where the apparent density becomes (at first order) $\delta_i=\delta^{\rm com}_i+ (5s-2) \kappa_i $ (with $s$ the logarithmic slope of the source number count, which is $s=0$ for our mass limited sample; in this peculiar case it is also called dilution bias). 
As a consequence, there is an additional lensing contribution to the matter-density angular power spectrum (see the second line of \cref{Eq:Cl_delta_delta_linmap}) that can be simplified as   \begin{equation}
   \label{eq:cldeltadelta}
   C_{\delta_i \delta_j}(l)\approx C_{\delta^{\rm com}_i \delta^{\rm com}_j}(l)-2 C_{\delta_i \kappa_j}(l)-2 C_{\kappa_i \delta_j}(l)+ 4 C_{\kappa_i \kappa_j}(l).  
   \end{equation}
   The RSD contributions to the multipoles of the 3D galaxy-galaxy two-point correlation function is important in the linear regime (the Kaiser effect, which corresponds to an amplification of the clustering due to coherent motions plus a quadrupole and hexadecapole component) and in the non-linear regime (the finger-of-God effect, which corresponds to a damping of the small-scale clustering due to incoherent motions). These contributions are however much smaller when using large spherical shells ($>100~h^{-1}$~Mpc) centred around the observer. Redshift perturbations mostly modify the apparent radial distance of the sources and therefore do not have a strong effect on the angular power spectra. There is however, a subtle and non negligible amplification (see the first term of the third line of \cref{Eq:Cl_delta_delta_linmap}) due to the Kaiser effect visible at large scale below $\ell \sim 100$. For the shell thickness we use, it can reach $5$\% at $\ell=20$ (for the $z=0.7\pm0.2$ spectra) and $20$\% at $\ell=90$ (for the $z=1.8\pm 0.1$ spectra). While the geometry makes the effect non-trivial, it is well reproduced by the simulation on agreement with \textsc{Class} predictions. The MB effect is also well captured up to $\ell\approx 1000$ using our MB estimate based on Born inverse magnification weight (we discuss the origin of the small discrepancy at large $\ell$ and low redshift in the lensing section). For both shells, the RSD effect is too small at large $\ell$ and we cannot see the damping behaviour of the spectra. In  \cref{subsec:redshift_evol}, we performed again the same exercise but at lower redshift, for narrower shells using the full-sky data. The damping of the angular spectra (related to the fingers-of-God effect) now becomes apparent and is not well reproduced by \textsc{Class}.  
   
   The bottom-right panel shows the relativistic contribution to the density cross-spectrum between two shells. Because the real-space spectrum is nearly zero, the relativistic contribution is compatible with $100\%$ for both \textsc{Class} and \textsc{RayGal} (between $\ell=50$ and $\ell=5000$). We note that we normalised the spectrum variation by the \textsc{Class} spectrum (even for \textsc{RayGal} spectrum). The good agreement between \textsc{Class} and \textsc{RayGal} means that the amplitudes of the total spectra including relativistic effects are close to each other. The amplitude is also in agreement with our MB estimate based on Born inverse magnification weight. Here again the relativistic effect can be understood following \cref{eq:cldeltadelta} where now the contribution of $C_{\delta \kappa}(l)$ becomes the dominant one compared to $C_{\delta^{\rm com} \delta^{\rm com}}(l)$ (and $C_{\kappa \kappa}(l)$). It means that the matter cross-spectra between two widely separated shell is a good estimate of (-2 times) the density-convergence spectrum. We now investigate these two spectra ($C_{\kappa \kappa}(l)$ and $C_{\delta \kappa}(l)$ ) in detail in the next subsections.  
   
   {Overall, we conclude that it is important to consider both RSD and weak-lensing effects including all the corrections from the linear to non-linear regime: our simulations provide a good `laboratory' to test these effects by naturally including all these effects in a natural common framework.}

\subsection{Lensing: Convergence-convergence angular spectra}
\label{subsec:result_kappakappa}
\begin{figure*}
   \centering
   \begin{tabular}{cc}
      \includegraphics[width=0.5\hsize]{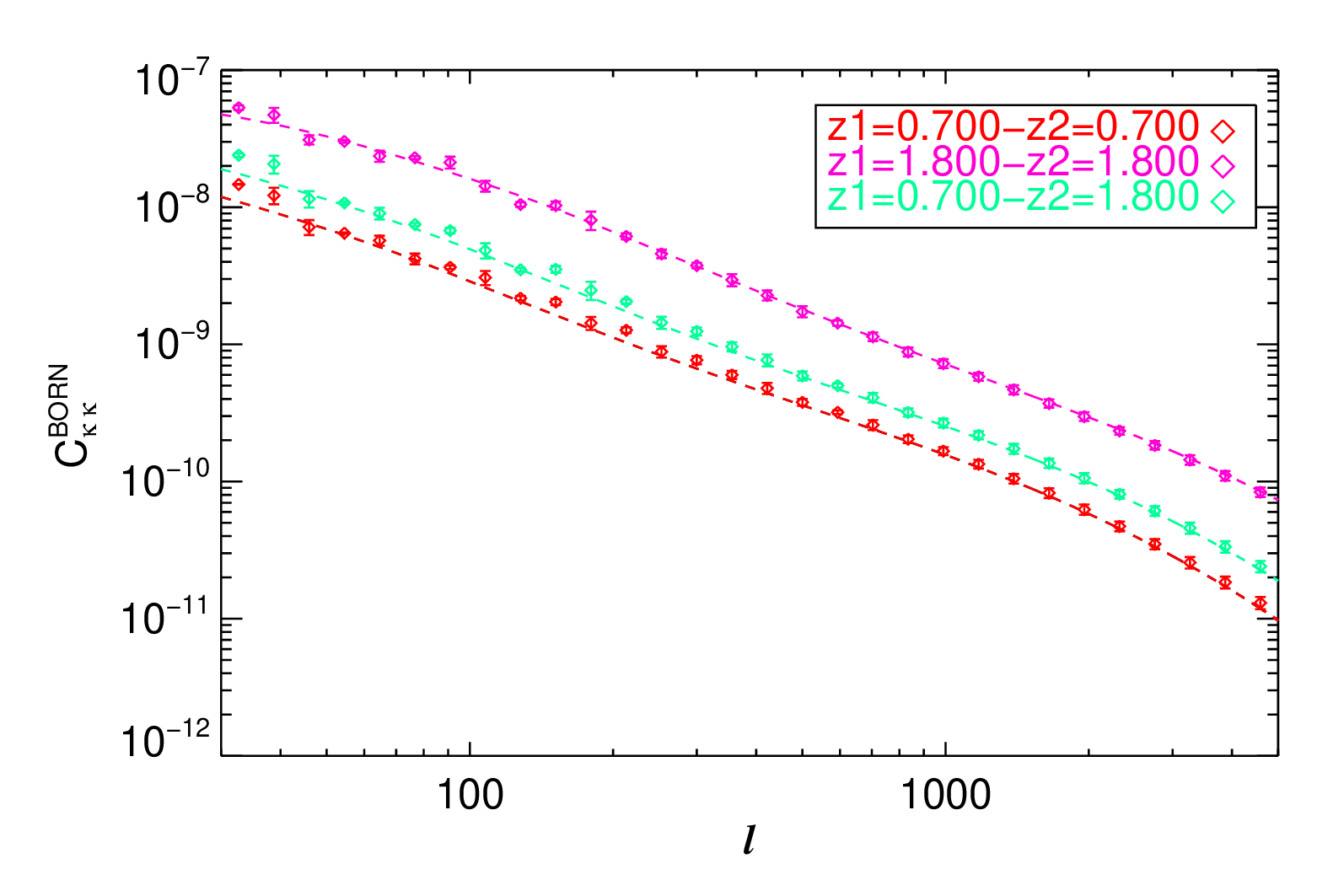}  & \includegraphics[width=0.5\hsize]{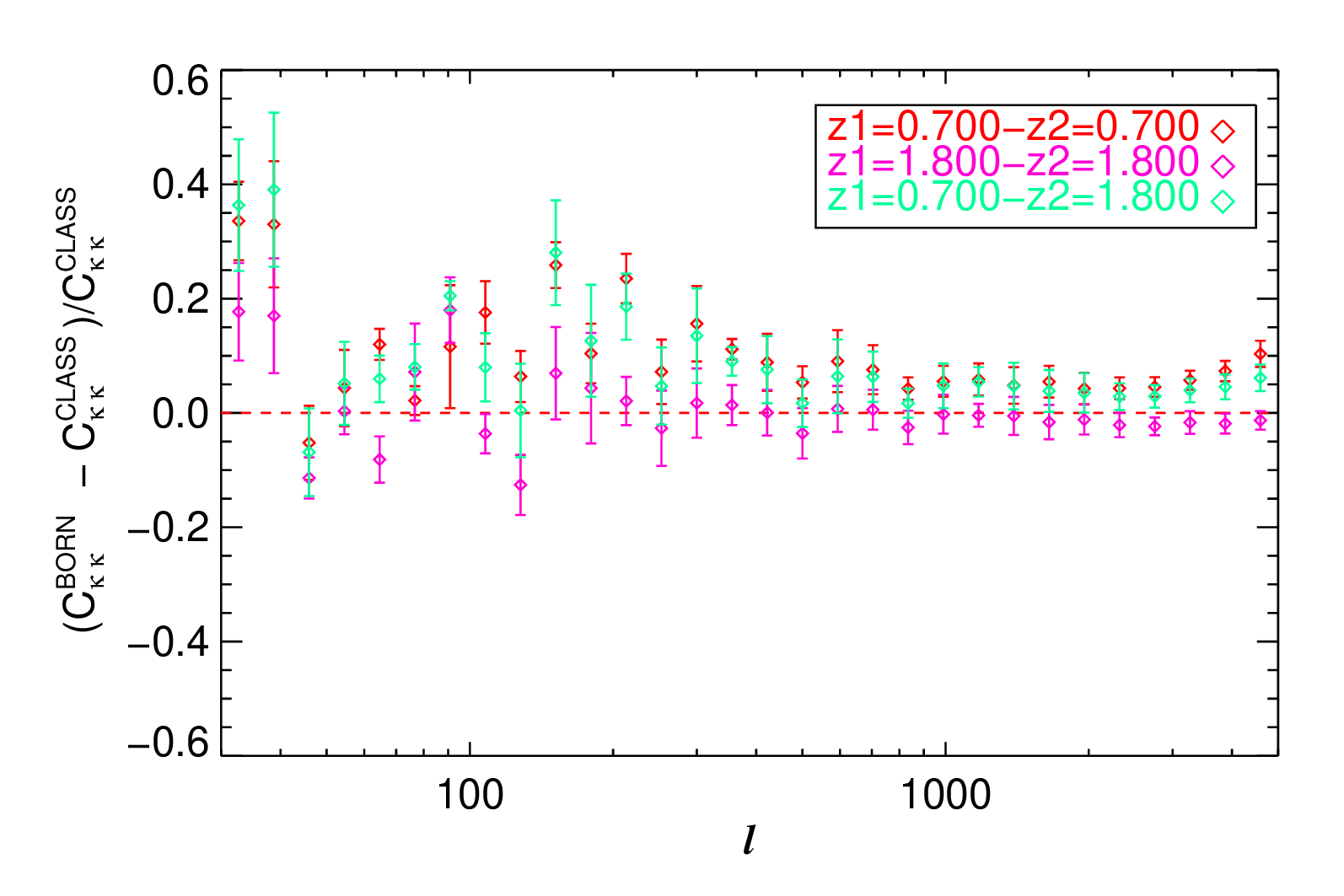} \\
   \end{tabular}
      \caption{Gravitational convergence angular auto-spectrum in a single shell at $z=0.7$ (red) and $z=1.8$ (pink) in $\Lambda$CDM cosmology assuming the {Born approximation}. The cross-spectrum between two shells at $z=0.7$ and $z=1.8$ is shown in green. Left: Spectrum from the \textsc{RayGal} 2500~deg$^2$ light cone (diamonds with error bars) and spectrum from \textsc{Class} (dashed line). Right: Relative deviation from the \textsc{Class} prediction. Overall, the measured power spectra are in agreement with \textsc{Class} predictions within the error bars except at large scales (small $\ell$) due to the finite area of the light cone. We also note a shift at the $\sim$5-10\% level at large $\ell$ and low redshifts, which vanishes at larger redshifts. 
      }
         \label{Fig:cl_lcdmw7_narrow_00002_kappa_born_kappa_born}
   \end{figure*}
\begin{figure*}
   \centering
     \begin{tabular}{cc}
      \includegraphics[width=0.48\hsize]{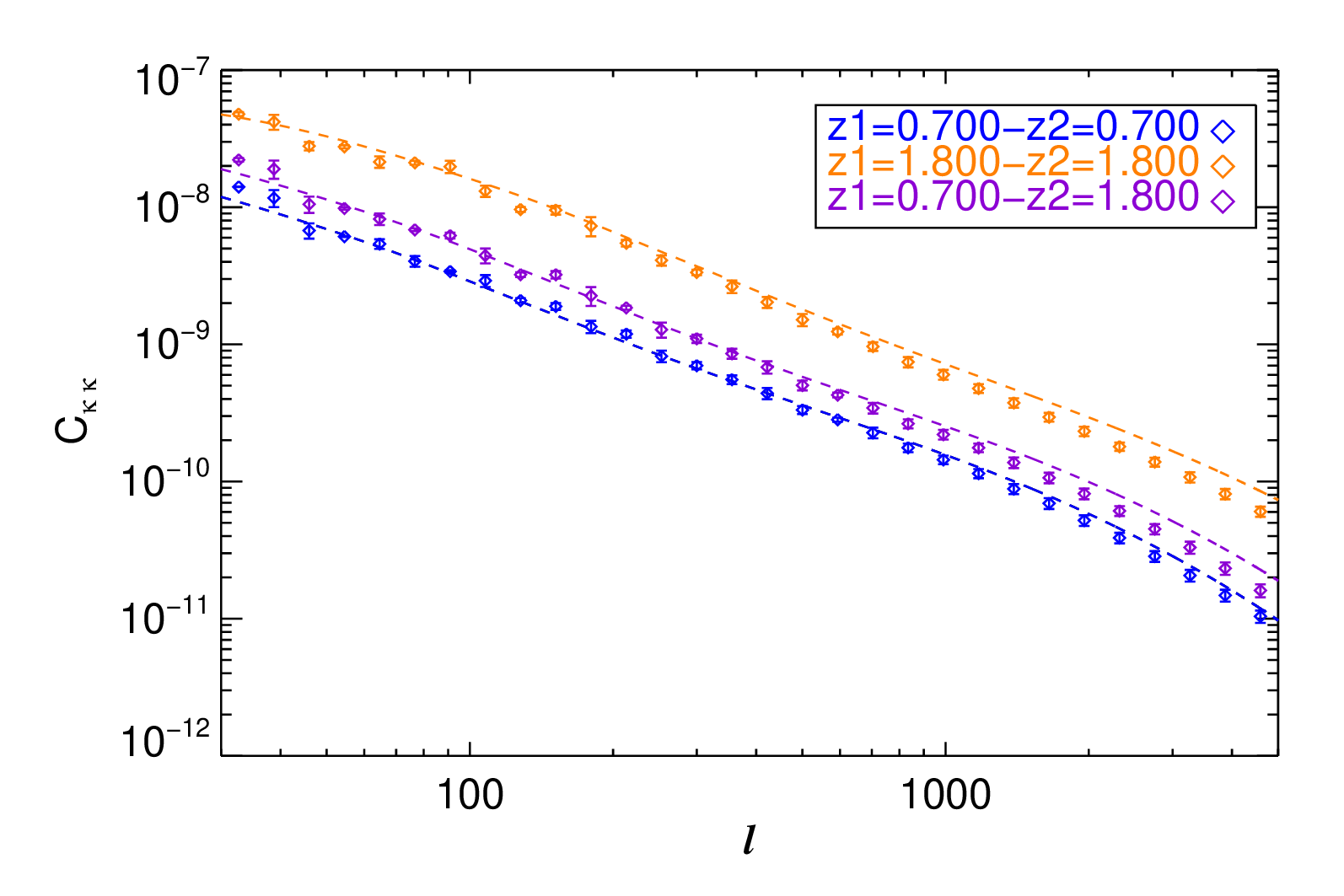}& \includegraphics[width=0.48\hsize]{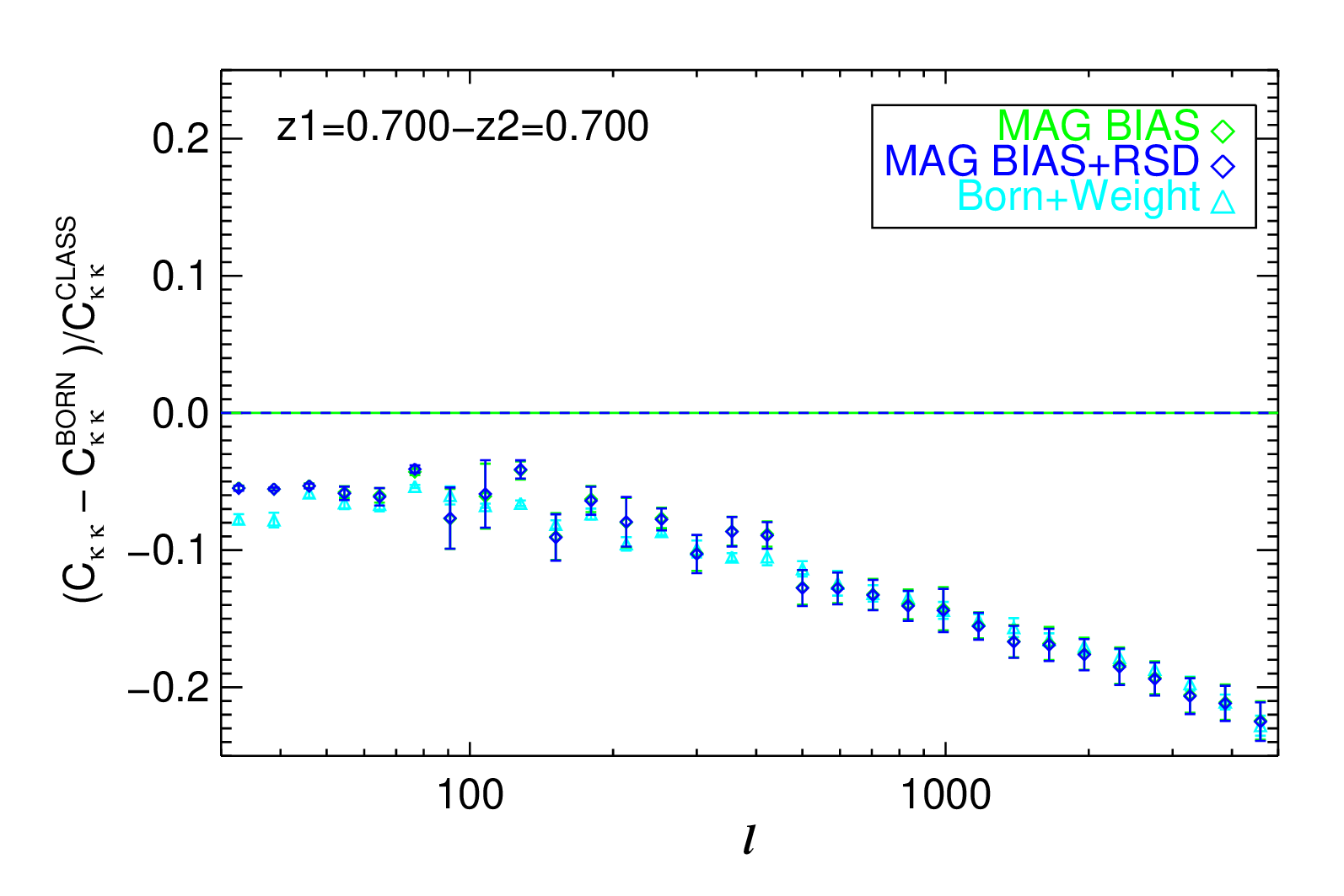}\\
      \includegraphics[width=0.48\hsize]{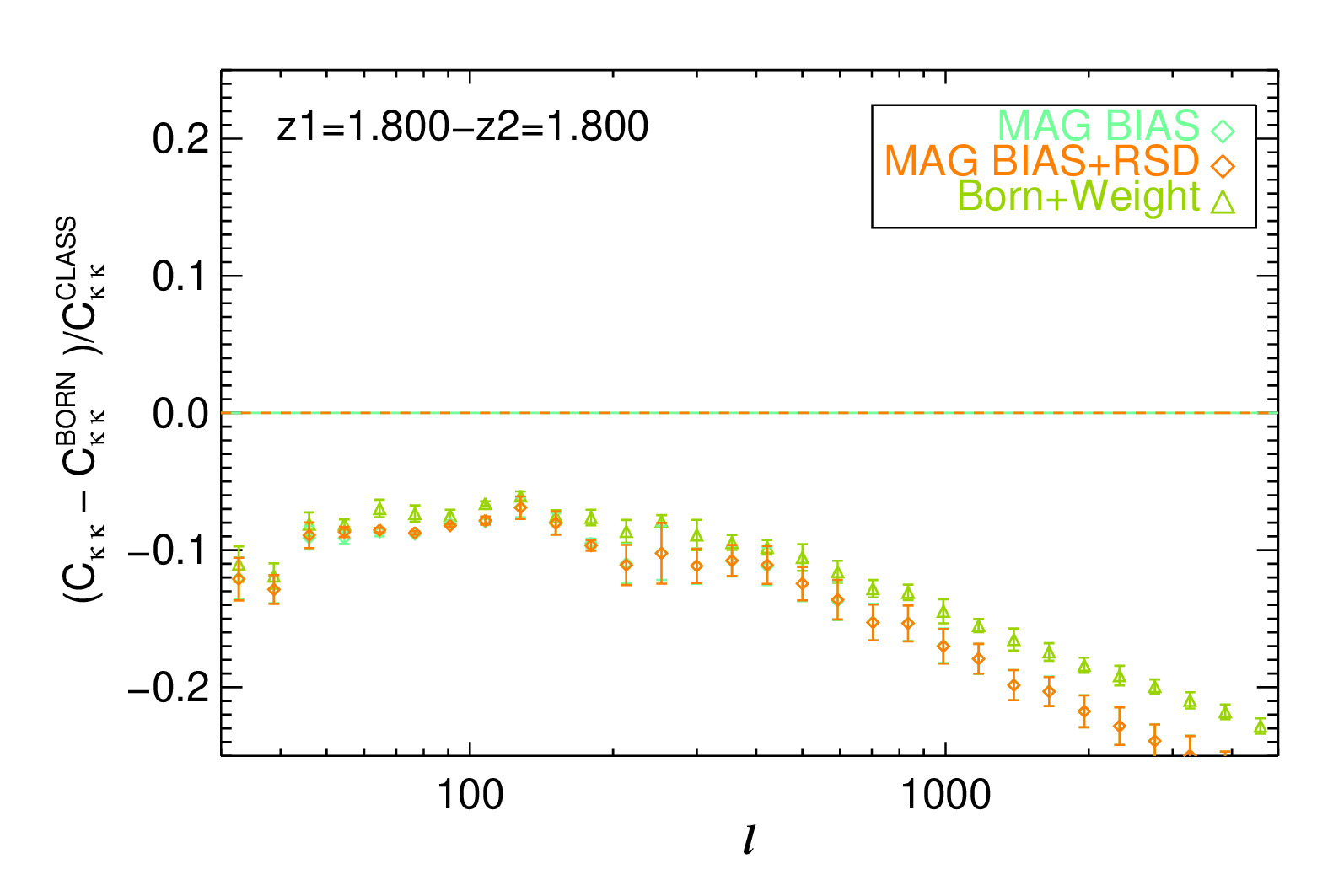}&
      \includegraphics[width=0.48\hsize]{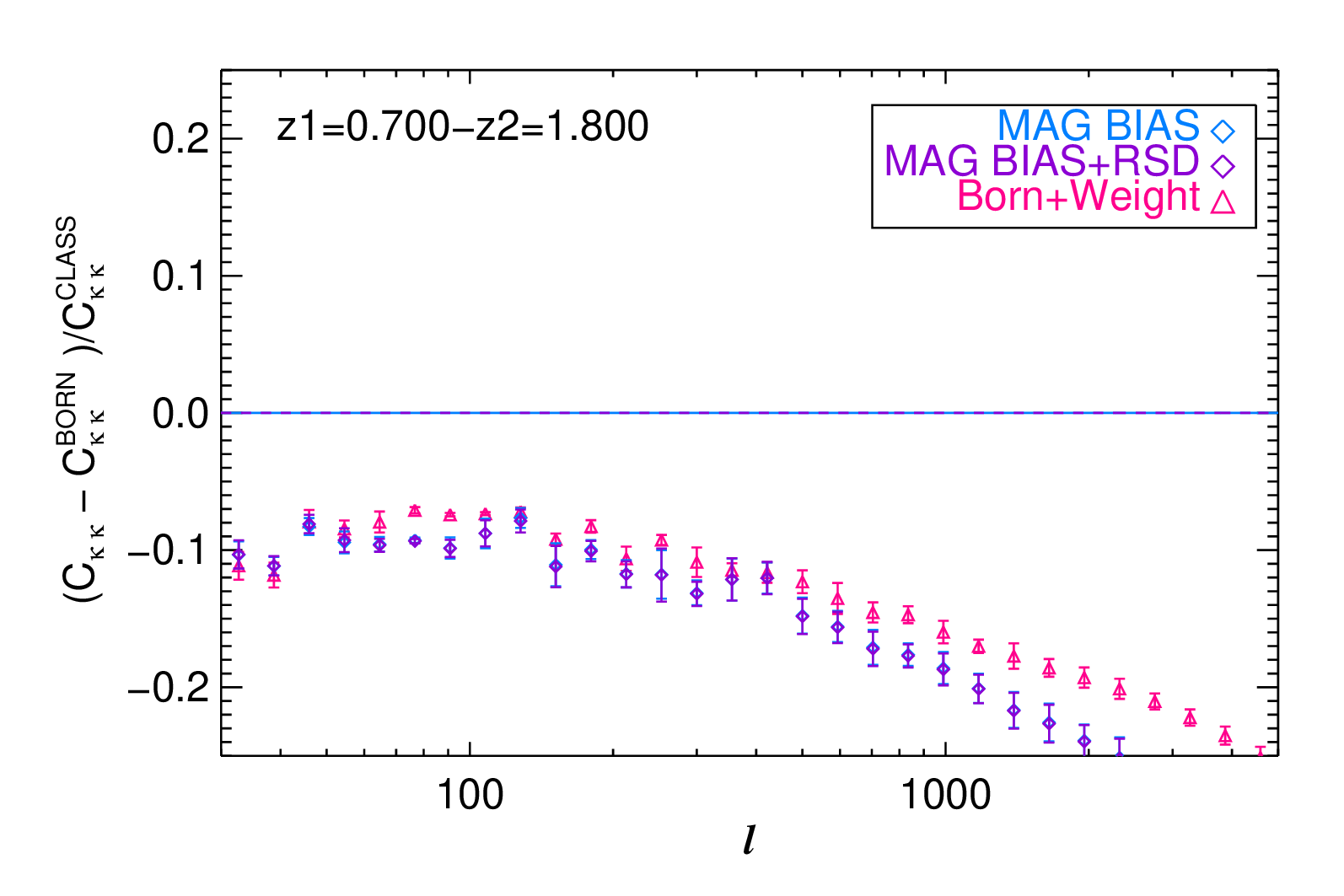}\\
     \end{tabular}
      \caption{Measurements from the \textsc{RayGal} 2500~deg$^2$ light cone (diamonds) and the \textsc{Class} predictions (dashed lines). Top left: Gravitational convergence auto-spectrum $C_{\kappa \kappa}^l$ {with relativistic corrections (MB and RSDs)} in a single shell at $z=0.7$ (blue) and $z=1.8$ (orange) in $\Lambda$CDM cosmology. The cross-spectrum between two shells at $z=0.7$ and $z=1.8$ is shown in purple. 
  Top right:  Relative deviation from gravitational convergence angular auto-spectrum (under the Born approximation) at $z=0.7$ in $\Lambda$CDM cosmology due to {non-trivial relativistic effects}. The full effects of MB (with $s=0$) are shown in green, while the effect of RSDs(+MB) is shown in blue. An estimate of the MB effect using an inverse magnification ($|\mu_{\rm Born}|^{-1}$) weight is shown as light blue triangles (see text for details). Bottom left: Same but for $z=1.8$ with MB in light green, RSDs(+MB) in orange, and the  $|\mu_{\rm Born}|^{-1}$ weight MB estimate in green. Bottom right: Same but for the $z=0.7-z=1.8$ cross-spectrum with MB in light blue, RSDs(+MB) in purple, and the $|\mu_{\rm Born}|^{-1}$ weight MB estimate in pink. MB effects are not accounted for in \textsc{Class} and cause $\sim 10-30\%$ deviations. This confirms that the dominant effect is the dilution term of MB related to source averaging (as opposed to angular averaging).
  }
\label{Fig:cl_lcdmw7_narrow_00002_kappa_kappa}
\end{figure*}
   
In this section we focus on the weak-lensing spectra, $C_{\kappa_1 \kappa_2} (\ell, z_1, z_2)$, for shells 1 and 2 at redshifts $z_1 \in \{0.7, 1.8\}$ and $z_2 \in \{0.7, 1.8\}$. We follow exactly the same methodology as for the clustering. We remind that we focus on the convergence defined from \cref{eq:distortion_matrix}.

\subsubsection{Born approximation}
   
We investigate the gravitational convergence angular spectra under the Born approximation in \cref{Fig:cl_lcdmw7_narrow_00002_kappa_born_kappa_born}. The spectra decrease with the multipole but now increase with redshifts (unlike the overdensity spectra). This is because the convergence is an integral of the overdensity (weighed by the lensing kernel) along the line of sight. A farther away source is therefore more distorted than a closer one. The cross-spectrum behaviour is similar to the auto-spectra one since $\kappa$ is not a local quantity but an integrated one: all the shells are therefore correlated.  Here again we find a good agreement between \textsc{Class} prediction and \textsc{RayGal} measurements in the linear and non-linear regime. The right panel showing the relative deviation between the two indicates that they are compatible within the error bars between $\ell=50$ and $\ell=2000$. The deviation below $\ell=50$ is related to the limited angular size of the cone. Between $\ell=2000$ and $\ell=5000$ there is a $\sim 5-10$\% deviation at low redshift, the simulation measurement is slightly higher than the prediction from \textsc{Class}. We are not sure of the origin of the deviation, which might be related to shot noise. 
Despite such a discrepancy at large $\ell$ that seems to vanish at large redshift, this provides a cross-validation of \textsc{RayGal} Born convergence maps and the \textsc{Class} predictions.  
  
\subsubsection{Relativistic effects}

We now move to \cref{Fig:cl_lcdmw7_narrow_00002_kappa_kappa}, which includes relativistic effects. The upper left plot shows a similar trend as with the Born approximation. However, a careful inspection shows that the points have moved from above \textsc{Class} predictions to below. We now focus on the three other panels (upper right, bottom left and bottom right), which show the relative deviations from the Born convergence spectra due to relativistic effects. It is interesting to note that RSDs do not seem to play any role in both \textsc{Class} predictions and \textsc{RayGal} measurements. This is not trivial since redshift perturbations modify the apparent distribution of sources (and thus the observed densities). However, these changes do not affect the lensing itself since they are mostly radial. Instead, MB plays an important role on the convergence spectra, which is not predicted by \textsc{Class} (\textsc{Class} relies on Born approximation, it can only account for the effect of MB on the overdensity but not on the convergence). The first contribution one can think of is the post-Born effect. These contributions are at the percent level and play a role mostly beyond $\ell=2000$ (see e.g. \citealt{petri2017}). The second (and main) contribution comes from (the dilution term of) MB: magnified regions have fewer sources than de-magnified ones. The decrease in the convergence spectra (computed from the observed position of sources) is not negligible at all since it grows from $5$\% to $25$\% between $\ell=100$ and $\ell=5000$. The decrease is also more important at larger redshifts. This is because the convergence is weighed by the density, which itself is de-magnified by lensing. At first order (see \cref{eq:kappa_decomposition}), $\kappa_i=\kappa_i^{\rm Born} (1+(5s-2)\kappa_i^{\rm Born})$ (with  $s=0$ in our mass-limited sample), and as a consequence, the convergence two-point spectrum 
(see \cref{Eq:Cl_kappa_kappa_linmap}) is approximated by
\begin{equation}
   \label{eq:clkappakappa}
   C_{\kappa_i \kappa_j}(l)\approx C_{\kappa^{\rm Born}_i \kappa^{\rm Born}_j}(l)-2 C_{\kappa_i \kappa_j^2}(l)-2 C_{\kappa_i^2 \kappa_j}(l)+ 4 C_{\kappa_i^2 \kappa_j^2}(l). 
\end{equation}

\noindent The convergence spectrum depends on some $\langle\kappa_i \kappa_j^2\rangle$ terms that are not necessarily small, especially at high redshifts. 

Alternatively, it is possible to emulate such an effect using the Born approximation and weighting the convergence by the inverse magnification for $s=0$. This is what we call Born + Weight in \cref{Fig:cl_lcdmw7_narrow_00002_kappa_kappa}. The trend is the same as with the catalogues, with small discrepancies at large $\ell>1000$. These most likely come from the fact that the value of the magnification (computed with he Born approximation) is not exactly the same as the density ratio in a pixel between the lensed and un-lensed matter distribution. However, this is a cross-check of the large effect due to lensing on convergence angular spectra.

Such a lensing bias has been investigated by \citet{schmidt2009}, who focused on the shear power spectrum and found similar trend (in their notation, $q=2$ corresponds to the opposite of our $s=0$ correction). They found the correction to increase with $\ell$ and with redshift. For a source at $z=1$ the (norm of the) correction increase from $2$\% at $\ell=500$ to $7$\% at $\ell=5000$ in their article, while we find a correction about four times larger (at $z=0.7$). More recently the same trend was found but with smaller amplitude in \citet{Deshpande2020magbias}\footnote{At first order, all the results can be compared by re-normalising the correction so that $5s-2 \rightarrow -2$.}.  Our result is not in contradiction with these findings since the effect of MB is much stronger on the convergence than on the shear (see \cref{appendix:2PCF_convergence_z1p8}). The origin of the lensing convergence bias is the same as the lensing shear bias: the magnified regions are less sampled than the de-magnified ones leading to a bias. However, the magnification is directly related at first order to $\kappa$ (and not to $\gamma$) and the effect on the convergence power spectrum is therefore larger. The bi-spectrum $\langle\gamma \gamma \kappa\rangle$ is indeed much smaller than the bi-spectrum $\langle\kappa \kappa \kappa\rangle$. {An important consequence is therefore that the convergence power spectrum (or correlation function) can differ from the shear power spectrum (or correlation function) by up to $10-30\%$ due to MB.}

{To conclude, we find that we reproduce basic weak-lensing results under the Born approximation and we capture non-trivial effects, such as MB, which turn out to play an important role in the resulting spectra.}

\subsection{Galaxy-galaxy lensing: Density-convergence angular cross-spectra}
\label{subsec:result_ggl}

In this section we focus on the galaxy-galaxy lensing spectra, $C_{\delta_1 \kappa_2} (\ell, z_1, z_2)$, for shells 1 and 2 at redshift $(z_1, z_2) \in \{0.7, 1.8\}^2$.
We follow exactly the same methodology as for the clustering and the weak-lensing parts. Investigating galaxy-galaxy lensing in the non-linear regime is challenging since it involves non-trivial relativistic effects on both the overdensity and the convergence. To our knowledge an accurate study of such effects down to the non-linear regime is still lacking and we now try to fill in the gap. 

\subsubsection{Comoving density and Born approximation}
\begin{figure*}
   \centering
   \begin{tabular}{cc}
      \includegraphics[width=0.48\hsize]{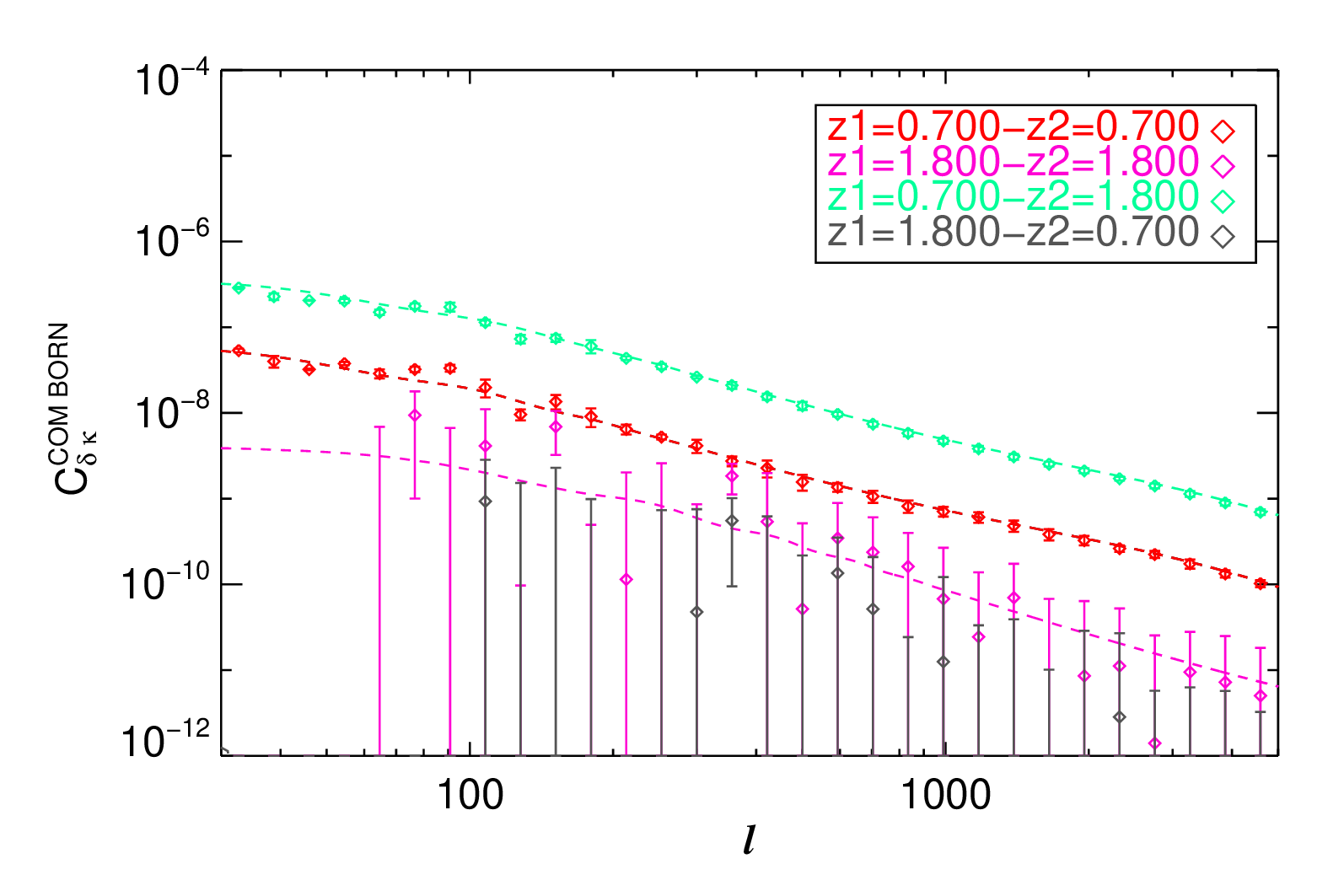}
      & \includegraphics[width=0.48\hsize]{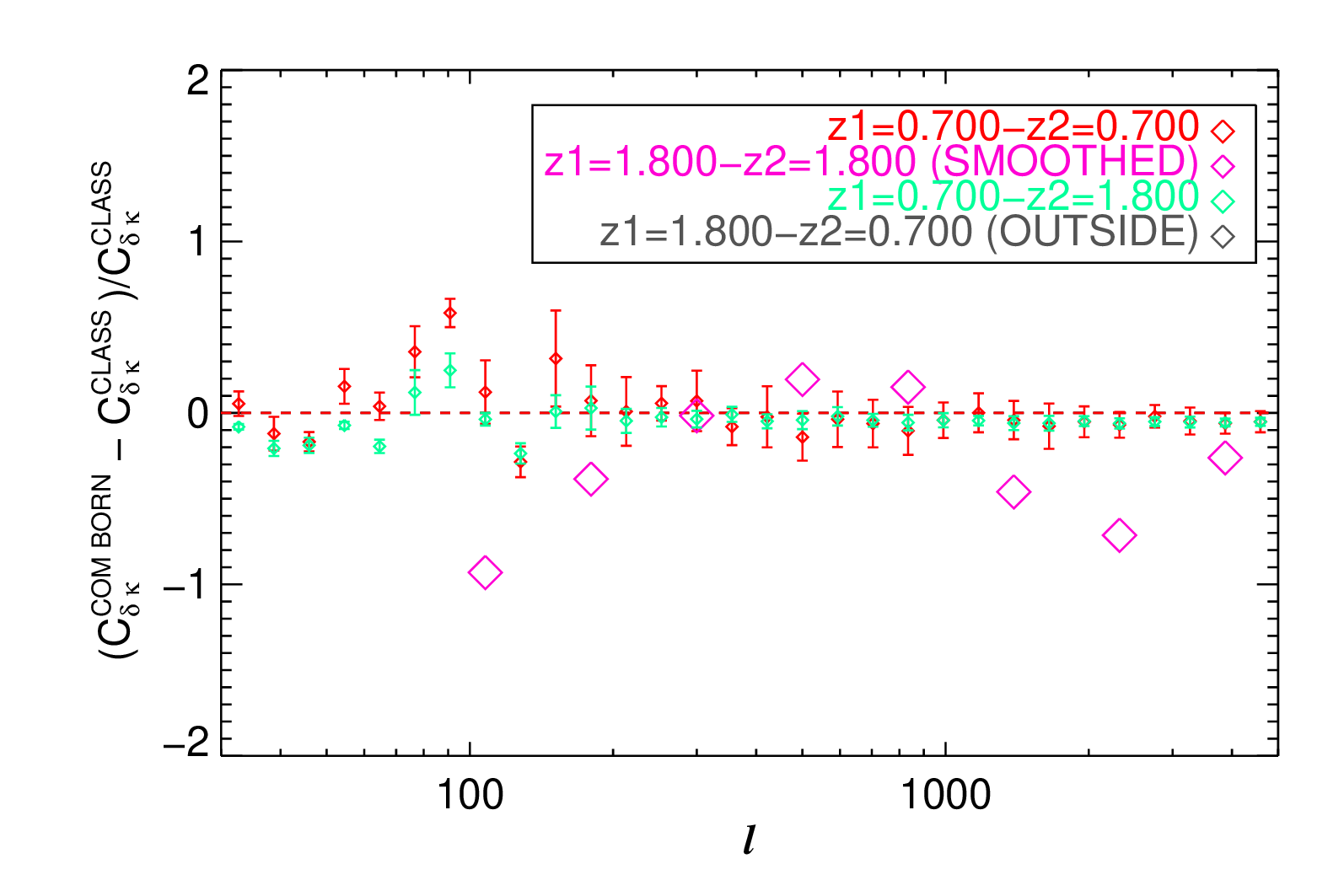} \\
   \end{tabular}
      \caption{{Comoving} matter density--gravitational convergence angular cross-spectrum in a single shell at $z=0.7$ (red) and $z=1.8$ (pink) in $\Lambda$CDM cosmology assuming  the {Born approximation}. The cross-spectrum between two shells at $z=0.7$ (density) and $z=1.8$ (convergence) is shown in green, and the non-trivial cross-spectrum between two shells at $z=0.7$ (convergence) and $z=1.8$ (density) is shown in black. Left: Spectrum from the \textsc{RayGal} 2500~deg$^2$ light cone (diamonds with error bars) and the spectrum from \textsc{Class} (dashed line). Right: Relative deviation from the \textsc{Class} prediction. The cross-spectrum for a single shell at $z=1.8$ being very noisy, the number of bins has been reduced and the large error bars have been omitted (for readability). The non-trivial cross-spectrum is nearly zero in real space, so the data points are outside the plot. Overall, the measured power spectra are in agreement with \textsc{Class} predictions within the error bars, except at large scales (small $\ell$) due to the finite area of the light cone.}
         \label{Fig:cl_lcdmw7_narrow_00002_delta_com_kappa_born}
   \end{figure*}

\begin{figure*}
   \centering
       \begin{tabular}{cc}
      \includegraphics[width=0.48\hsize]{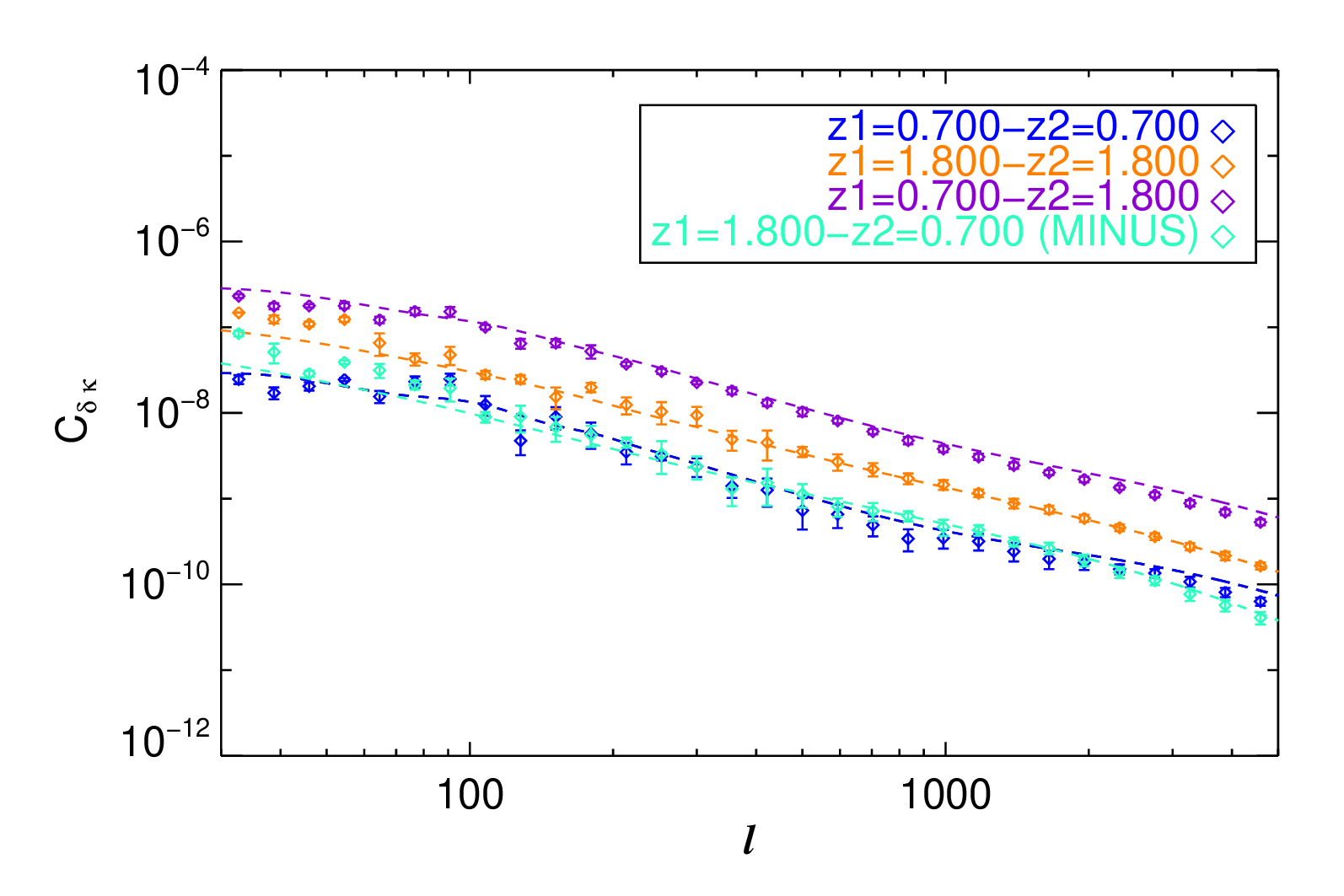}& \includegraphics[width=0.48\hsize]{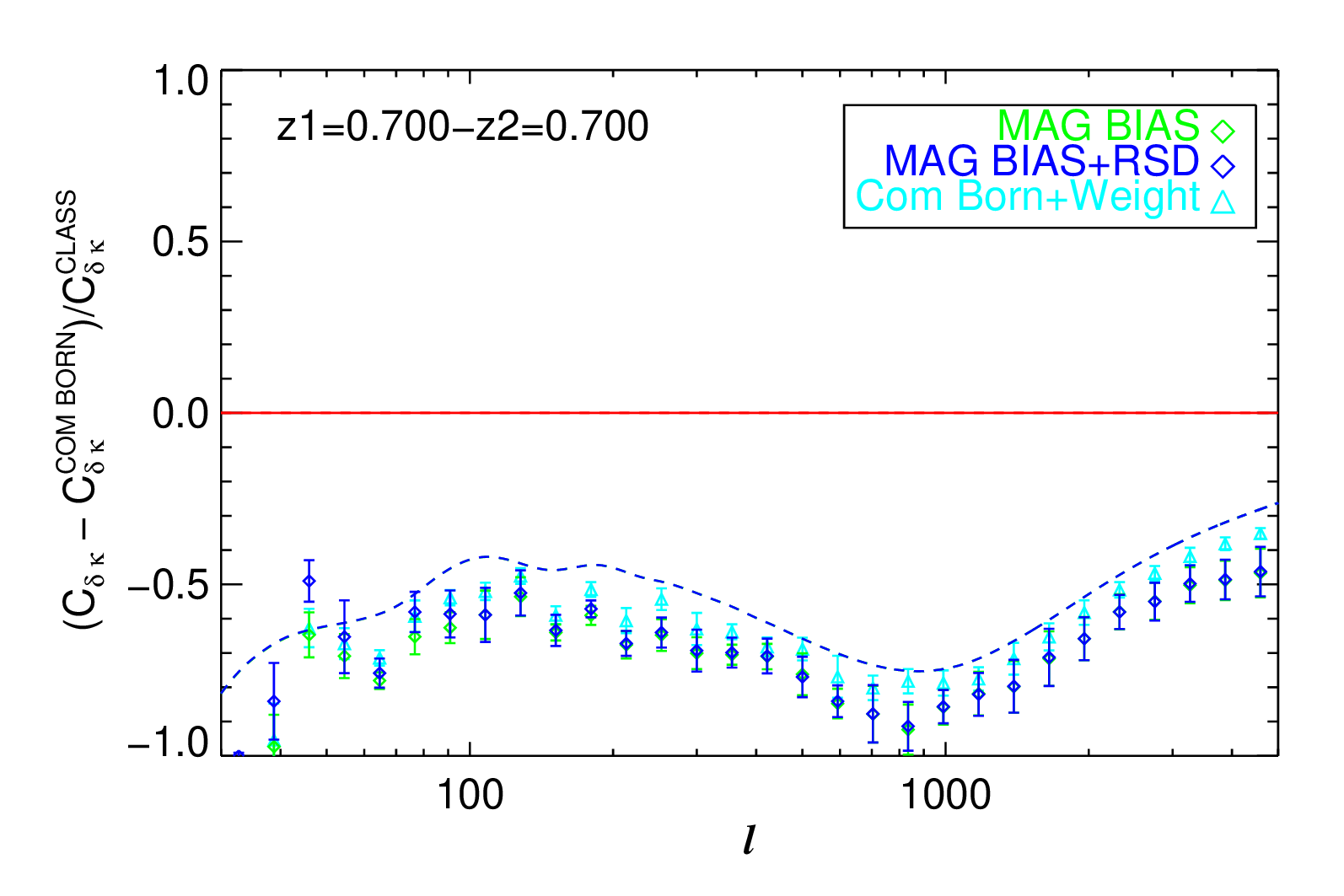}\\
      \includegraphics[width=0.48\hsize]{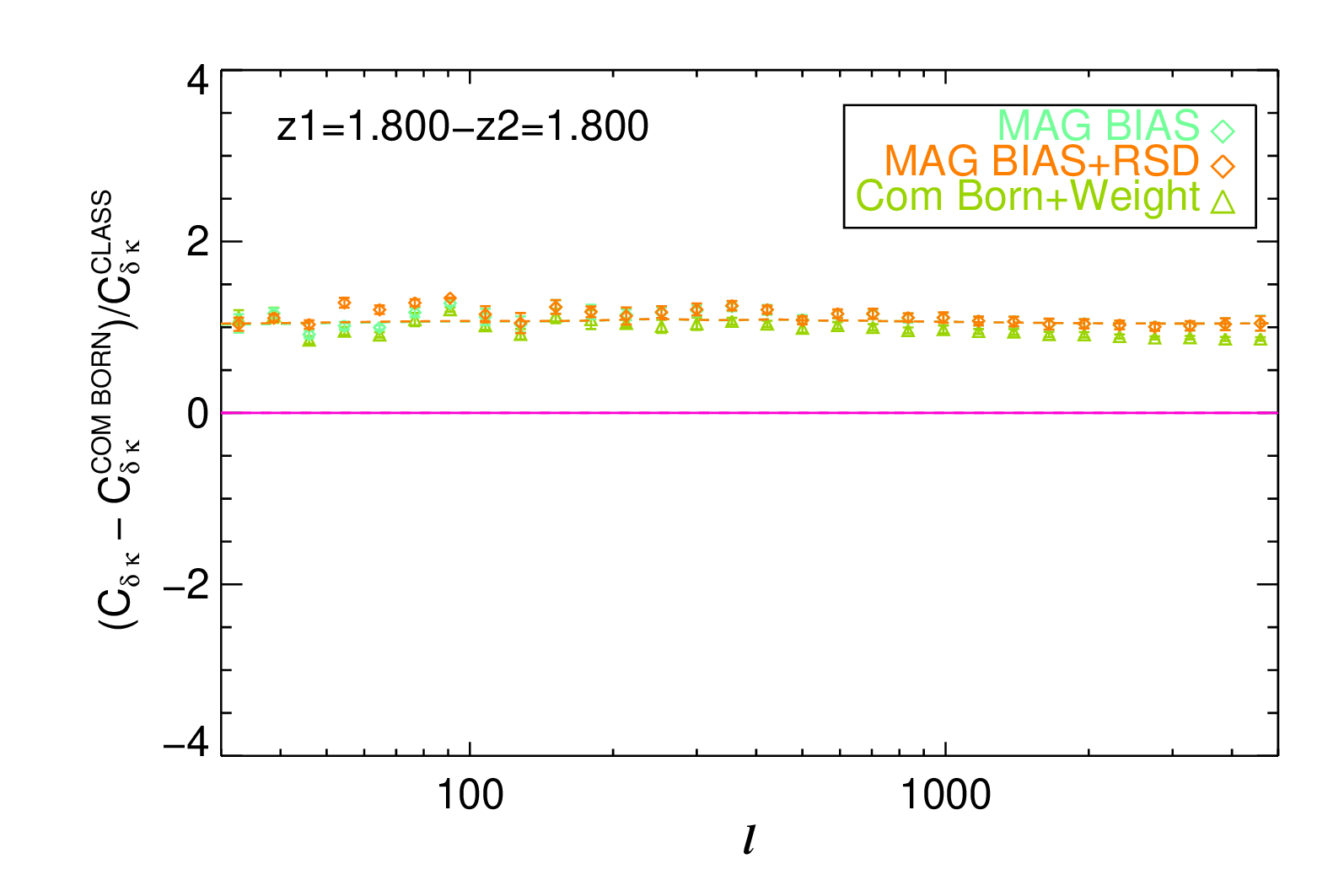}&\includegraphics[width=0.48\hsize]{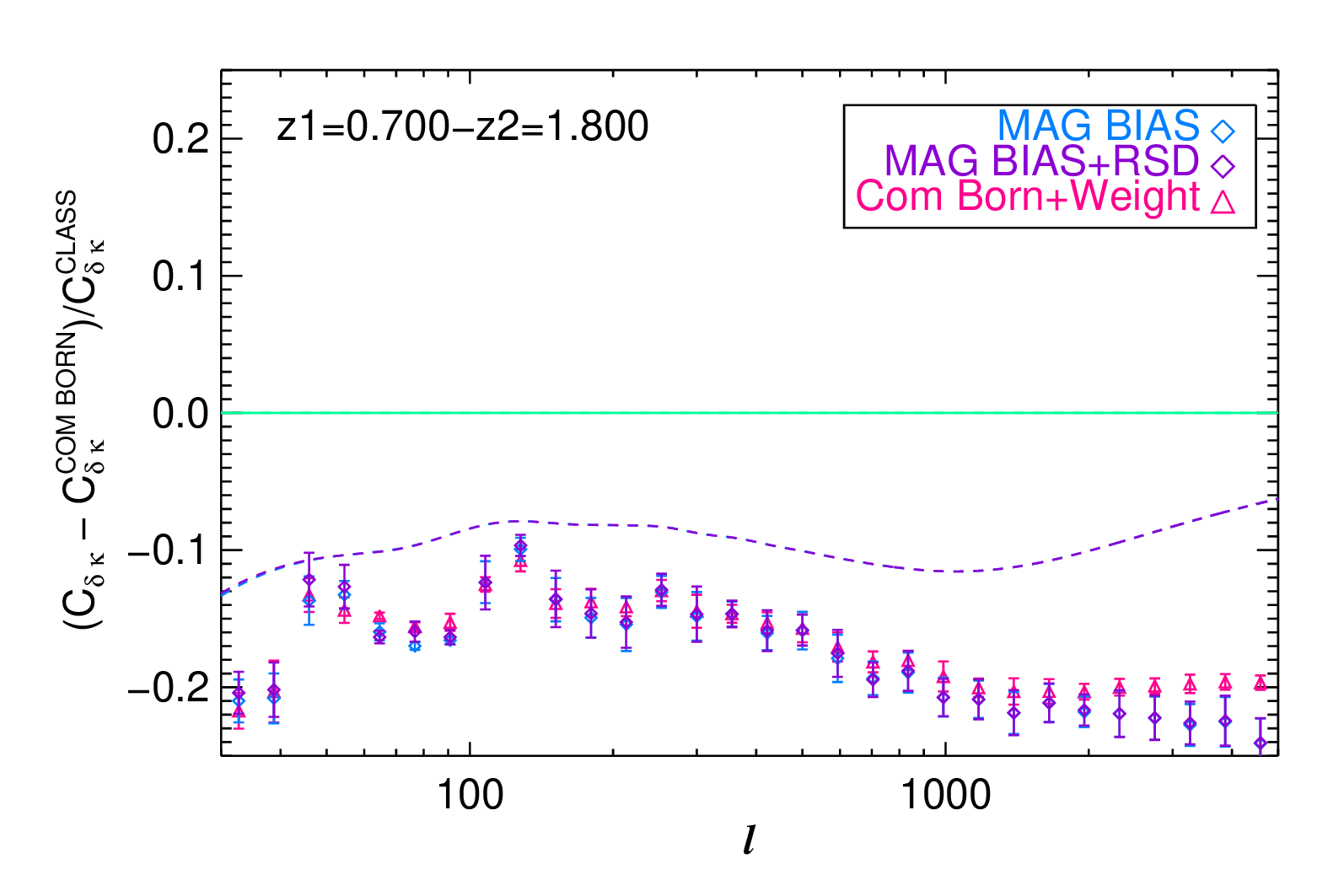}\\
     \end{tabular}
      \caption{ Measurements from the \textsc{RayGal} 2500~deg$^2$ light cone (diamonds) and the \textsc{Class} predictions (dashed lines). Top left: Matter density--gravitational convergence cross-spectra {with relativistic corrections (MB+RSDs)} in a single shell at $z=0.7$ (blue) and $z=1.8$ (orange) in $\Lambda$CDM cosmology. The (usual) cross-spectrum between two shells at $z=0.7$ (density) and $z=1.8$ (convergence) is shown in purple, and the (non-trivial) reverse one (cross-spectra between convergence at $z=0.7$ and density at $z=1.8$) is shown in green. 
      Top right:  Relative deviation from comoving matter density--Born gravitational convergence cross-spectrum at $z=0.7$ in $\Lambda$CDM cosmology due to {relativistic effects}. MB is shown in green, while the effect of RSDs(+MB) is shown in blue. An estimate of the MB effect using an inverse magnification ($|\mu_{\rm Born}|^{-1}$) weight is shown as light blue triangles (see text for details)  Bottom left: Same but for $z=1.8$ with MB in green, RSDs(+MB) in orange, and the $|\mu_{\rm Born}|^{-1}$ weight MB estimate in green. Bottom right: Same but for the $z=0.7$ (density)-$z=1.8$ (convergence) cross-spectrum with MB in light blue, RSDs(+MB) in purple, and the $|\mu_{\rm Born}|^{-1}$ weight MB estimate in pink.  Relativistic corrections are between $50$ and $100$\% for the cross-spectrum at a single redshift. Overall, there is a good agreement with \textsc{Class} at the $20$\% level. Here again the overestimation of the power spectrum is related to the dilution term of the MB effect on the convergence. 
      }
    \label{Fig:cl_lcdmw7_narrow_00002_delta_kappa}
   \end{figure*}

The cross-spectra between the gravitational convergence (computed with Born approximation) and the (comoving) matter overdensity are presented in \cref{Fig:cl_lcdmw7_narrow_00002_delta_com_kappa_born}. Unlike for lensing and clustering there are now four cross-spectra since $C_{\delta_1 \kappa_2} (\ell, z_1=0.7, z_2=1.8)$ is not equal to $C_{\delta_1 \kappa_2} (\ell, z_1=1.8, z_2=0.7)$. The dominant spectrum is the one with the overdensity at low redshift, which acts as a lens for the convergence at high redshift, $C_{\delta_1 \kappa_2} (\ell, z_1=0.7, z_2=1.8)$.  This spectrum is in good agreement with the \textsc{Class} expectation. The cross-spectra with the density and the convergence at the same redshift $C_{\delta_1 \kappa_2} (\ell, z_1=z_2)$ are smaller but non-zero. This is because our Born calculation accounts for the shell thickness. As a consequence, the density in the inner part of the shell acts as a lens for the convergence in the outer part of the shell. This contribution vanishes in the limit of infinitesimal shells (Dirac radial selection function). While the signal is clearly visible for the shell at $z=0.7$, it becomes smaller and very noisy for the shell at $z=1.8$ because the shell thickness is much smaller than the distance to the observer. Finally, the non-trivial spectrum correlating the convergence at $z=0.7$ with the density at $z=1.8$, $C_{\delta_1 \kappa_2} (\ell, z_1=1.8, z_2=0.7)$, is nearly zero both in \textsc{RayGal} and \textsc{Class} expectations.  The right panel shows the relative deviation from \textsc{Class} predictions. Again, we observe a good agreement between the analytical prediction and the numerical data points from $\ell=50$ to $\ell=5000$ for the main spectrum $C_{\delta_1 \kappa_2} (\ell, z_1=0.7, z_2=1.8)$. The deviations at lower $\ell$ are related to sample variance. The spectrum $C_{\delta_1 \kappa_2} (\ell, z_1=0.7, z_2=0.7)$ is noisy but compatible with \textsc{Class} within the error bars. The spectra $C_{\delta_1 \kappa_2} (\ell, z_1=1.8, z_2=1.8)$ seems roughly in agreement with the expectation between $\ell=100$ and $\ell=5000$ but the oscillations are extremely large and can reach almost $100$\%. Finally, the spectrum $C_{\delta_1 \kappa_2} (\ell, z_1=1.8, z_2=0.7)$ is outside the range of the graph because of a division by nearly zero. Overall, there is a good agreement between \textsc{Class} predictions and \textsc{RayGal} measurements.

\subsubsection{Relativistic effects}

We include relativistic effects in \cref{Fig:cl_lcdmw7_narrow_00002_delta_kappa}. The upper left panel shows that the 4 cross-spectra are now different from zero and they are even of the same order of magnitude (i.e. within a factor of ten in absolute value). Moreover, the order has changed since the second largest spectrum is now $C_{\delta_1 \kappa_2} (\ell, z_1=1.8, z_2=1.8),$ which has been seriously boosted by relativistic effects. The most striking spectrum is $C_{\delta_1 \kappa_2} (\ell, z_1=1.8, z_2=0.7),$ which is now of the same order as $C_{\delta_1 \kappa_2} (\ell, z_1=0.7, z_2=0.7)$ although with a minus sign. Even though the changes are important, all the \textsc{RayGal} spectra seem at first sight in agreement with \textsc{Class} predictions (which include MB and RSDs) within a few $\sigma$ as we will see later. This is a good cross-check of the results. We now investigate the amplitude of relativistic effects to better understand the origin of the changes. The relativistic contributions in galaxy-galaxy lensing are presented in \citet{Ghosh2018} (using analytical predictions from \textsc{Class}). In their work and in all our panels, the dominant effect is the MB\ effect, which modifies the density (and, in a subtler way, the convergence, as shown previously). While RSDs are negligible here, they cannot always be neglected since they play a (small) role at low redshifts and low multipoles for some cross-spectra within thin shells (e.g. we checked that they are non-negligible but noisy contributions to $C_{\delta_1 \kappa_2} (\ell, z_1=0.45, z_2=0.45)$ of order $10$\% for $\ell<80$). The dominant contribution is again MB,
where the apparent density becomes (at first order) $\delta_i=\delta^{\rm com}_i+ (5s-2) \kappa_i $ and the convergence becomes $\kappa_i=\kappa_i^{\rm Born} (1+(5s-2)\kappa_i^{\rm Born})$ (with $s=0$ here). 
As a consequence, there is an additional lensing contribution in the cross-spectrum (see \cref{Eq:Cl_delta_kappa_linmap}): 

\begin{equation}
   \label{eq:cldeltakappa}
   C_{\delta_i \kappa_j}(l)\approx C_{\delta^{\rm com}_i \kappa^{\rm com}_j}(l)-2 C_{\kappa_i \kappa_j}(l)-2 C_{\delta_i \kappa_j^2}(l)+ 4 C_{\kappa_i \kappa_j^2}(l).  
   \end{equation}
The first two terms (as well as the RSD contribution) are included in \textsc{Class} but not the two last ones (nor the post-Born effects).

\begin{figure}
   \centering
   \includegraphics[width=\hsize]{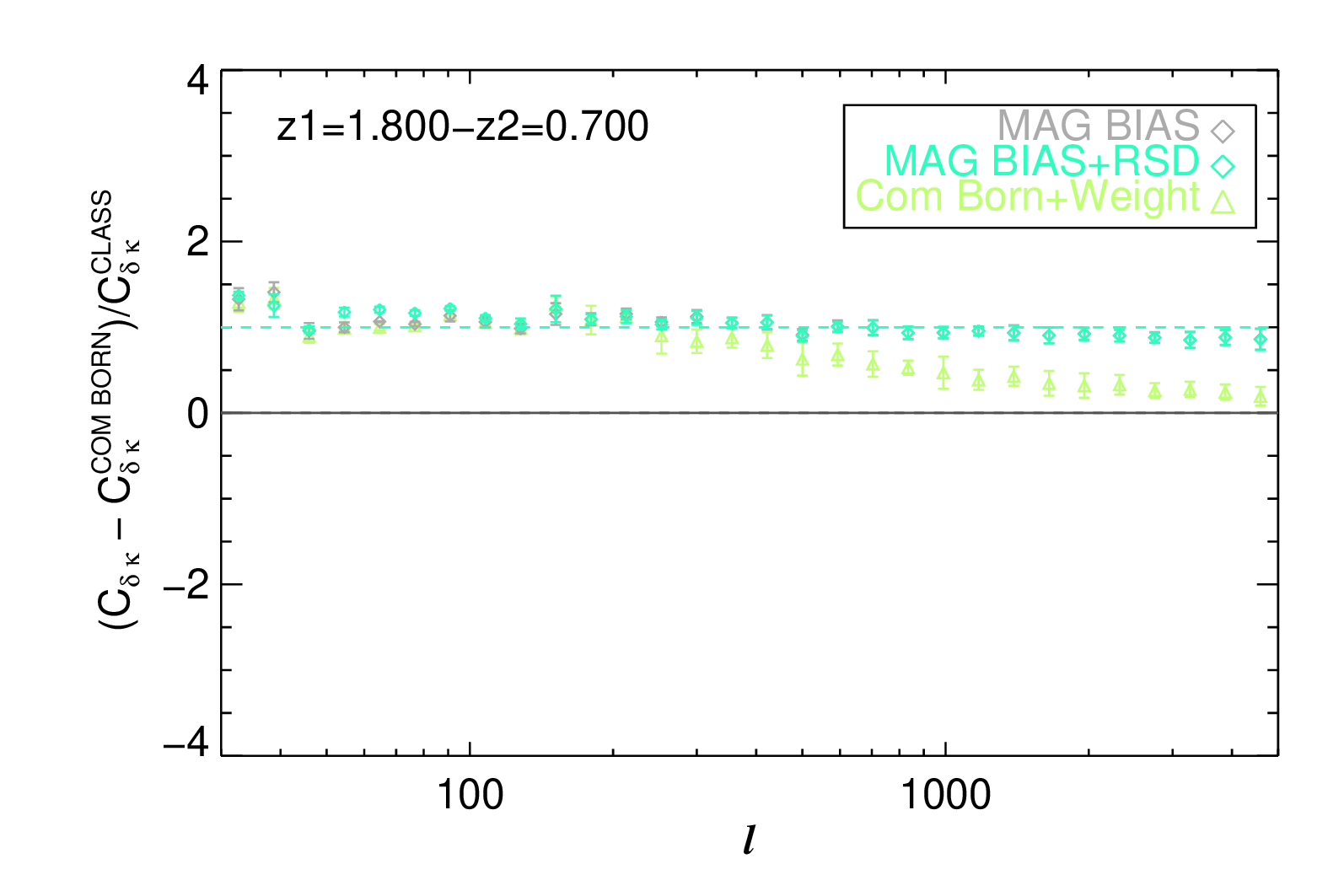}
   \caption{Non-trivial reverse configuration of gravitational convergence at $z=0.7$ and matter density at $z=1.8$, similar to the bottom-right panel of \cref{Fig:cl_lcdmw7_narrow_00002_delta_kappa}. MB is in grey, MB+RSDs in cyan, and the $|\mu_{\rm Born}|^{-1}$ weight MB estimate in light green. Relativistic effects almost reach 100\%, in agreement with \textsc{Class}. This configuration turns out to be a sensitive probe of the lensing convergence spectrum.
   }
   \label{Fig:relative_cl_lcdmw7_narrow_00002_delta_kappa_vs_comborn_21}
\end{figure}

The relativistic contribution to $C_{\delta_1 \kappa_2} (\ell, z_1=0.7, z_2=1.8)$ is presented in the bottom-right panel. The dominant relativistic contribution is MB\ with $-15\%$ between $\ell=50$ and $\ell=500$ and $-20\%$ between $\ell=500$ and $\ell=5000$. \textsc{Class} predicts a negative effect of order $10\%$ beyond $\ell=50$. Given the size of the error bars, the deviation is of order $3\sigma$ below $\ell=500$ and $5\sigma$ above. The discrepancy is likely to be related to the effect of MB on the convergence (i.e. the two last terms of \cref{eq:cldeltakappa}, which also decrease the $\kappa - \kappa$ cross-spectra (as seen in \cref{subsec:result_kappakappa}) and is not accounted for in \textsc{Class}. This is confirmed by our MB estimate from Born inverse magnification weight. In this configuration, the relativistic contributions to galaxy-galaxy lensing cannot be neglected.
The other galaxy-galaxy lensing spectra are dominated by relativistic effects. The upper right panel shows the relativistic contributions to $C_{\delta_1 \kappa_2} (\ell, z_1=0.7, z_2=0.7)$. This negative effect ranges from $-50\%$ to $-100\%$. The shape is highly non-trivial with a bump near $\ell=100$, a minimum near $\ell=1000$ and another bump near $\ell=5000$. All these effects are well reproduced by \textsc{Class} (with deviations at the $2\sigma$ level ) and our $|\mu^{-1}|$weight MB spectra below $\ell=1000$. The bottom-left panel shows the contributions to $C_{\delta_1 \kappa_2} (\ell, z_1=1.8, z_2=1.8),$ which are about $+100\%$ since $C_{\delta_{\rm com} \kappa_{\rm Born}} (\ell, z_1=1.8, z_2=1.8)$ was small because of the small thickness of the shell. The relativistic contribution agrees with the \textsc{Class} prediction. Finally, as expected, the cross-spectrum $C_{\delta_1 \kappa_2} (\ell, z_1=1.8, z_2=0.7)$ is $100\%$ explained by relativistic effects, and is compatible with the \textsc{Class} prediction (see \cref{Fig:relative_cl_lcdmw7_narrow_00002_delta_kappa_vs_comborn_21}).  This non-trivial contribution can be understood using \cref{eq:cldeltakappa}. It is dominated by $-2 C_{\kappa_1 \kappa_2} (\ell, z_1=1.8, z_2=0.7)$. The corresponding $\kappa - \kappa$ cross-spectrum presented in \cref{Fig:cl_lcdmw7_narrow_00002_kappa_kappa} is indeed very similar (when multiplied by a factor -2). For thin shells or when the density shell is beyond the convergence shell, it is therefore possible to probe lensing spectra from galaxy-galaxy lensing spectra. {To conclude, the relativistic contributions to galaxy-galaxy lensing play an important role that cannot be neglected. The contributions of the MB effect (on the density) are well captured by \textsc{Class} while we found additional contributions from the MB (on the convergence). The actual impact of the MB depends on the slope $s$, and the shell configurations, but it can give a $20-100\%$ effect}.

\subsection{Evolution and cosmological dependence}
\subsubsection{Redshift evolution}
\label{subsec:redshift_evol}

\begin{figure}
   \centering
    \begin{tabular}{ccc}
   \includegraphics[width=\hsize]{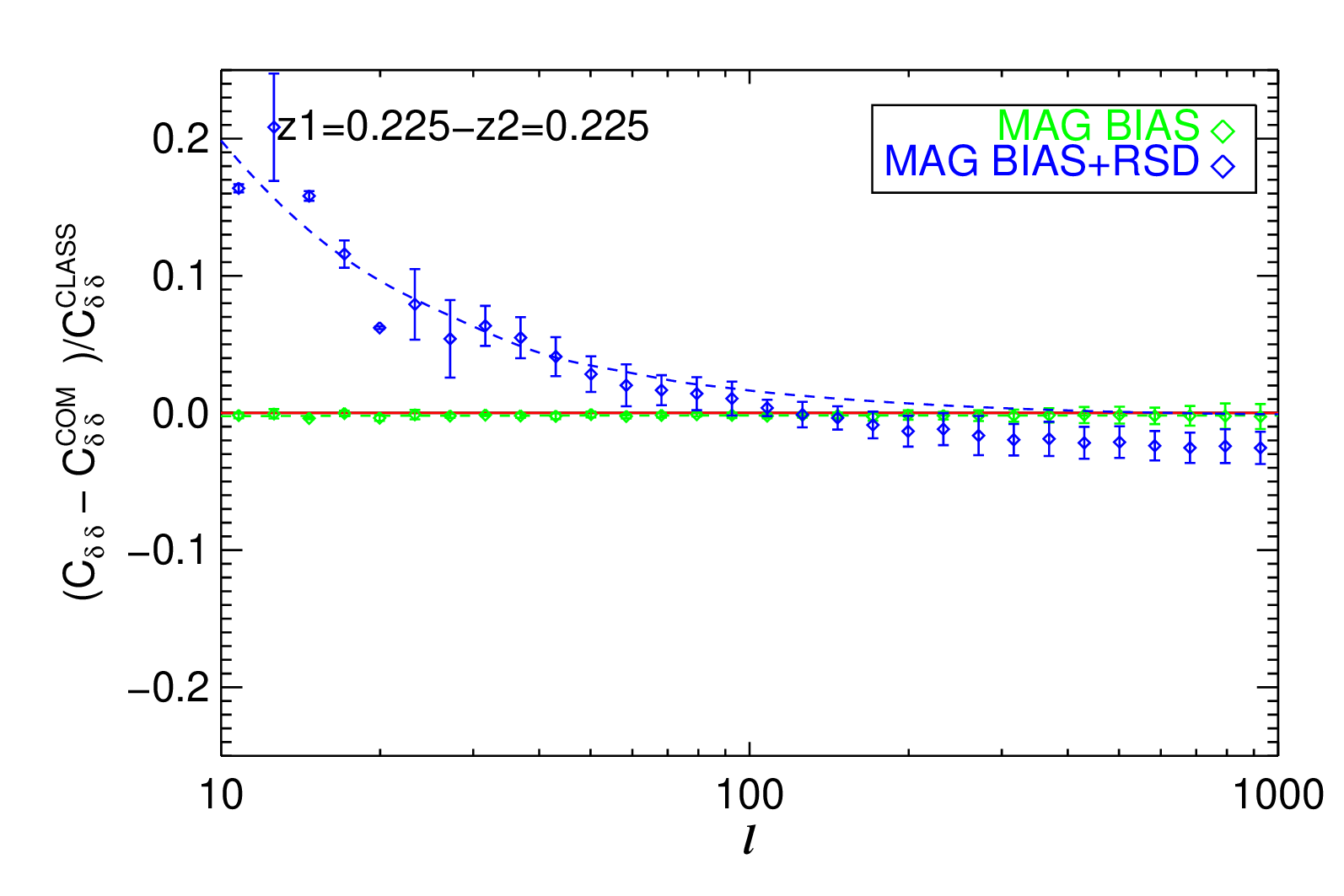}\\
   \end{tabular}
   \caption{Relativistic effects at low redshift on matter power spectrum $C_{\delta_1 \delta_2} (\ell, z_1=0.225, z_2=0.225)$ (symbols are \textsc{RayGal} measurements, and lines are \textsc{Class} predictions).  The MB effect (green) and RSDs(+MB) effect (blue) are shown. The trends are similar compared to higher redshifts, but the RSD effect plays a dominant role and the MB effects are smaller. Finger-of-God effects (ignored by \textsc{Class}) are also present.  
   }
   \label{fig:low_redshift_subset_delta_delta}
\end{figure}

\begin{figure}
   \centering
    \begin{tabular}{ccc}
   \includegraphics[width=\hsize]{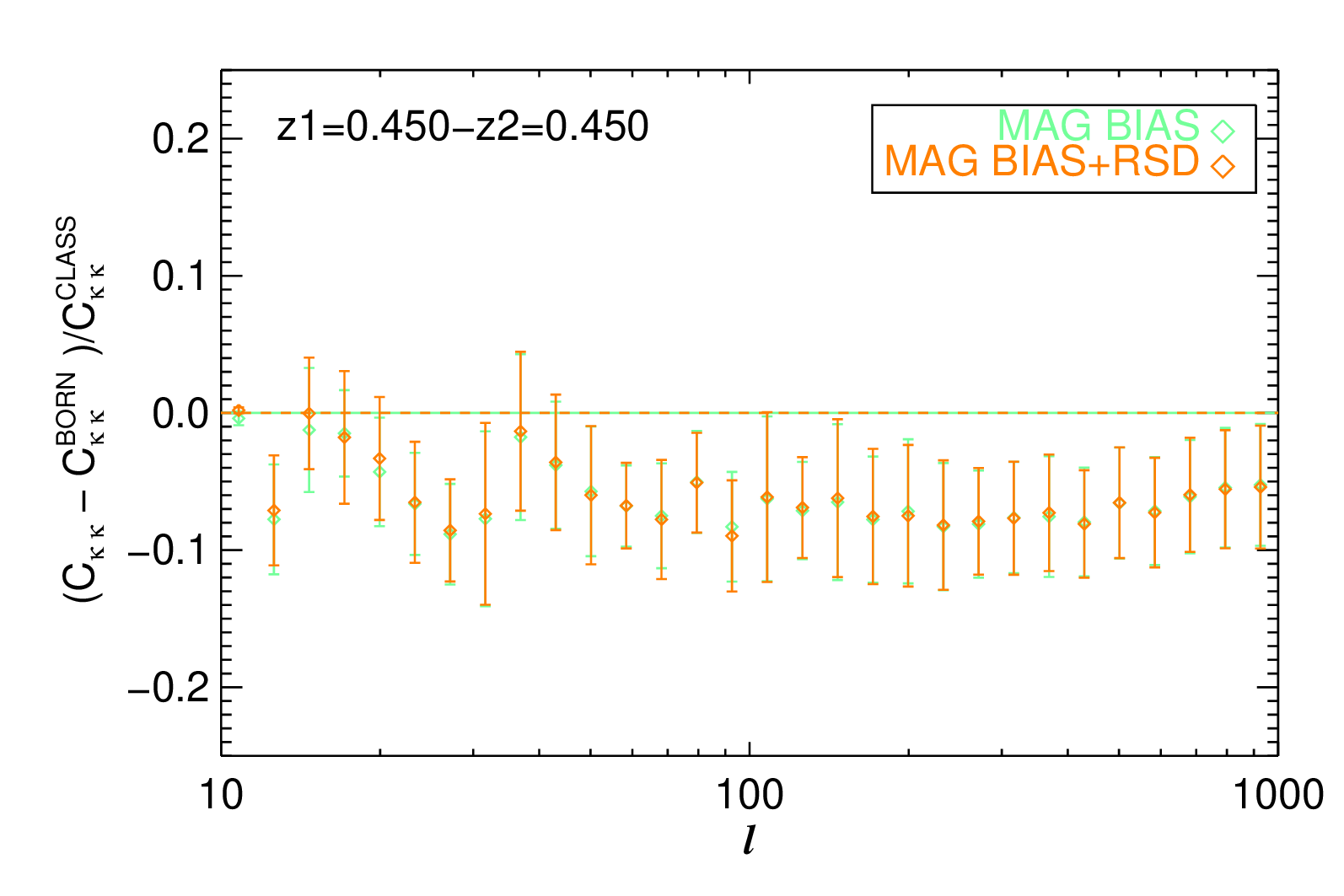}\\
   \end{tabular}
   \caption{Relativistic effects at low redshift on convergence power spectrum $C_{\kappa_1 \kappa_2} (\ell, z_1=0.45, z_2=0.45)$ (symbols are \textsc{RayGal} measurements, and lines are \textsc{Class} predictions). MB effects are in green and RSDs(+MB) in orange. The trends are similar compared to at higher redshifts, but MB effects are smaller.  
   }
   \label{fig:low_redshift_subset_kappa_kappa}
\end{figure}

\begin{figure}
   \centering
    \begin{tabular}{ccc}
   \includegraphics[width=\hsize]{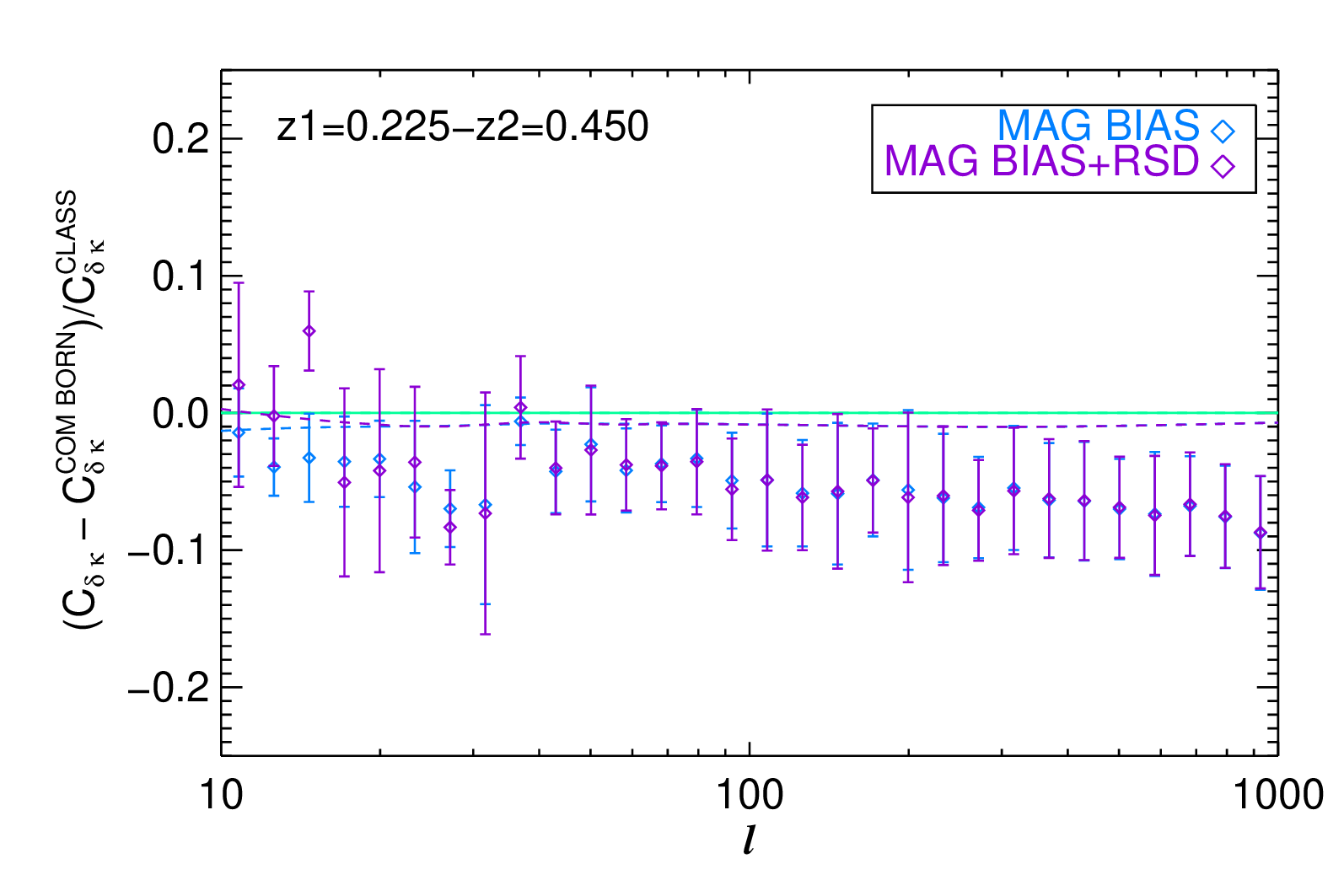}\\
   \end{tabular}
   \caption{Relativistic effects at low redshift on density-convergence spectrum $C_{\delta_1 \kappa_2} (\ell, z_1=0.225, z_2=0.45)$ (symbols are \textsc{RayGal} measurements, and lines are \textsc{Class} predictions). The MB effect is in blue and RSDs(+MB) in purple. The trends are similar compared to at higher redshifts, but the MB effects are smaller. 
   }
   \label{fig:low_redshift_subset_delta_kappa}
\end{figure}

We now investigate the redshift evolution of relativistic effects. We therefore perform the same analysis with two shells at low redshifts $z_1=0.225\pm 0.045$ and $z_2=0.45\pm 0.025$ extracted from the full-sky catalogue. The resolution of the \textsc{Healpix} maps built from the catalogue is $N_{\rm side}=1024$ while the resolution of the Born map is $N_{\rm side}=2048$. These maps are then degraded to $N_{\rm side}=1024$. The estimation of the power spectrum is straightforward since the cone is full-sky (i.e. no need of apodisation to account for large masked region). We find that the global trend is similar to the one of high redshift measurements. There is also a good agreement with \textsc{Class} in most cases, while the deviations are well understood.  

We now highlight the main differences between the high- and low-redshift cross-spectra -- we illustrate some of them in \cref{fig:low_redshift_subset_delta_delta}, \cref{fig:low_redshift_subset_kappa_kappa}, and \cref{fig:low_redshift_subset_delta_kappa}. It shows the impact of relativistic effects on three spectra. \cref{fig:low_redshift_subset_delta_delta} shows the impact of relativistic effects on $C_{\delta_1 \delta_2} (\ell, z_1=0.225, z_2=0.225)$. We notice that the influence of the MB is now negligible, while RSDs play a more important role. This is due to the weak-lensing effect, which decreases with redshift. \textsc{Class} performs a good prediction at large scales; however, beyond $\ell \sim 100$ (corresponding roughly to $k\sim 0.15$~h~Mpc$^{-1}$), the effects of RSDs on the power spectrum become negative. As expected, this effect is not well reproduced by \textsc{Class} because of the assumption of linear mapping between real-space and redshift-space. Analytical predictions could be made using perturbation theory. A recent work by \citet{grasshorn20} about non-linear RSDs in the harmonic space galaxy power spectrum finds similar trend thus reinforcing our claim of the damping of the angular matter power spectrum at large $\ell$ due to the fingers-of-God effect. 

We now turn to the lensing power spectrum. We verified that the agreement between the power spectrum of Born convergence map and \textsc{Class} expectation is better than $5\%$. The impact of relativistic effects on $C_{\kappa_1 \kappa_2} (\ell, z_1=0.45, z_2=0.45)$ is illustrated in \cref{fig:low_redshift_subset_kappa_kappa}. The RSD effects are still negligible. The effect of MB is also much smaller, being of order $-2\%$ over a wide range of multipoles from $\ell\sim 10$ to $\ell\sim 400$. We also noted that the MB\ effect remains large ($5$ to $15\%$) for lensing cross-spectra.

We then investigated galaxy-galaxy lensing at low redshifts. We checked that trends are similar and that the spectra are noisier (and some sign changes occurred, such as for $C_{\delta_1 \kappa_2} (\ell, z_1=0.45, z_2=0.45)$).  The impact of relativistic effects on the overdensity-convergence power spectrum $C_{\delta_1 \kappa_2} (\ell, z_1=0.225, z_2=0.45)$ is illustrated in \cref{fig:low_redshift_subset_delta_kappa}. The MB effect is negative of order $-2\%$ to $-5\%$ between $\ell=100$ and $\ell=1000$. The decrease is smaller than at higher redshift but still larger than the \textsc{Class} expectation (which ignores the MB effect on the convergence). Though RSD effects are small, we noted that, according to \textsc{Class} and our (noisy) measurement, they contribute to $\sim 10\%$ of $C_{\delta_1 \kappa_2} (\ell, z_1=0.45, z_2=0.45)$ for $\ell<80$. While RSD seem negligible in most galaxy-galaxy lensing studies, they might play a role in some configurations: it is therefore important to consider all relativistic effects and to verify their relative amplitude before neglecting some of these effects. 
{To conclude, at lower redshift the relative importance of RSDs increases with respect to those of MB effects. It is also interesting to note that fingers-of-God effects are also visible within the angular power spectra.}

\subsubsection{Cosmological dependence}
\begin{figure}
   \centering
    \begin{tabular}{ccc}
   \includegraphics[width=\hsize]{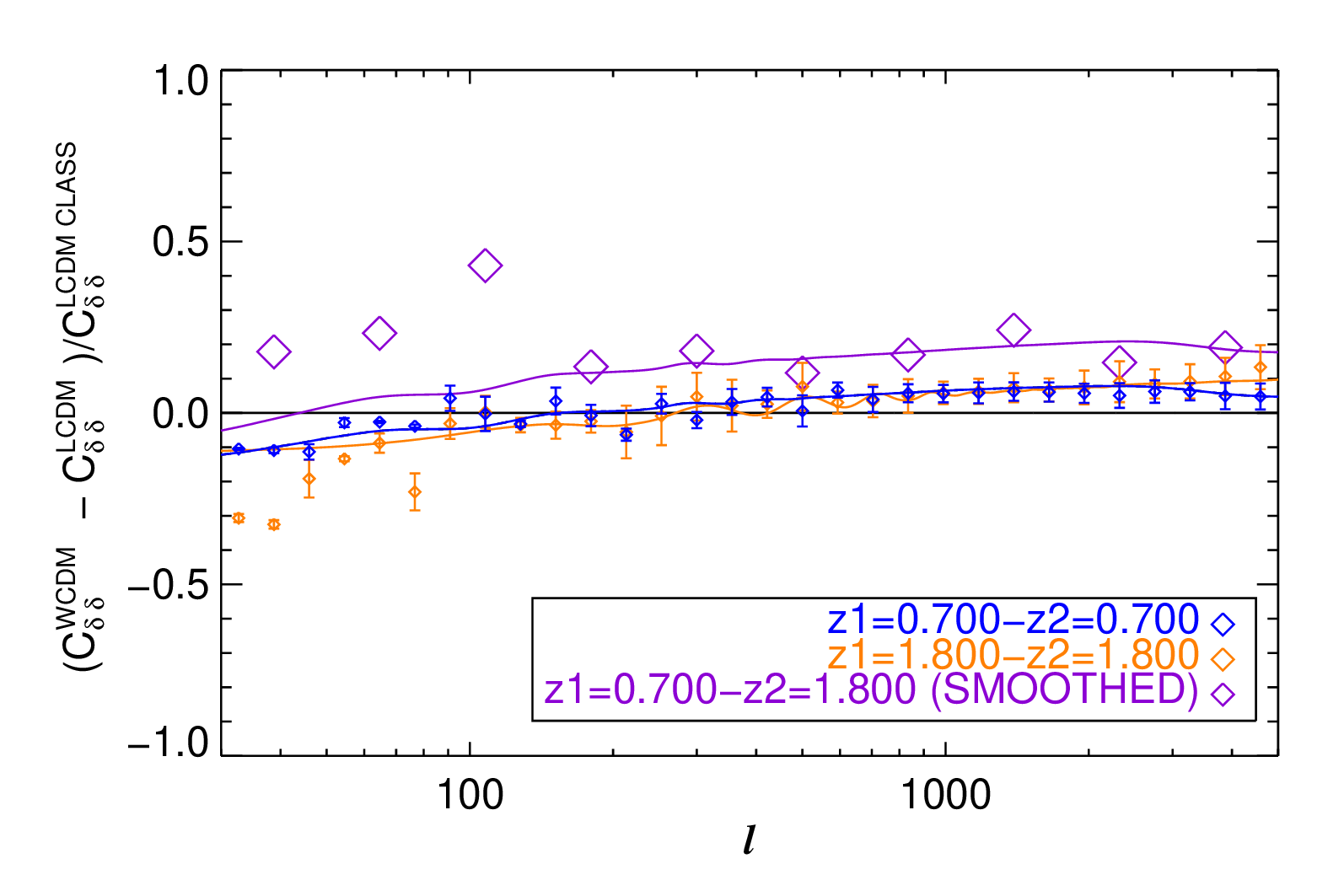}\\
   \end{tabular}
   \caption{Relative deviations of the matter-density spectra between $w$CDM and $\Lambda$CDM spectra (including all relativistic effects) considering one shell at $z=0.7$ (blue), one shell at $z=1.8$ (orange), and two shells at $z=0.7$ and $z=1.8$. Diamonds are measurements from the \textsc{RayGal} 2500~deg$^2$ light cone, and dashed lines are \textsc{Class} predictions. This plot shows the interest of considering all cross-spectra since they are more or less sensitive to cosmology. Even though the two models were calibrated on CMB data, the differences can reach $20$\%. The agreement with \textsc{Class} is good. 
   }
   \label{Fig:relative_cl_lcdmw7_vs_wcdmw7_narrow_00002_delta_delta}
\end{figure}

\begin{figure}
   \centering
    \begin{tabular}{ccc}
   \includegraphics[width=\hsize]{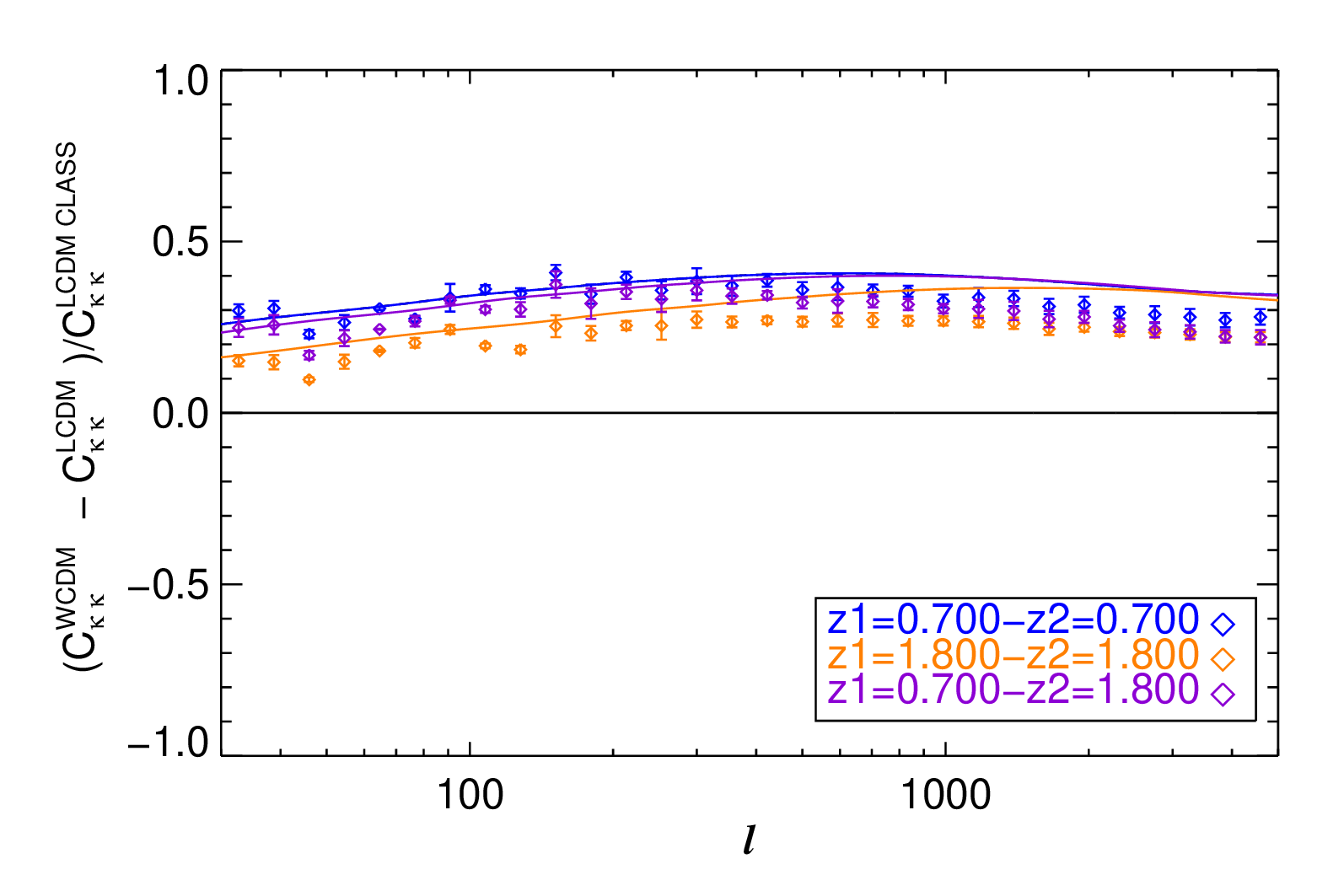}\\
   \end{tabular}
   \caption{Relative deviations of the convergence spectra between $w$CDM and $\Lambda$CDM spectra (including all relativistic effects) considering one shell at $z=0.7$ (blue), one shell at $z=1.8$ (orange), and two shells at $z=0.7$ and $z=1.8$. Diamonds are measurements from the \textsc{RayGal} 2500~deg$^2$ light cone, and dashed lines are \textsc{Class} predictions. This plot shows the interest of considering all cross-spectra since they are more or less sensitive to cosmology. Even though the two models were calibrated on CMB data, the differences can reach $40$\%. The agreement with \textsc{Class} is good except for $\ell >500$, which shows a discrepancy of $\sim 10\%$ . 
   }
   \label{Fig:relative_cl_lcdmw7_vs_wcdmw7_narrow_00002_kappa_kappa}
\end{figure}

\begin{figure}
   \centering
    \begin{tabular}{ccc}
   \includegraphics[width=\hsize]{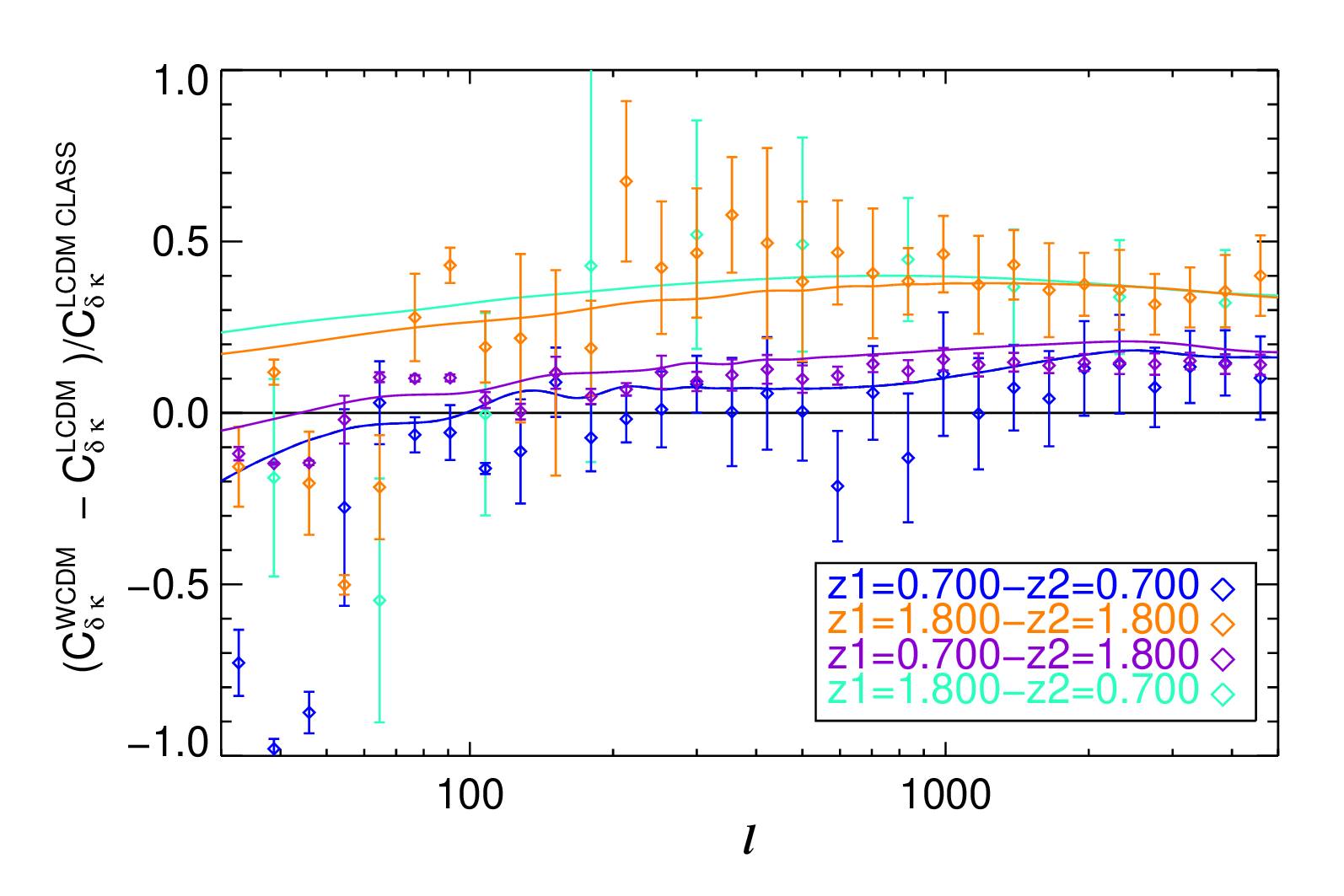}\\
   \end{tabular}
   \caption{Relative deviations of the density-convergence spectra between $w$CDM and $\Lambda$CDM spectra (including all relativistic effects) considering one shell at $z=0.7$ (blue), one shell at $z=1.8$ (orange), and two shells at $z=0.7$ and $z=1.8$ (purple for the standard configuration and green for the reverse one). Diamonds are measurements from \textsc{RayGal} 2500~deg$^2$ light cone, and dashed lines are \textsc{Class} predictions. This plot shows the interest of considering all cross-spectra since they are more or less sensitive to cosmology. Even though the two models were calibrated on CMB data, the differences can reach $40$\%. The agreement with \textsc{Class} is good. 
   }
   \label{Fig:relative_cl_lcdmw7_vs_wcdmw7_narrow_00002_delta_kappa}
\end{figure}

A detailed study of the full cosmological dependence of all cross-spectra and all relativistic effects is beyond the scope of this paper. Here we would like to broadly address the questions of whether the $w$CDM cross-spectra are valid; what the relative variation in the ten cross-spectra (including all relativistic effects) is when changing cosmology from $\Lambda$CDM to $w$CDM; if \textsc{Class} can predict this relative variation in the non-linear regime; and whether these spectra allow us to discriminate between these two competitive scenarios, both of which are compatible with CMB constraints at the $2\sigma$ level. 

The analysis of the $w$CDM simulation is similar to the one of the $\Lambda$CDM. We consider two shells at the same redshift and we build maps with the same resolution. The $\Lambda$CDM and $w$CDM cosmologies do not differ strongly since they are both calibrated on CMB data. As a consequence the relativistic effects behave qualitatively in a similar way and we therefore did not plot again the same figures as in $\Lambda$CDM cosmology. In \cref{Fig:relative_cl_lcdmw7_vs_wcdmw7_narrow_00002_delta_delta}, \cref{Fig:relative_cl_lcdmw7_vs_wcdmw7_narrow_00002_kappa_kappa} and \cref{Fig:relative_cl_lcdmw7_vs_wcdmw7_narrow_00002_delta_kappa} we investigate the relative variation in the ten density-convergence auto- and cross-spectra (including relativistic effects) when moving from $\Lambda$CDM to $w$CDM.  The first global remark is that there is a good agreement below $\ell=5000$ between our results and \textsc{Class} calculation, which is an encouraging validation of our work. In \cref{Fig:relative_cl_lcdmw7_vs_wcdmw7_narrow_00002_delta_delta}, the matter-density auto-spectrum in $w$CDM cosmology differs from the one of $\Lambda$CDM cosmology by $\sim -10\%$ near $\ell\sim 30$, $\sim 0\%$ near $\ell\sim 250$ and $\sim +10\%$ near $\ell\sim 3000$. The same trend can be found in the 3D matter power spectrum at $z=1$ in \cref{Fig:pk_raygalgroupsims_vs_coyote}. The relative variation in the cross-spectrum (which is dominated by relativistic effect) reach larger values of order $+20\%$ near $\ell\sim 5000$. All the dependence (including the non-trivial cross-spectrum) are well reproduced by \textsc{Class} (since the 3D power spectra were calibrated to \textsc{RayGal} simulations). In \cref{Fig:relative_cl_lcdmw7_vs_wcdmw7_narrow_00002_kappa_kappa}, the convergence power spectrum shows a larger relative difference of $15-25\%$ at $\ell \sim 30$, $30-40\%$ at $\ell \sim 1000$ and $30\%$ at $\ell \sim 5000$. There is a good agreement with \textsc{Class} expectation up to $\ell\sim 500$. Beyond this multipole, the relative variation is overestimated by \textsc{Class}. This is likely to be related to the MB effect, which is not included in the analytical prediction. The difference remains reasonable of order $10\%$. In \cref{Fig:relative_cl_lcdmw7_vs_wcdmw7_narrow_00002_delta_kappa}, the density-convergence cross-spectra are presented. The cosmological dependence of the stronger signal $C_{\delta_1 \kappa_2} (\ell, z_1=0.7, z_2=1.8)$ (with smaller error bars in purple) rises from $0\%$ near $\ell\sim 40$ up to $\sim 20\%$ at $\ell\sim 1000$ and beyond. The agreement with \textsc{Class} is good at the 2-$\sigma$ level. The spectrum $C_{\delta_1 \kappa_2} (\ell, z_1=0.7, z_2=0.7)$ shows a similar trend rising from $0\%$ near $\ell\sim 100$ up to $\sim 20\%$ at $\ell\sim 2000$ and beyond. Finally, the spectra $C_{\delta_1 \kappa_2} (\ell, z_1=1.8, z_2=1.8)$ and $C_{\delta_1 \kappa_2} (\ell, z_1=1.8, z_2=0.7)$ dominated by relativistic effects show a larger variation rising from $20\%$ at $\ell \sim 30$ to $40\%$ near $\ell \sim 1000$ and beyond. It indicates that even though these spectra are noisier, they carry a strong cosmological signal. The cosmological dependence is well captured by \textsc{Class} for overdensity-convergence spectra.

 {To conclude, if one assumes that the cross-spectra are accurately known  in $\Lambda$CDM, the cosmological dependence for the models in the vicinity of $\Lambda$CDM can be captured at first order by \textsc{Class}. To go further the modelling of MB and non-linear RSD should be improved. In any case, the cross-spectra carry complementary cosmological information (the cosmological dependences are different). For the studied $w$CDM and $\Lambda$CDM cosmologies, the weak-lensing power spectra and the galaxy-galaxy spectra dominated by relativistic effect seems the most sensitive probes followed by the relativistic matter power spectrum and the standard density-convergence spectrum. Relativistic 3$\times$ 2 points spectra are therefore a promising cosmological probe.}

\section{Conclusion}\label{conclusions}

In this work we have introduced a simulation suite for the study of relativistic effects in cosmology and illustrated its use with an application to lensing-matter clustering statistics.

We have described the \textsc{RayGal} suite. It consists of two large and well-resolved N-body simulations of two different cosmological models ($\Lambda$CDM and $w$CDM) calibrated on CMB data. For each simulation, several snapshots have been stored in addition to (wide, intermediate, or deep) particles and gravity light cones. Halos have been detected in snapshots and light cones according to a wide range of possible definitions. The specificity of this work lies in the relativistic ray-tracing technique used to generate the data: light rays are propagated from the observer to the sources according to weak-field geodesic equations of GR. Several hundred million particles and tens of millions of halos have been replaced in redshift space. Within our intermediate light cone ($2500$~deg$^{2}$ up to $z=2$), the source (or halo) angular densities are similar to those expected from present and future photometric (or spectroscopic) surveys, such as DES, KIDS, HSC, Euclid, and LSST (or BOSS, eBOSS, DESI, Euclid, and SKA). These relativistic particle and halo catalogues contain information about the lensing deflection angle, the weak-lensing distortion matrix, and the apparent redshift, including all contributions at first order in metric perturbation (comoving, Doppler, potential, transverse Doppler, and ISW-RS). Relativistic \textsc{Healpix} maps were also produced. {All these catalogues and maps are publicly available\footnote{ \href{https://cosmo.obspm.fr/public-datasets/}{https://cosmo.obspm.fr/public-datasets/}}}.  

The catalogues and maps are very generic: they allow a wide range of applications. We illustrate the strength of the approach with an application to the study of lensing-matter clustering statistics, which is a hot topic in modern cosmology. Since the matter overdensity and the gravitational convergence are the fundamental quantities of clustering and weak lensing, we focus on the position-position, position-convergence, and convergence-convergence 3$\times$2 points correlations. The extension to more observable quantities, such as the usual 3$\times$2 points involving the shear instead of the convergence (see e.g. \citealt{DES2021cosmo}), can be investigated in a future work. We have considered two shells extracted from our intermediate light cone: a low-redshift shell ($z_1=0.7\pm 0.2$) and a high-redshift shell ($z_2=1.8 \pm 0.1)$. We computed the matter overdensity, $\delta$, and gravitational convergence, $\kappa$. We then evaluated the ten auto- and cross-spectra between overdensity and convergence at these two redshifts (including the non-trivial ones) up to $\ell=5000$. Finally, we investigated the relative amplitude of all the contributions beyond the standard one (evaluated from comoving overdensity, $\delta^{\rm com}$, and Born convergence, $\kappa^{\rm Born}$): MB and RSDs. These contributions were compared to \textsc{Class} predictions. Our {main results} are the following:
\begin{itemize}
    \item {Validation}: There is an agreement at the few percent level between the standard cross-spectra (without relativistic effects) and \textsc{Class} predictions once the 3D real-space matter power spectra are properly calibrated to the spectrum of \textsc{RayGal} simulations (which itself is in agreement at the percent level with that of \textsc{Cosmicemu}).
    \item {Clustering}: Both RSDs and MB contribute to angular density spectra. Magnification bias  is a nearly scale-independent contribution at the $2-5\%$ level, while the RSD contribution is small for $\ell>100-500$ (depending on redshift) but can reach $20-40\%$ at small $\ell<20-100$. Magnification bias  also correlates distant shells ($\sim 100\%$ contribution to the cross-spectra). In this configuration, the matter cross-power spectrum becomes a sensitive probe of the density-convergence spectrum. The agreement with \textsc{Class} predictions (which include both MB and RSDs) is excellent at these redshifts, but not at lower ones.
    \item {Weak lensing}: The dominant effect is MB, whose contribution rises from $\sim 5-10\%$ to $\sim 30\%$ between $\ell=200$ and $\ell=5000$. Magnification bias is not included in \textsc{Class}. Since the MB contribution to the shear power spectrum is small, it indicates that the gravitational convergence and shear power spectrum can differ by $\sim 20-30\%$ at large $\ell$. The reason for the lensing  bias is that highly magnified regions are less sampled than de-magnified ones. However, the magnification is directly linked at first order to the convergence. The convergence is more correlated to itself than to the shear, and thus the convergence lensing bias is much larger than the shear lensing bias. This bias can have observational implications for the interpretation of surveys that probe sources of known size or known luminosity.
    \item {Galaxy-galaxy lensing}: The dominant non-trivial contribution to overdensity-convergence cross-spectra is MB, which plays a role in both the overdensity and the convergence. The contribution is of order $-20\%$ in $C_{\delta_1 \kappa_2} (\ell, z_1=0.7, z_2=1.8)$, $\sim -50\%$ in $C_{\delta_1 \kappa_2} (\ell, z_1=0.7, z_2=0.7)$, and $\sim 100\%$ in $C_{\delta_1 \kappa_2} (\ell, z_1=1.8, z_2=1.8)$ and  $C_{\delta_1 \kappa_2} (\ell, z_1=1.8, z_2=0.7)$. The three last cross-spectra are usually neglected in cosmological analysis (and in particular the last one), but their amplitude is not negligible compared to $C_{\delta_1 \kappa_2} (\ell, z_1=0.7, z_2=1.8)$. For thin shells or when the density shell is behind the convergence shell, the density-convergence cross-spectrum becomes a sensitive probe of the gravitational convergence power spectrum. \textsc{Class} captures only the effect of MB on the overdensity. Redshift-space distortions  are usually negligible but may play a role at the $\sim 10\%$ level at low $\ell$ for some specific configurations.
    \item {Redshift evolution}: At low redshift, $z<0.5$, we observe the same trends, although MB plays a smaller role compared to RSDs. Even though we consider angular spectra, fingers-of-God effects are visible at large $\ell$ as a damping of the matter power spectra. These effects are not included in \textsc{Class}. 
    \item {Cosmological dependence}: By changing the cosmology to $w$CDM, we observe a relative variation in the ten cross-spectra. However, the amplitudes of the variations are different for each spectrum. For the specific model we considered, the largest variations are observed for convergence spectra and for the density-convergence spectra dominated by relativistic effects, followed by the relativistic matter-density cross-spectrum and the standard density-convergence spectrum. This indicates that relativistic 3$\times$2 points spectra are powerful cosmological probes.
    \item {Correlation function}: There is a good agreement between the convergence angular correlation function estimated from convergence maps weighted by the observed density (i.e. the inverse transform of the spectra presented in this article) and the direct pair-based estimate from catalogues. The agreement is not observed without density weight nor when removing the monopole and dipole. However, MB plays an important role in all cases (including with a pixel-based estimate without a density weight). 
\end{itemize}

The publicly available catalogues are very generic (they have not been built for a specific community, but instead from first principle). They may help when investigating any N-point cross-correlation between several observables (each being sensitive to a particular combination of relativistic effects). We identify some examples of applications of \textsc{RayGal} data, but the reader can probably find many others. We also point to some illustrative articles on these topics that have already made use of \textsc{RayGal} data:
\begin{itemize}
    \item {Galaxies (weak-lensing studies)}: This work can be extended to shear and ellipticities, more realistic selection functions,  and higher-order statistics.    \item {Galaxies (clustering and RSD studies)}: The impact of relativistic effects on higher-order statistics, the impact on baryon acoustic oscillations, and two-points correlation functions (e.g. \citealt{Breton2019imprints,didio19,taruya2020,didio20,saga2020}) can be studied.\     \item {Galaxy cluster, group, and void studies}: The profile (e.g. \citealt{corasaniti2017probing}), mass function, gravitational redshift, weak lensing, and relativistic cross-correlation.\ 
    \item {Supernovae, standard candles, and standard rulers}: The fluctuation of the cosmic distances (e.g. \citealt{breton2020}).\ 
    \item {CMB studies}: The ISW-RS effect, weak lensing, and  CMB cross-correlation.    \item {Other sources}: Relativistic effects in the Lyman-$\alpha$ forest, in intensity mapping, and so on. 
\end{itemize}

In the future, we plan to expand our source catalogue (to include galaxies and active galactic nuclei), to explore alternative cosmological models (modified gravity, etc.), to extend our work to strong lensing (i.e. by implementing a multiple root finder since our current version finds one image in the case of strong lensing), to consider alternative messengers (gravitational waves), and to investigate the role of baryonic physics. Relativistic ray tracing is a promising tool of modern cosmology: it provides a natural framework for combining multiple cosmological probes and may pave the way to shedding light on the nature of the dark sector.

\begin{acknowledgements}
        We acknowledge financial support from the DIM ACAV of the Région Île-de-France, the Action f\'ed\'eratrice {Cosmologie et structuration de l'univers} as well as the JSPS Grant L16519.  This work was granted access to HPC resources of TGCC/CINES through allocations made by GENCI (Grand \'Equipement National de Calcul Intensif) under the allocations 2016-042287, 2017-A0010402287, 2018-A0030402287, 2019-A0050402287 and 2020-A0070402287. 
        AT acknowledges the support from MEXT/JSPS KAKENHI Grant Nos. JP17H06359, JP20H05861, JP21H01081, and JST AIP Acceleration Research Grant No. JP20317829, Japan. We thank Enea Di Dio for pointing out the "hidden" \textsc{Class} options to tune, S. Prunet for help on \textsc{MPGRAFIC} and \textsc{Polspice}, R. Teyssier for help on \textsc{Ramses}, B. Li for sharing his TSC routine, C. Murray for useful comments on the catalogues, J. Adamek for tips about \textsc{Healpix}. We also thank J.-M. Alimi, S. Colombi, I. Achitouv and A.Le Brun for fruitful discussions. We deeply thank Stéphane Mene for tremendeous efforts in making the local servers working at LUTH. S. A. was supported in part by the project “Combining Cosmic Microwave Background and Large Scale Structure data: An Integrated Approach for Addressing Fundamental Questions in Cosmology,” funded by the MIUR Progetti di Rilevante Interesse Nazionale (PRIN) Bando 2017, Grant No. 2017YJYZAH.
        
\end{acknowledgements}

\bibliographystyle{aa} 
\bibliography{biblio}

\appendix
\section{Power-spectrum expressions under the linear mapping assumption}
\label{appendix:linear_mapping}

Under the linear mapping assumption (which is only used for analytical predictions), the matter overdensity $\delta_i$ of a shell $i$ can be decomposed as
\begin{equation}
    \delta_i \approx \delta_i^{\rm com}+\delta_i^{\rm MB}+\delta_i^{\rm RSD},
\end{equation}
where $\delta_i^{\rm com}$ is the comoving overdensity, $\delta_i^{\rm MB}=\frac{1+\delta^{\rm com}_i}{\mu_i} -(1+\delta^{\rm com}_i) \approx -2 \kappa_i$ is the MB\ contribution dominated by the dilution term ($\mu_i$ is the magnification and we assumed a slope s=0), and $\delta_i^{\rm RSD}$ is the contribution from redshift perturbations (see \citealt{Breton2019imprints} and reference therein). The matter-density auto- and cross-power spectra are given by
\begin{equation}
\label{Eq:Cl_delta_delta_linmap}
\begin{split}
C_{\delta_i \delta_j}(l)&\approx&C_{\delta_i^{\rm com} \delta_j^{\rm com}}(l)\\
   &+& C^{\rm Perm}_{\delta_i^{\rm com} \delta_j^{\rm MB}}(l)+C_{\delta_i^{\rm MB} \delta_j^{\rm MB}}(l)\\
   &+& C^{\rm Perm}_{\delta_i^{\rm com} \delta_j^{\rm RSD}}(l)+  C^{\rm Perm}_{\delta_i^{\rm MB} \delta_j^{\rm RSD}}(l)+C_{\delta_i^{\rm RSD} \delta_j^{\rm RSD}}(l),
   \end{split}
\end{equation}
where $C^{\rm Perm}_{X_i^{\rm A} Y_j^{\rm B}}(l)=C_{X_i^{\rm A} Y_j^{\rm B}}(l)+ C_{X_i^{\rm B} Y_j^{\rm A}}(l) $. The first line is the comoving power spectrum. The second line adds the weak-lensing contribution (i.e. MB with $s=0$). The third line adds the RSD contributions.

The convergence can be approximated at first order as
   \begin{equation}
    \kappa_i \approx \kappa_i^{\rm Born},
\end{equation}
where $\kappa_i^{\rm Born}$ is the convergence according to Born approximation. However, in cosmological analysis, one needs to perform some averaging procedure to extract some statistics. While the above expression is recovered with direction averaging, it is not with source averaging (which is used in the analysis of observational data, in the calculation of power spectra and correlation functions). The average convergence depends on the apparent overdensity and is therefore sensitive to both MB and RSDs.  The average becomes $\langle\kappa_i\rangle=(\Sigma_{\rm sources} (1+\delta_i) \kappa_i) (\Sigma_{\rm sources} (1+\delta_i))^{-1}$ where $\Sigma_{\rm sources}$ is the sum over the sources. As a consequence, the (source) average convergence can be written as
\begin{equation}
    \langle\kappa_i\rangle_s \approx \kappa_i^{\rm Born}+\kappa_i^{\rm MB}+\kappa_i^{\rm RSD},
\end{equation}
where $\kappa_i^{\rm MB}\approx \frac{\kappa_i^{\rm Born}}{\mu_i}-\kappa_i^{\rm Born}+\kappa_i^{\rm PB}\approx -2 (\kappa_i^{\rm Born})^2+\kappa_i^{\rm PB}$ is the MB contribution (assuming $s=0$; see \citealt{schmidt2009}) and $\kappa_i^{\rm RSD}\approx (\Sigma_{\rm sources} (1+\delta_i^{\rm com}+\delta_i^{\rm RSD}) \kappa^{\rm Born}_i) (\Sigma_{\rm sources} (1+\delta_i^{\rm com}+\delta_i^{\rm RSD}))^{-1} -\kappa^{\rm Born}_i$ is the contribution from redshift perturbation. The MB contribution includes the sub-dominant post-Born correction $\kappa_i^{\rm PB}$, which plays a percent level role at large $\ell>2000$. It is dominated by a geometric contribution (i.e. the integral of second derivative of the potential along the true light-ray path) and a lens-lens coupling contribution (see \citealt{petri2017}).

In the following we omit the average to make the equations more readable. The resulting power spectrum for the gravitational convergence can be expressed as
\begin{equation}
\label{Eq:Cl_kappa_kappa_linmap}
\begin{split}
   C_{\kappa_i \kappa_j}(l)&\approx&C_{\kappa_i^{\rm Born} \kappa_j^{\rm Born}}(l)\\
   &+& C^{\rm Perm}_{\kappa_i^{\rm Born} \kappa_j^{\rm MB}}(l)+C_{\kappa_i^{\rm MB} \kappa_j^{\rm MB}}(l)\\
   &+& C^{\rm Perm}_{\kappa_i^{\rm Born} \kappa_j^{\rm RSD}}(l)+ C^{\rm Perm}_{\kappa_i^{\rm MB} \kappa_j^{\rm RSD}}(l)+C_{\kappa_i^{\rm RSD} \kappa_j^{\rm RSD}}(l),
   \end{split}
\end{equation}
where the first line is the convergence spectrum under the Born approximation, the second line adds adds MB\ (assuming $s=0$) and the third line adds the RSD contribution.

Given our decomposition of the overdensity and convergence, we can proceed similarly to evaluate the galaxy-galaxy lensing power spectrum,
\begin{equation}
\label{Eq:Cl_delta_kappa_linmap}
\begin{split}
C_{\delta_i \kappa_j}(l)&\approx&C_{\delta_i^{\rm com} \kappa_j^{\rm Born}}(l)\\
   &+& C_{\delta_i^{\rm com} \kappa_j^{\rm MB}}(l)+C_{\delta_i^{\rm MB} \kappa_j^{\rm Born}}(l)+C_{\delta_i^{\rm MB} \kappa_j^{\rm MB}}(l)\\
   &+& [ C_{\delta_i^{\rm RSD} \kappa_j^{\rm Born}}(l)+C_{\delta_i^{\rm RSD} \kappa_j^{\rm MB}}(l)+C_{\delta_i^{\rm RSD} \kappa_j^{\rm RSD}}(l)+\\
   &&C_{\delta_i^{\rm com} \kappa_j^{\rm RSD}}(l)+C_{\delta_i^{\rm MB} \kappa_j^{\rm RSD}}(l) ],
   \end{split}
\end{equation}
where the first line is the density-convergence spectrum under the comoving and Born approximation, the second line adds MB (assuming $s=0$) and the third line and fourth line (within the square bracket) adds the RSD contribution.

In \textsc{Class}, all the weak-field relativistic effects and wide angle effects are included. All the terms are evaluated from the knowledge of the 3D matter power spectrum. The approximations we identified are the following: the code makes use of the above linear mapping assumption (which is an approximation in particular for $\delta_i^{\rm RSD}$),  the matter power spectrum is evaluated from \textsc{Halofit}, the Born approximation is used, the Doppler contribution to $\delta_i^{\rm RSD}$ is evaluated approximately and the terms $\kappa_i^{\rm MB}$, $\kappa_i^{\rm RSD}$ are not included.

\section{Convergence two-point correlation function}
\label{appendix:2PCF_convergence_z1p8}
In this section we perform configuration-space measurements for the convergence two-point correlation function (the harmonic-space measurements can be found in \cref{subsec:result_kappakappa}). This allows us to test the methodology used in this paper, based on the \textsc{Healpix} maps and angular power spectrum estimators.

To do so, we used the code \textsc{Athena} \citep{schneider2002analysis}.
First, we verified that the configuration-space measurement on the \textsc{Healpix} map containing the convergence using the Born approximation gave the same results as its counterpart using \textsc{Polspice}.

\begin{figure}
   \centering
      \includegraphics[width=\hsize]{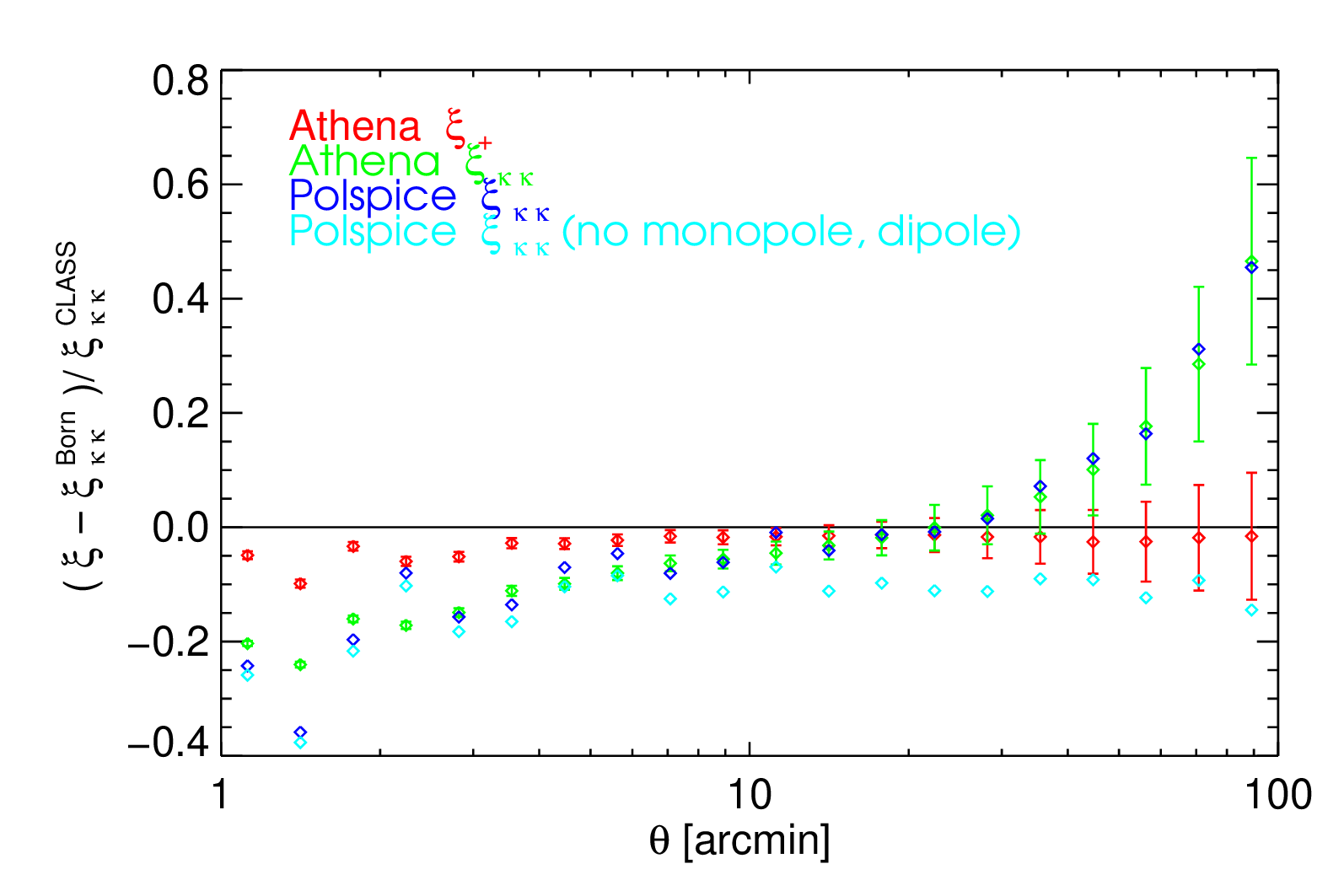}
      \caption{Relative difference between lensing angular two point correlation function on the source catalogue accounting for the dilution bias and the Born convergence angular two point correlation function. In red and green diamonds we show the measurements of cosmic shear and convergence correlation funtion using \textsc{Athena}, and in blue and light blue we show the results using the same methodology as in \cref{sec:results_application_3x2pt}, keeping and removing the monopole and dipole, respectively.}
         \label{Fig:2PCF_convergence_z1p8}
   \end{figure}

In \cref{Fig:2PCF_convergence_z1p8}, we show the relative difference between the measurements made on the catalogue containing the dilution bias with respect to Born at $z = 1.8$.
We see that the measurements from \textsc{Athena} and \textsc{Polspice} agree extremely well, and yield a $\sim 20\%$ difference with respect to Born at the arcminute scale and $\sim 50\%$ at 100 arcmin. This confirms that the dilution bias has a strong impact on the convergence two-point statistics as in \cref{Fig:cl_lcdmw7_narrow_00002_kappa_kappa}, where we used the same \textsc{Polspice} measurement. We note that had we used \textsc{Polspice} and subtract the average and dipole to the data (which is common practice), then the results would vastly differ, especially at larger angular scales. This most likely comes from the fact that when averaging over sources (which differs from averaging over directions; \citealt{kibble2005average, bonvin2015cosmological,breton2020}), $\langle\kappa\rangle = -2\langle\kappa^2\rangle \neq 0$. This effect, in the form of $\langle\kappa\rangle\langle\kappa\rangle$ gains a relative importance as the signal decreases with angular scale, hence our results. Moreover we expect some leakage from small $\ell$ to large and intermediate angles. It is interesting to see that there is already a $\sim 10\%$ discrepancy at 10 arcmin.

Finally, we show the relative difference when computing the {cosmic shear}. We see that in this case, the results are very close to the Born approximation. This shows that the very large effect of lensing bias mostly impacts the convergence, and only to a lesser the extent cosmic shear. This comes from the fact that the source-averaged convergence is non-zero (while the averaged shear is, neglecting intrinsic alignment) and the convergence three-point correlation function contribution is large (i.e. $\langle\kappa \kappa \kappa\rangle/\langle\kappa \kappa\rangle$ is larger than $ \langle\gamma \gamma \kappa\rangle/\langle\gamma \gamma\rangle$).

\end{document}